\documentclass[preprintnumbers, floatfix, letterpaper, twocolumn,aps,prd,epsfig,nofootinbib,natbib,
longbibliography
]{revtex4-1}
\usepackage{bm,graphicx,dcolumn,epstopdf,epsf, latexsym,mathbbol, amssymb,amsmath,color,slashed, mathrsfs,mathcomp,simplewick}
\usepackage[center]{subfigure}
\usepackage{multirow}
\usepackage{makecell}
\usepackage[colorlinks,linkcolor=blue,citecolor=blue,urlcolor=blue]{hyperref}

\begin{document}
\allowdisplaybreaks
 \newcommand{\bq}{\begin{equation}}
 \newcommand{\eq}{\end{equation}}
 \newcommand{\bqn}{\begin{eqnarray}}
 \newcommand{\eqn}{\end{eqnarray}}
 \newcommand{\nb}{\nonumber}
 \newcommand{\lb}{\label}
\newcommand{\f}{\frac}
\newcommand{\p}{\partial}
\newcommand{\PRL}{Phys. Rev. Lett.}
\newcommand{\PLB}{Phys. Lett. B}
\newcommand{\PRD}{Phys. Rev. D}
\newcommand{\CQG}{Class. Quantum Grav.}
\newcommand{\JCAP}{J. Cosmol. Astropart. Phys.}
\newcommand{\JHEP}{J. High. Energy. Phys.}
\title{Pre-inflationary universe  in loop quantum cosmology}

\author{Tao Zhu, Anzhong Wang}
\affiliation{Institute for Advanced Physics $\&$ Mathematics, Zhejiang University of Technology, Hangzhou, 310032, China\\
GCAP-CASPER, Physics Department, Baylor University, Waco, TX 76798-7316, USA}

\author{Gerald Cleaver}
\affiliation{EUCOS-CASPER, Physics Department, Baylor University, Waco, TX 76798-7316, USA}

\author{Klaus Kirsten and Qin Sheng}
\affiliation{GCAP-CASPER, Mathematics Department, Baylor University, Waco, TX 76798-7328, USA}

\date{\today}

\begin{abstract}

The evolutions of the flat FLRW universe and its linear perturbations are studied systematically  in  the dressed metric approach of LQC. When  it 
is dominated by the kinetic energy of the inflaton at the quantum bounce, the evolution of  the background can be divided into three different phases  
prior to the preheating, {\em bouncing, transition and slow-roll inflation}. During the bouncing phase,  the evolution is  independent of not only  the 
initial  conditions, but also the inflationary potentials. In particular, the expansion factor can be well described by the same exact solution in all the 
cases considered. In contrast, in the potential dominated case  such a universality is lost. It is because of this universality that  the linear perturbations 
are also  independent of the inflationary models and  obtained exactly. During the transition phase,   the evolutions of the background and its linear
perturbations are  found explicitly, and then matched to the ones given in the other two phases. Hence, once the initial conditions are imposed, the
linear scalar and tensor perturbations will be uniquely determined. Considering two different sets of  initial conditions,  one imposed during the 
contracting phase and the other  at the bounce, we calculate the Bogoliubov coefficients and find that the two sets yield the same results and all lead 
to  particle  creations at the onset of the inflation. Due to the pre-inflationary dynamics,  the scalar and tensor power spectra become scale-dependent. 
Comparing  with the Planck 2015  data, we find constraints on the total e-folds  that the universe must have expanded since the bounce, in order to be 
consistent with current observations.

\end{abstract}


\maketitle

\section{Introduction}
\renewcommand{\theequation}{1.\arabic{equation}} \setcounter{equation}{0}

The inflationary paradigm   not only  solves elegantly the problems of the standard big bang
cosmology, but also predicts the primordial power spectra whose evolutions determine the temperature fluctuations in cosmic microwave background
(CMB) and the formation of the large-scale structure of the universe \cite{guth_inflationary_1981, sato_first-order_1981, starobinsky_new_1980}
(see \cite{baumann_tasi_2009} for an updated review). This prediction explains the power spectrum of the galaxy distribution, and has been remarkably
confirmed by  CMB measurements with unprecedented precisions \cite{komatsu_seven-year_2011, planck_collaboration_planck_2014-1,planck_collaboration_planck_2015-4}.

However, the inflationary scenario is sensitive to the ultraviolet (UV) physics, and its successes are tightly contingent on the understanding of such UV physics \cite{baumann_tasi_2009}.
In particular,   the underlying quantum field theory on a classical spacetime becomes questionable  for a large class of inflationary
models, in which  the e-folds of the expansion of the universe  are more than 70 \cite{MRV}. This is because  in these models  the sizes of the current universe are less than that of Planck  at the onset of inflation,
then the treatment of the spacetime as classical becomes invalid. This is the well-known trans-Planckian problem \cite{martin_trans-planckian_2001, brandenberger_trans-planckian_2013}. In addition,
general relativity (GR) inevitably leads  to an initial singularity \cite{borde_eternal_1994, borde_inflationary_2003}, with which it is not clear how to impose the initial conditions. Instead,
one usually ignores the pre-inflationary dynamics and sets the Bunch-Davies (BD) vacuum at the time when the perturbation modes are inside the Hubble horizon during inflation. However, from
the beginning of inflation (which is normally believed to start at the energy scale about  $10^{16}$ GeV) to the Planckian scale, an energy gap of at least three-order exists. Once this
pre-inflationary dynamics  is taken into account, it is not clear how such a picture will be altered, as it is quite reasonable to expect that particles are created generically during the pre-inflationary phases, and
non-BD states could be created even when the modes were well inside the Hubble horizon during inflation.

To address these important issues,  loop quantum cosmology (LQC) provides an interesting framework, in which the big bang singularity is simply replaced by a quantum bounce in the deep Planck era, due to
 the quantum gravitational effects \cite{bounce,MB15}, and a large number of cosmological models  has been investigated
 \cite{bojowald_absence_2001, ashtekar_quantum_2006-2, ashtekar_quantum_2006, ashtekar_quantum_2006-1, ashtekar_robustness_2008,yang_alternative_2009,AKL09,
 agullo_quantum_2012,agullo_extension_2013,agullo_pre-inflationary_2013,%
 BHKS09,MCBG12,CMBG12,CBGV12,CLB14,BBCGK15,Bojowald2011a,Bojowald2011b,Zhu1,Zhu2,Zhu3,bonga_inflation_2016,bonga_phenomenological_2016}, including the flat  Friedmann-Lema\v{\i}tre-Robertson-Walker (FLRW) universe,
 the model  that we shall  focus on in this paper.
In such a framework,  the universe that was dominated by the kinetic energy of the inflaton at the bounce can eventually evolve to the desired slow-roll inflation
\cite{ashtekar_loop_2010, ashtekar_probability_2011, singh_nonsingular_2006, mielczarek_inflation_2010, zhang_inflationary_2007, chen_loop_2015, bolliet_comparison_2015, schander_primordial_2016,AS17}.

An important question now is whether the quantum bounce and its subsequent pre-inflationary dynamics can leave any observational signatures to the current and/or forth-coming experiments, so LQC can be placed directly under tests.
Such considerations have attracted a great deal of attention lately, see,  for example, the special issue of International Journal of Modern Physics D on {\em Loop Quantum Cosmology}  \cite{PS16}, and the  monograph of World Scientific,
{\em Loop Quantum Gravity: The First 30 Years} \cite{AP17}.  A crucial step for connecting the quantum bounce with observations is the understanding of  the evolutions of the background as well as the cosmological perturbations during the
pre-inflationary period.  Extensive studies have been carried out, but mainly numerical \cite{PS16,AP17}. In this paper, one of our goals is to study the quantum bounce and its subsequent pre-inflationary dynamics {\em analytically}, with the hope that
 it will provide deeper insights into the physics involved.

 The studies of the pre-inflationary dynamics  in the framework of LQC have been carried out by following mainly two different approaches \cite{barrau_conceptual_2016},  {\em  dressed metric}
 \cite{AKL09,agullo_quantum_2012,agullo_extension_2013,agullo_pre-inflationary_2013} and
  {\em deformed algebra} \cite{BHKS09,MCBG12,CMBG12,CBGV12,CLB14,BBCGK15} \footnote{Other approaches,  such as separate universe \cite{EWE16}, hybrid models \cite{NMBM16}
  and consistent histories \cite{Craig16},  are still in their developments.}. In the latter, with some reasonable assumptions and the choice of initial conditions, it has been already shown
  that  the resulting cosmological perturbations are in conflict with current observations \cite{bolliet_observational_2016,Grain16}. In the former, among other things,
it was argued that  the pre-inflationary  effects could produce a way to relieve the tension between the standard spectra obtained in GR and observations at large scales \cite{ashtekar_quantum_2017}.

In this paper, we shall  provide a systematical  study of the effects of the quantum bounce and its subsequent pre-inflationary dynamics on the background evolution and
primordial perturbations in the framework of  {\em  the dressed metric approach}
\cite{AKL09,agullo_quantum_2012, agullo_extension_2013, agullo_pre-inflationary_2013}. Earlier works on the subjects  are mainly  numerical
\cite{agullo_quantum_2012, agullo_pre-inflationary_2013, agullo_detailed_2015, bonga_inflation_2016, bonga_phenomenological_2016, ashtekar_quantum_2017}, and often requires time and memory intensive
computations by using high-performance computing,  in order  to explore the most interesting region of the parameter space \cite{agullo_detailed_2015}.
Our purpose in this paper  is two-fold: First, we shall carry out an {\em analytical} investigation on the evolutions of both the background of the universe and its linear scalar and tensor perturbations.
Second, we shall focus mainly on universal properties of the
evolutions during the pre-inflationary period. That is, properties that do not depend on the inflationary potentials, so that they hold for any inflationary models.
This is tightly related to the previous results that   the universe that was dominated at the quantum bounce by the kinetic energy of the inflaton will eventually evolve to a desired slow-roll inflation
\cite{ashtekar_loop_2010, ashtekar_probability_2011, singh_nonsingular_2006, mielczarek_inflation_2010, zhang_inflationary_2007, chen_loop_2015, bolliet_comparison_2015, schander_primordial_2016}.

Based on the above observations, in this paper we shall mainly consider the models that are dominated at the quantum bounce by the  kinetic energy of the inflaton. In particular, we shall show that under this assumption
not only the evolution of the background of the  universe is universal, but also the evolutions of its linear perturbations during the pre-inflationary period. We also study models that are dominated at the bounce
by the potential of the inflaton, and show explicitly that such universalities are lost.  In particular, the rest of the paper is organized  as follows. In Sec. \ref{numerical} and \ref{analytical}, we present a detailed analysis of the
background evolution first  numerically (Sec. II) and then analytically (Sec. III), and show that {\em the evolution of the background is universal and independent of the form of the inflationary potentials, as long as it is
dominated by the kinetic energy of the inflaton at the quantum bounce}.  Based on the understanding of the background evolution, in Sec. \ref{general_solution} we turn to study
the cosmological scalar and tensor perturbations. In particular, we show that the effective potentials of the perturbations of both scalar and tensor can be mimicked very well by a P\"{o}schl-Teller (PT) potential during the bouncing phase, whereby
we obtain analytically the mode functions of the perturbations. In Sec. \ref{initial_conditions},    we consider two different sets of  initial
conditions for the cosmological perturbations, one is  {\em the Bunch-Davies (BD) vacuum imposed at the contracting phase right before the quantum bounce}, and  the other is {\em the fourth-order adiabatic vacuum state imposed
at the bounce}.  By using our analytical solution we show explicitly that the BD vacuum state imposed at the contracting phase reduces to the fourth-order adiabatic vacuum state at the bounce. In addition, with these initial
conditions, we also derive explicitly the analytical expressions of primordial power spectra and discuss the associated bouncing and pre-inflationary effects. In Sec. \ref{MCMC}, we perform the CosmoMC code to study
cosmological constraints by using the Planck2015 data \cite{planck_collaboration_planck_2015-4}. Our main conclusions and discussions  are presented in Sec. \ref{conclusions}.
Two appendices are also included. Part of the results have been already  reported in \cite{tao_zhu_universal_2016}.

\section{Background Evolution: Numerical}
\renewcommand{\theequation}{2.\arabic{equation}} \setcounter{equation}{0}
\lb{numerical}

In this section, let us begin to consider the evolution of the flat FLRW background coupled with a single scalar field $\phi$, the inflaton,  in the framework of LQC,
\bqn
ds^2=-dt^2 +a(t)\delta_{ij}dx^i dx^j,
\eqn
where $a(t)$ is the cosmological scale factor with $t$ being the cosmic time. In LQC, the FRLW spacetime can be quantized by using the canonical quantization framework of loop quantum gravity. The quantized background spacetime together with the scalar field $\phi$, is described by a quantum state $\Psi_0(a, \phi)$ which is a complex function of the scale factor $a(t)$ and scalar field $\phi$. The evolution of this quantum state is governed by the LQC quantum Hamiltonian constraint, and a remarkable feature is that it is nonsingular. Among many states $\Psi_{0}(a, \phi)$ in the LQC Hilbert space, one is in general interested in a state that is sharply peaked around a classical trajectory at late times, when the curvature of the Universe is well below the Planck scale and the classical GR is an excellent  approximation \cite{ashtekar_quantum_2006-2, ashtekar_quantum_2006, ashtekar_quantum_2006-1, ashtekar_robustness_2008}. Evolving this state by using the LQC quantum Hamiltonian constraint, it has been shown that it remains sharply peaked during the whole dynamical trajectory, even in the deep Planck era \cite{ashtekar_quantum_2006, ashtekar_quantum_2006-1}. As a result, the evolution of the peak of such states can be accurately described by an effective trajectory that governed by its effective equations.

These effective equations for a flat FLRW background has been derived in Refs.
\cite{taveras_corrections_2008, bojowald_effective_2006, bojowald_effective_2009} (see also \cite{yang_alternative_2009} for an alternative approach), from which the modified Friedmann equation takes the form
\bqn
\lb{friedmann}
H^2=\frac{8\pi}{3m_{\text{Pl}}^2}\rho\left(1-\frac{\rho}{\rho_\text{c}}\right),
\eqn
where $H\equiv \dot a/a$ denotes the Hubble parameter and the dot represents the derivative with respect to the cosmic time $t$, $m_\text{Pl}=1/\sqrt{G}$ is the Planck mass,
$\rho$ is the energy density of the universe, and $\rho_\text{c}$ is the critical energy density which represents the maximum value of the energy density in LQC and is about $\rho_\text{c} \simeq 0.41 m_\text{Pl}^4$.
For a single scalar field $\phi$ with a potential $V(\phi)$ in the FLRW background, the effective equation of motion in LQC takes the same form of the Klein-Gordon equation as in GR,
\bqn
\lb{klein-gordon}
\ddot \phi +3 H \dot \phi +V_{,\phi}=0,
\eqn
where $V_{,\phi}= dV(\phi)/d\phi$.

A robust prediction of the above effective dynamics is the occurrence of a non-singular quantum bounce, which removed the initial singularity in the early stage of the classical universe
(see \cite{bojowald_absence_2001, ashtekar_quantum_2006-2, ashtekar_quantum_2006, ashtekar_quantum_2006-1, ashtekar_robustness_2008, yang_alternative_2009}
and references therein). Eq.~(\ref{friedmann}) shows that the quantum bounce occurs at $\rho=\rho_\text{c}$, where the energy density reaches the maximum value and the
Hubble parameter becomes zero. The background evolution with a bounce phase has been extensively studied, and one of the main results is that, right following the quantum bounce,
a desired slow-roll inflation phase is almost inevitable   \cite{bounce, singh_nonsingular_2006, mielczarek_inflation_2010,%
zhang_inflationary_2007, chen_loop_2015} (for recent considerations, see \cite{ashtekar_quantum_2017}).

It is remarkable to note that the modified Friedmann equation (\ref{friedmann}) and the Klein-Gordon equation (\ref{klein-gordon}) are also derived in {\em the deformed algebra approach} 
\cite{BHKS09,MCBG12,CMBG12,CBGV12,CLB14,BBCGK15}. Hence, all the results obtained in this section as well as the results obtained  in the next section  will be equally applicable to this approach, too. 

 In this section, we will study the ``bounce plus slow-roll inflation" scenario
by considering two typical inflationary potentials, the power-law potential and Starobinsky potential, as specified below:

\begin{itemize}
\item {\em Inflation with a power-law potential.} A power-law potential takes the form
\bqn
\lb{pl_potential}
V(\phi)=\frac{1}{2}m^{4-n}\phi^{n},
\eqn
where the parameter $m$ has dimension of mass. We consider two specific values of $n$: $n=2$ and $n=1/2$, respectively. The corresponding values of the mass for each potential used for numerical calculations are set to
\bqn
m=
\begin{cases}
1.3\times 10^{-6} m_\text{Pl}, & \;n=2, \\
7.4\times 10^{-4} m_\text{Pl}, & \; n=\frac{1}{2}.
\end{cases}
\eqn
Note that these values are chosen to be consistent with Planck 2015 data \cite{planck_collaboration_planck_2015-4}.

\item {\em $R^2$-inflation.} This is also known as the Starobinsky inflation, whose potential has the form
\bqn
\lb{SP}
V(\phi)=\frac{3}{32\pi}M^2 m_{\text{Pl}}^2 \left(1-e^{-\sqrt{\frac{16\pi}{3}}\frac{ \phi}{m_{\text{Pl}}}}\right)^2,
\eqn
where the parameter $M$ has dimension of mass, whose value used for numerical calculations is set to \cite{bonga_phenomenological_2016, planck_collaboration_planck_2015-4}
\bqn
\lb{MV}
M=2.51 \times 10^{-6} m_\text{Pl},
\eqn
to be consistent with Planck 2015 data. 
\end{itemize}

In this paper, we find that it is also convenient to use the conformal time,
\bqn
\eta=\int_{t_\text{end}}^{t} \frac{dt'}{a(t')},
\eqn
so that at the end of the inflation $t=t_{\text{end}}$ and at the bounce $t_\text{B}$, the corresponding conformal times are, respectively, given by
\bqn
\eta_{\text{end}}=0,\;\;\;\;\;\eta_{\text{B}}=\int^{t_\text{B}}_{t_{\text{end}}}\frac{dt'}{a(t')}.
\eqn

Let us first study the background evolution numerically for different inflationary potentials.  Eqs.~(\ref{friedmann}) and (\ref{klein-gordon})
can be solved numerically by imposing the initial conditions for $a(t)$, $\phi (t)$, and $\dot \phi(t)$ at a specific point.
A convenient choice of such a point is the bounce $t=t_\text{B}$,  at which we have the relations
\bqn
\frac{1}{2} \dot \phi^2(t_{\text{B}})+V(\phi(t_{\text{B}}))=\rho_\text{c}, \;\;\text{and}\;\;\; \dot a(t_{\text{B}})=0.
\eqn
Thus, if we consider $\rho_c$ as a given constant, using the first equation we can write $\dot{\phi}_B$ in terms of $\rho_c$ and $\phi_B$, once a potential $V(\phi)$ is specified. Therefore, now we only need to specify
$a(t_{\text{B}})$ and $\phi(t_\text{B})$ as the  initial conditions. For the sake of simplicity, we further rescale $a(t)$ by setting $a(t_\text{B})=1$ at the bounce. Then,
 the initial conditions finally reduce to specifying the value of   $\phi(t_\text{B})$ only. In the following, we shall consider the two classes $\dot \phi_\text{B}>0$ and $\dot \phi_\text{B}<0$, separately.

We shall pay particular attention to two important issues, namely how likely the occurrence of the slow-roll inflation is, and whether enough $e$-folds can be generated during the slow-roll inflation.
For these purposes, let us first introduce the following background quantities:

\begin{itemize}

\item[(1)] The equation of state $w(\phi)$, which is defined by
\bqn
\lb{EOS}
w(\phi)\equiv \frac{\dot \phi^2/2-V(\phi)}{\dot \phi^2/2+V(\phi)}.
\eqn
During the slow-roll inflation, $w(\phi)$ has to be very close to $-1$.

\item[(2)] The slow-roll parameter $\epsilon_H$,  which is defined by the derivative of the Hubble parameter,
\bqn
\lb{epsilonH}
\epsilon_{H}\equiv -\frac{\dot H}{H^2}.
\eqn
During the slow-roll inflation, $\epsilon_H$ is required to be very small, i.e., $\epsilon_H\ll 1$.

\item[(3)] The $e$-folds of the slow-roll inflation $N_\text{inf}$, which is defined as the $e$-folds between the onset of the slow-roll inflation until the end of it,
\bqn
N_\text{inf}\equiv \ln\left(\frac{a_\text{end}}{a_i}\right).
\eqn
In this paper, the onset of the inflation is defined by the time when the universe begins to accelerate, $\ddot a(t_i) = 0$, i.e., $\ddot a(t)$ first changes its sign right after the bouncing phase.
The end of the inflation is defined by the time when the accelerating expansion of the universe stops, that is, $w(\phi_\text{end})=-1/3$.

\end{itemize}

In the following, we shall study   the background evolution for each of the two potentials mentioned above  separately.

\subsection{Quadratic potential}

\begin{figure}
{\label{scalar_quadratic}
\includegraphics[width=8.1cm]{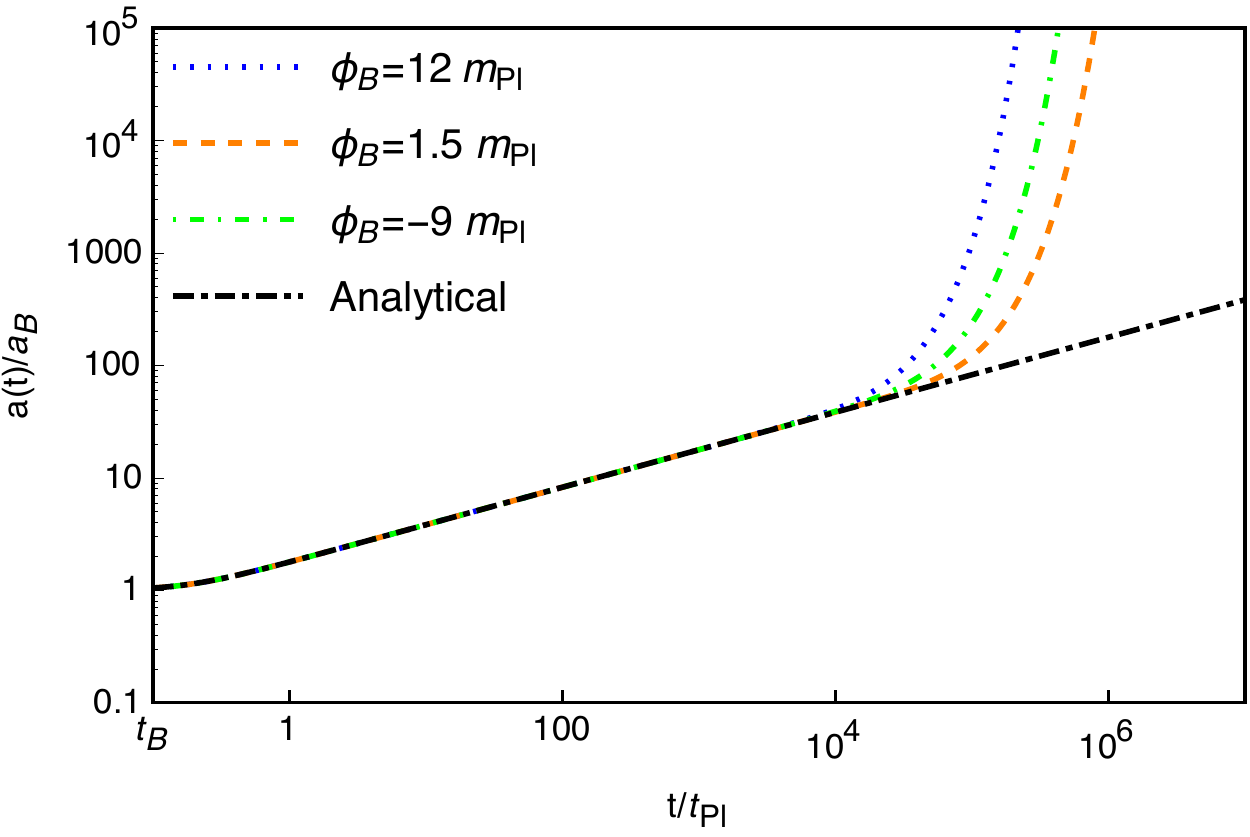}}\\
{\label{wphi_quadratic}
\includegraphics[width=8.1cm]{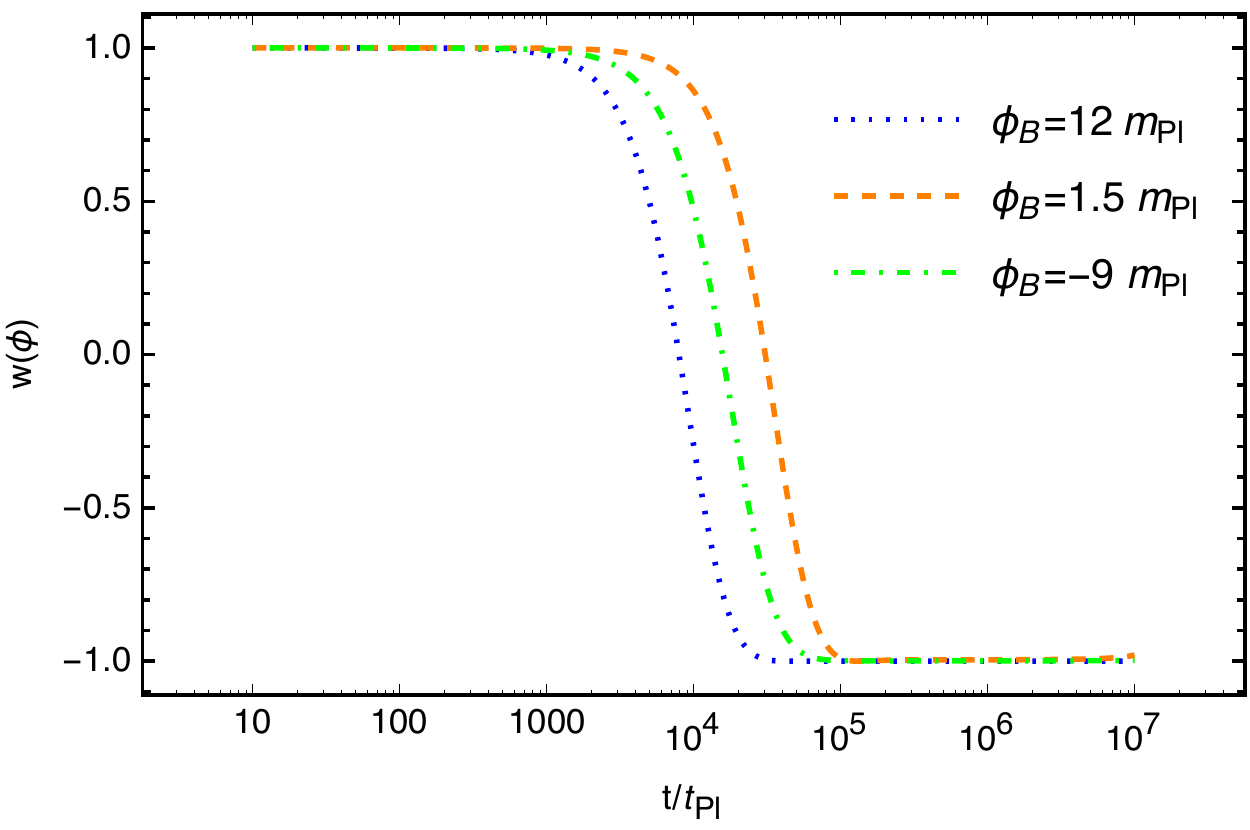}}\\
{\label{eH_quadratic}
\includegraphics[width=8.1cm]{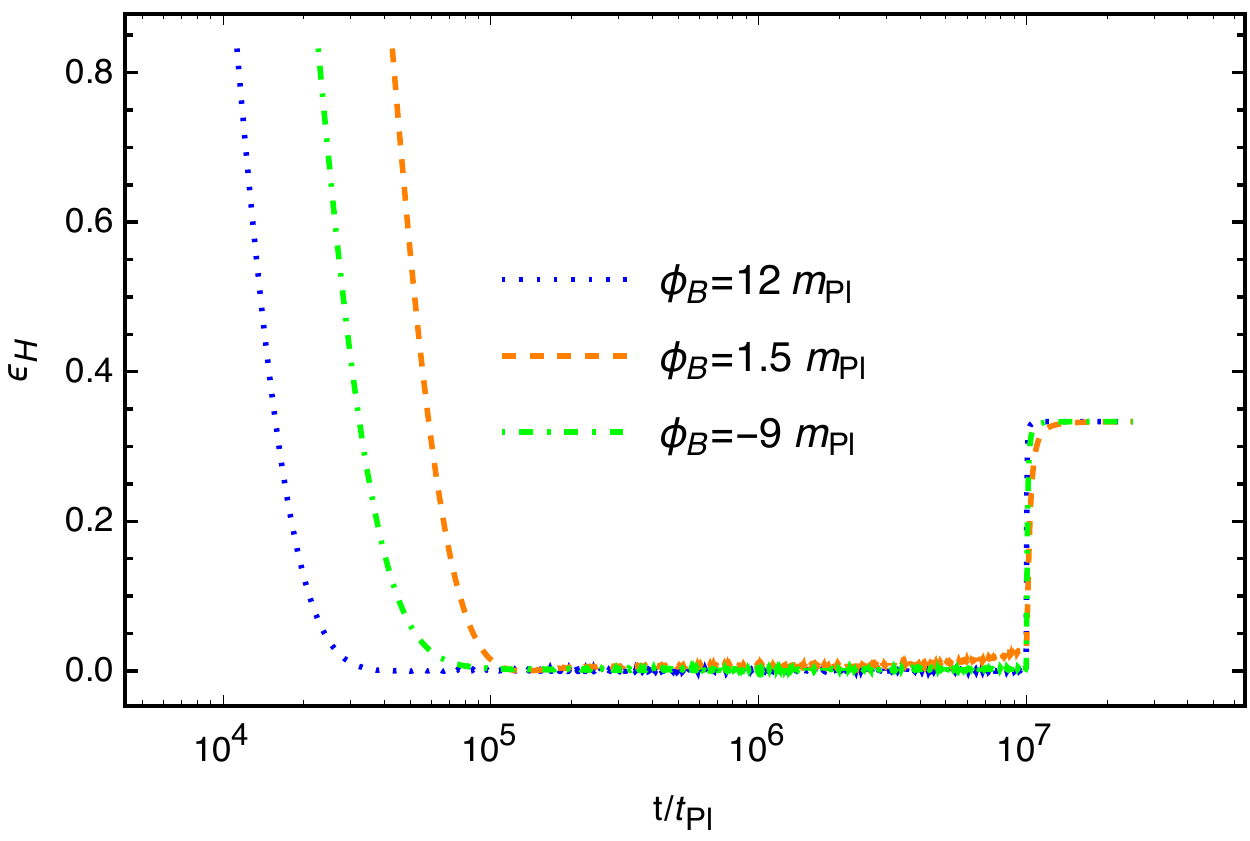}}\\
\caption{Numerical solution for the quadratic potential with kinetic energy dominated initial conditions at the bounce and $\dot\phi_\text{B}>0$. Top panel: the evolution of the scale factor $a(t)$ for different choices of the initial
conditions.  The analytical solution given by Eq.~(\ref{scalar_analytical}) is also shown in order to compare it with the numerical ones. Middle panel: the equation of state $w(\phi)$ for the same set of initial conditions.
Bottom panel: the slow-roll parameter $\epsilon_\text{H}$ during the transition and slow-roll inflationary phases.}
\label{quadratic_kinetic}
\end{figure}

 \begin{figure}
{\label{VK_quadratic_1}
\includegraphics[width=8.1cm]{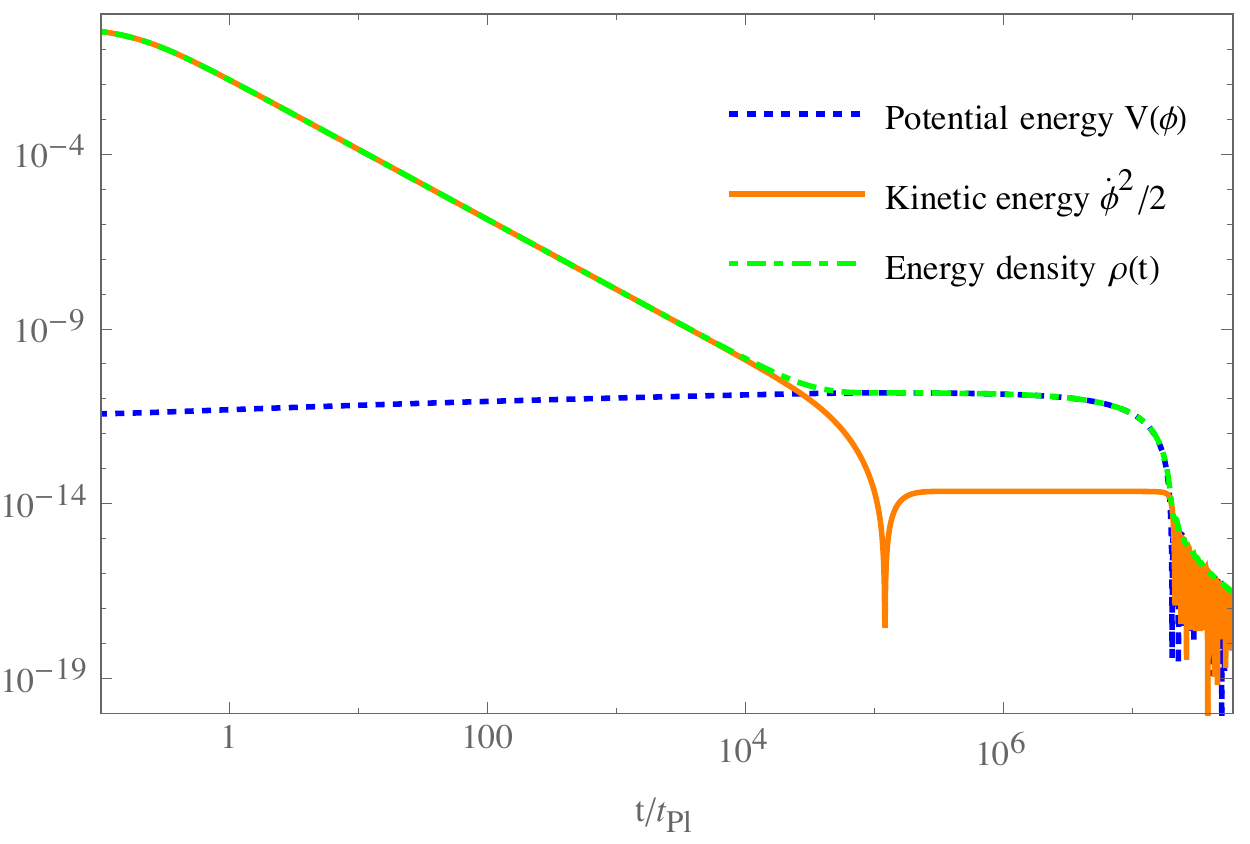}}\\
{\label{VK_quadratic_1}
\includegraphics[width=8.1cm]{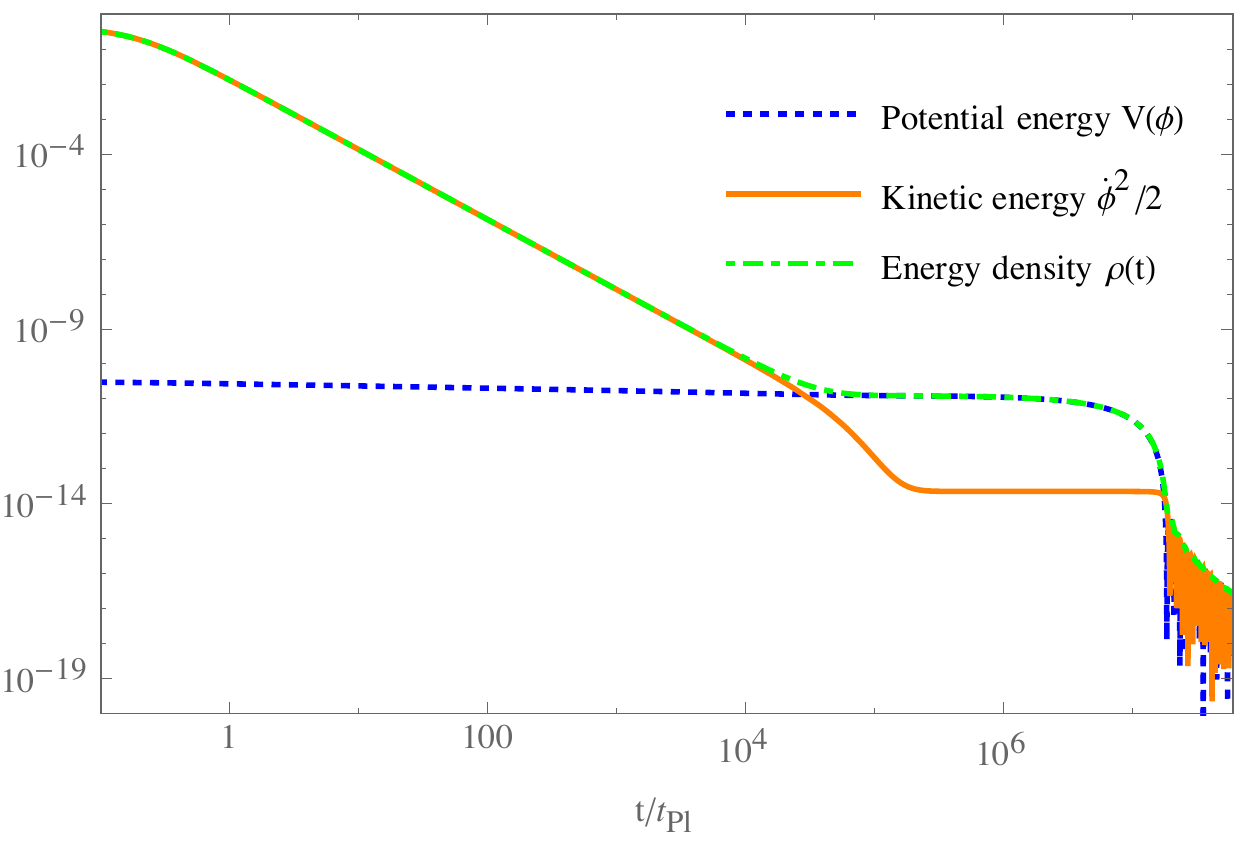}}
\caption{Comparison between the potential energy $V(\phi)$ and the kinetic energy $\dot \phi^2 /2$ for the quadratic potential.  The energy density $\rho = \dot \phi ^2 /2 + V(\phi)$ is also shown.
Top panel: for the initial condition $\phi_\text{B}=2 m_\text{Pl}$ with $\dot \phi_\text{B}>0$. Bottom panel: for the initial condition $\phi_\text{B}=6 m_\text{Pl}$ with $\dot \phi_\text{B} <0$. } \label{VK_quadratic}
\end{figure}

Let us begin by discussing the evolution of the background with the quadratic potential (i.e., Eq.~(\ref{pl_potential}) with $n=2$), which has already been discussed in detail in
Refs.~\cite{agullo_pre-inflationary_2013, agullo_detailed_2015}. Here we summarize some main results.

As the initial conditions for the quadratic potential at the bounce have the symmetry $(\phi_\text{B},\dot \phi_\text{B})\to (-\phi_\text{B}, -\dot \phi_\text{B})$, in this subsection we only need
to consider the case $\dot \phi_\text{B}>0$, and the results can be easily extended to the case $\dot \phi_\text{B}<0$ by using the above symmetry. We can further divide the initial conditions
into two subclasses, the kinetic energy dominated  and the potential energy dominated cases at the quantum bounce.

The background evolution for a set of kinetic energy dominated initial conditions is illustrated in Fig.~\ref{quadratic_kinetic}, in which the scale factor $a(t)$, the equation of state $w(\phi)$,
and the slow-roll parameters $\epsilon_H$ are all obtained numerically
 for the same set of the initial values of $\phi_\text{B}$. It is shown clearly that the desired slow-roll inflationary phase for these initial conditions is achieved. During  this phase,
    the scale factor is exponentially growing (c.f. the top panel of Fig.~\ref{quadratic_kinetic}), and $w(\phi)$ is very close to $-1$ (c.f. the middle panel of Fig.~\ref{quadratic_kinetic}),
    while the parameter $\epsilon_H \ll 1$ (c.f. the bottom panel of Fig.~\ref{quadratic_kinetic}). For initial conditions with $\dot \phi_\text{B}<0$, the replacement   $\phi_\text{B}$ by
 $ -\phi_\text{B}$ [so that now $\phi_\text{B}/m_\text{Pl}\in (-12, -1.5, 9)$] shall yield the same results.

 From the curves of the equation of state $w(\phi)$ [the middle panel of Fig. \ref{quadratic_kinetic})],  we can see clearly that the evolution of the universe before preheating can be divided into three different phases,
 {\em the bouncing, transition and slow-roll inflation}. During the bouncing
 phase, the kinetic energy of the inflaton is dominant, and $w(\phi) \simeq +1$.  At $t/t_\text{Pl} \simeq 10^4$, $w(\phi)$   suddenly decreases  from  $w(\phi) \simeq +1$   to  $w(\phi) \simeq -1$ at
 $t/t_\text{Pl} \simeq 10^5$. Comparing with the other two
 phases, this transition phase is rather short. Afterward, $w(\phi)$ remains $w(\phi) \simeq -1$ until the end of the slow-roll inflation.  It is remarkable to note that the evolution of the expansion
 factor $a(t)$ during the bouncing phase is independent of the choices of the initial
 values of $\phi_B$, and can be well described by the analytical solution given by Eq.~(\ref{scalar_analytical}) below.

In Fig. \ref{VK_quadratic}, the kinetic and potential energies, as well as the energy density of the inflationary field $\phi$, are illustrated for   both $\dot \phi_\text{B} >0$ and $\dot \phi_\text{B} <0$.
A remarkable feature is that the potential energy remains almost
the same during the three different  phases, while the kinetic energy starts at about the Planckian energy at the bounce and then drops about 12-orders before the slow-roll inflation starts,
whereby the potential energy starts to dominate the evolution of
the universe.

\begin{figure}
{\label{Ninf_quadratic}
\includegraphics[width=8.1cm]{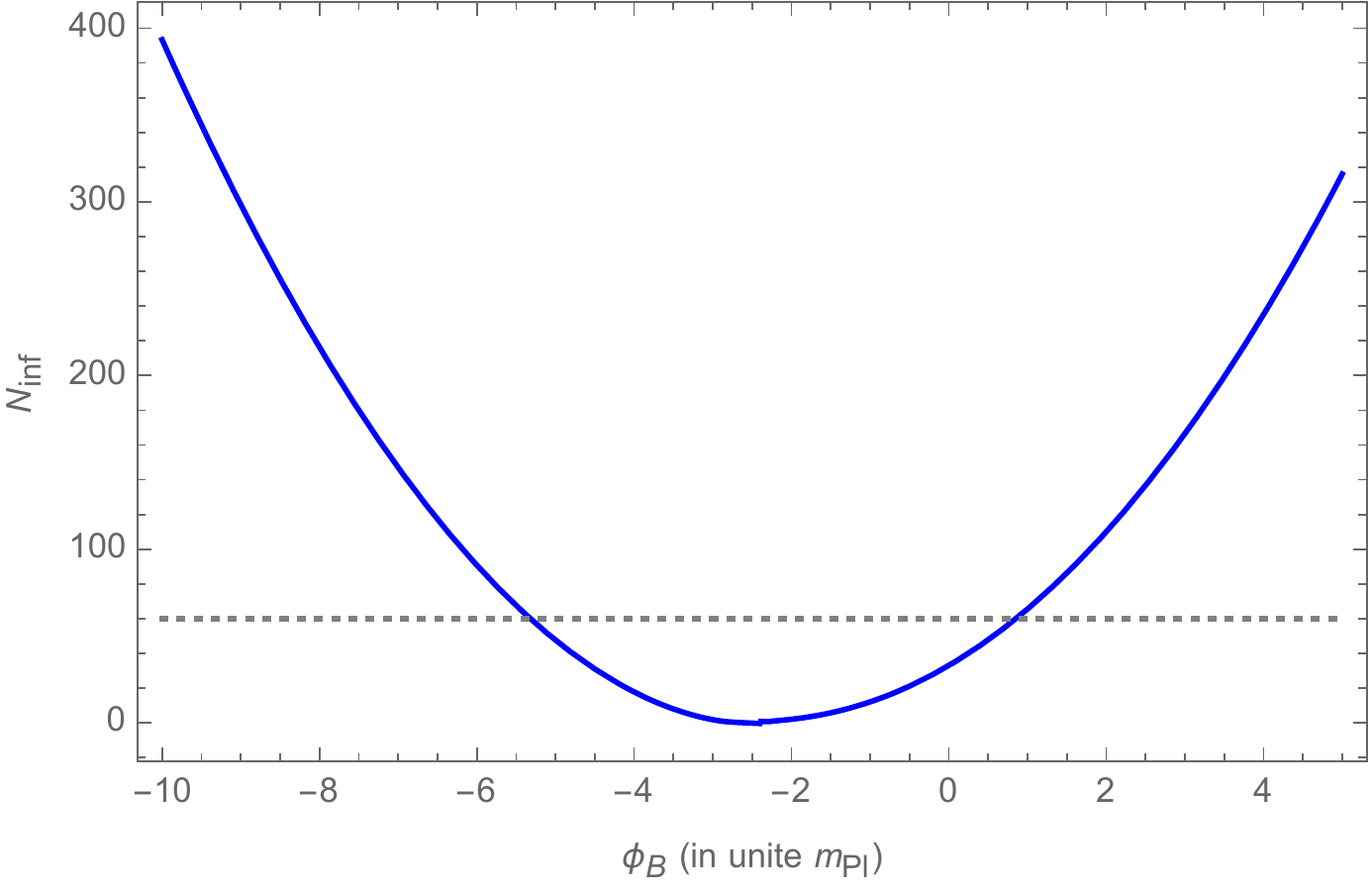}}\\
\caption{The e-folds $N_\text{inf}$ during the slow-roll inflationary phase for different choices of  the initial condition $\phi_\text{B}$ for the case $\dot \phi_\text{B}>0$.
The gray dotted line represents the minimum value of e-folds ($N_*=60$), in order to be consistent with observations.} \label{Ninf_quadratic}
\end{figure}

The corresponding  e-folds $N_\text{inf}$ during the slow-roll inflation as a function of $\phi_\text{B}$ is illustrated in Fig.~\ref{Ninf_quadratic}. In order to produce at least $60$ e-folds during the slow-roll inflation,  Fig.~\ref{Ninf_quadratic}
shows clearly  that one has to require
\bqn
\phi_\text{B} \in (-\phi_\text{max}, -5.3 m_\text{Pl}) \cup (0.83 m_\text{Pl}, \phi_\text{max}),
\eqn
for $\dot \phi_\text{B}>0$, where $\phi_\text{max}=\sqrt{2\rho_\text{c}/m}$. For the initial conditions with $\dot \phi_\text{B}<0$, using the symmetry $(\phi_\text{B}, \dot \phi_\text{B}) \to (- \phi_\text{B}, -\dot \phi_\text{B})$, one gets the constraints
\bqn
\phi_\text{B} \in (-\phi_\text{max}, -0.83) \cup (5.3 m_\text{Pl}, \phi_\text{max}).
\eqn
The e-folds $N_\text{inf}$ increases when the absolute values of $\phi_\text{B}$ is increasing, which implies that a larger value of the potential energy at the bounce can produce more e-folds during the slow-roll inflation than a smaller one.
Note that similar results  were already obtained in \cite{agullo_pre-inflationary_2013, agullo_detailed_2015}.

\begin{figure}
{\label{scalar_quadratic_Vdominated}
\includegraphics[width=8.1cm]{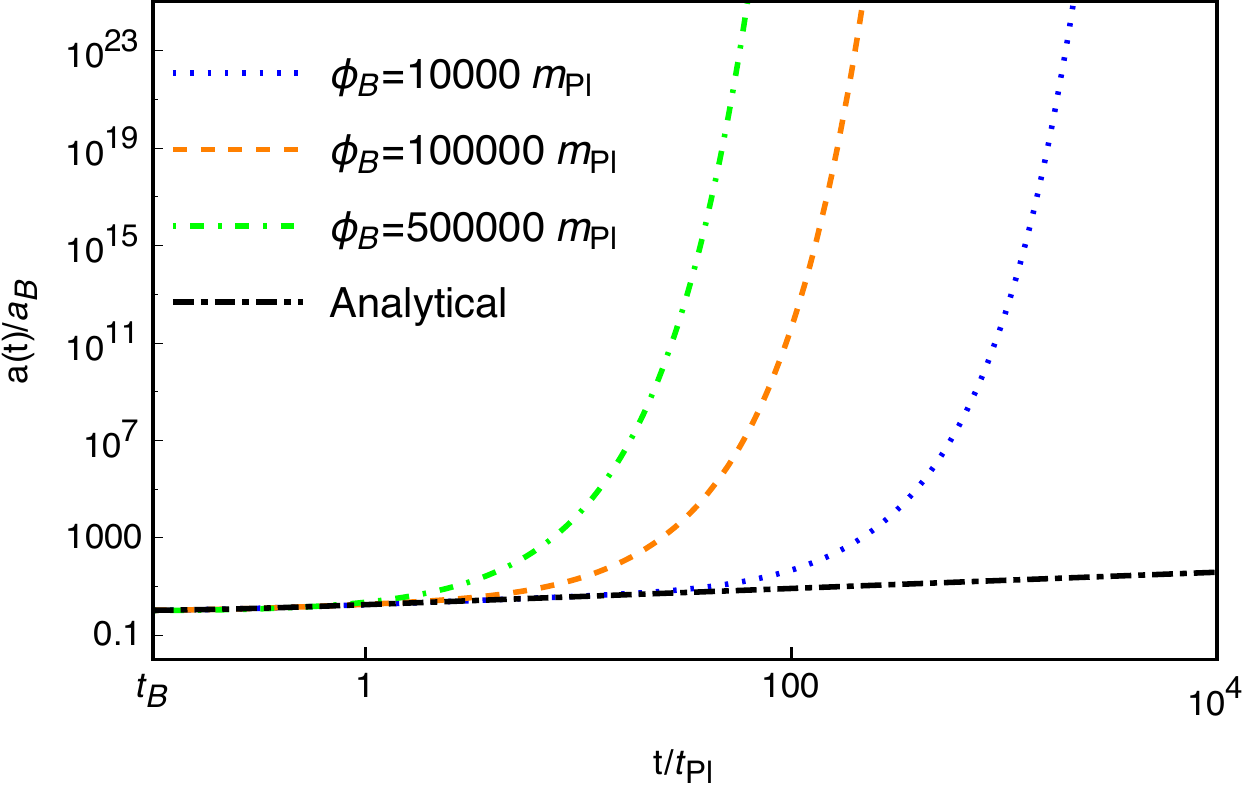}}\\
{\label{wphi_quadratic_Vdominated}
\includegraphics[width=8.1cm]{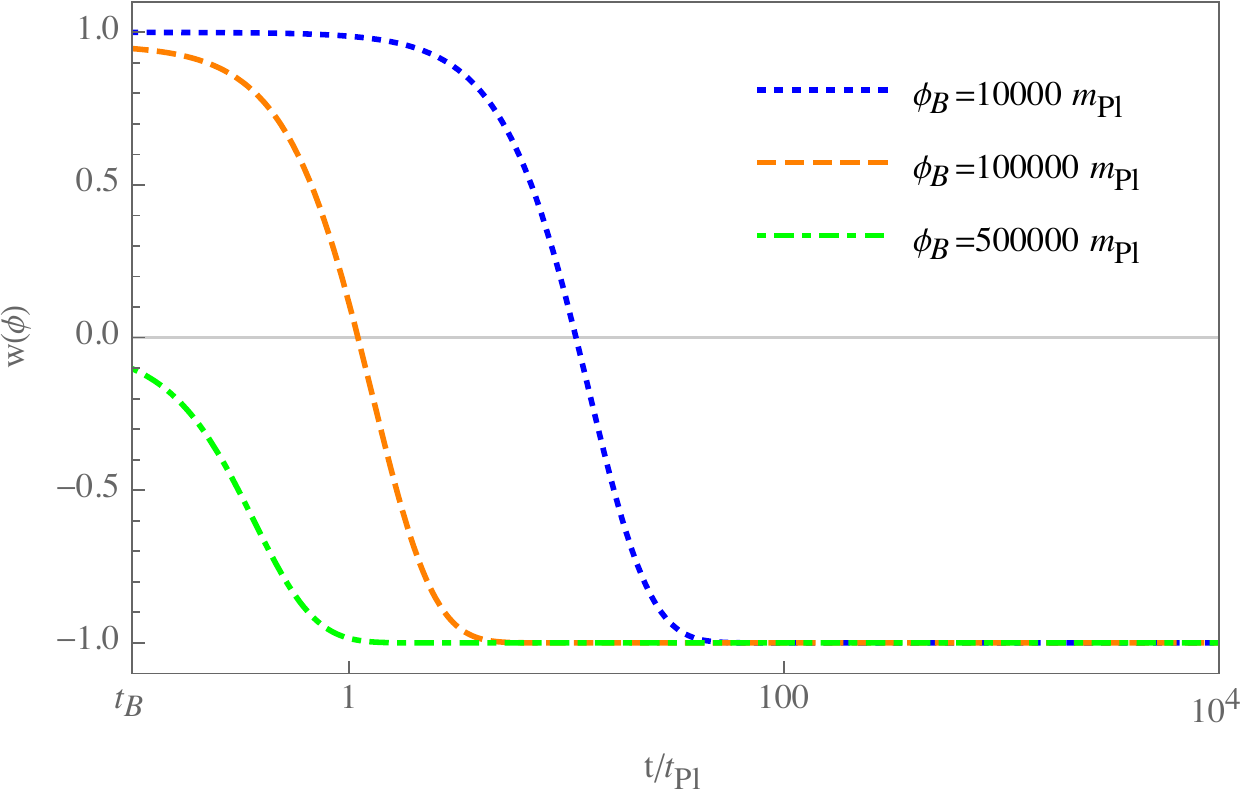}}\\
{\label{eH_quadratic_Vdominated}
\includegraphics[width=8.1cm]{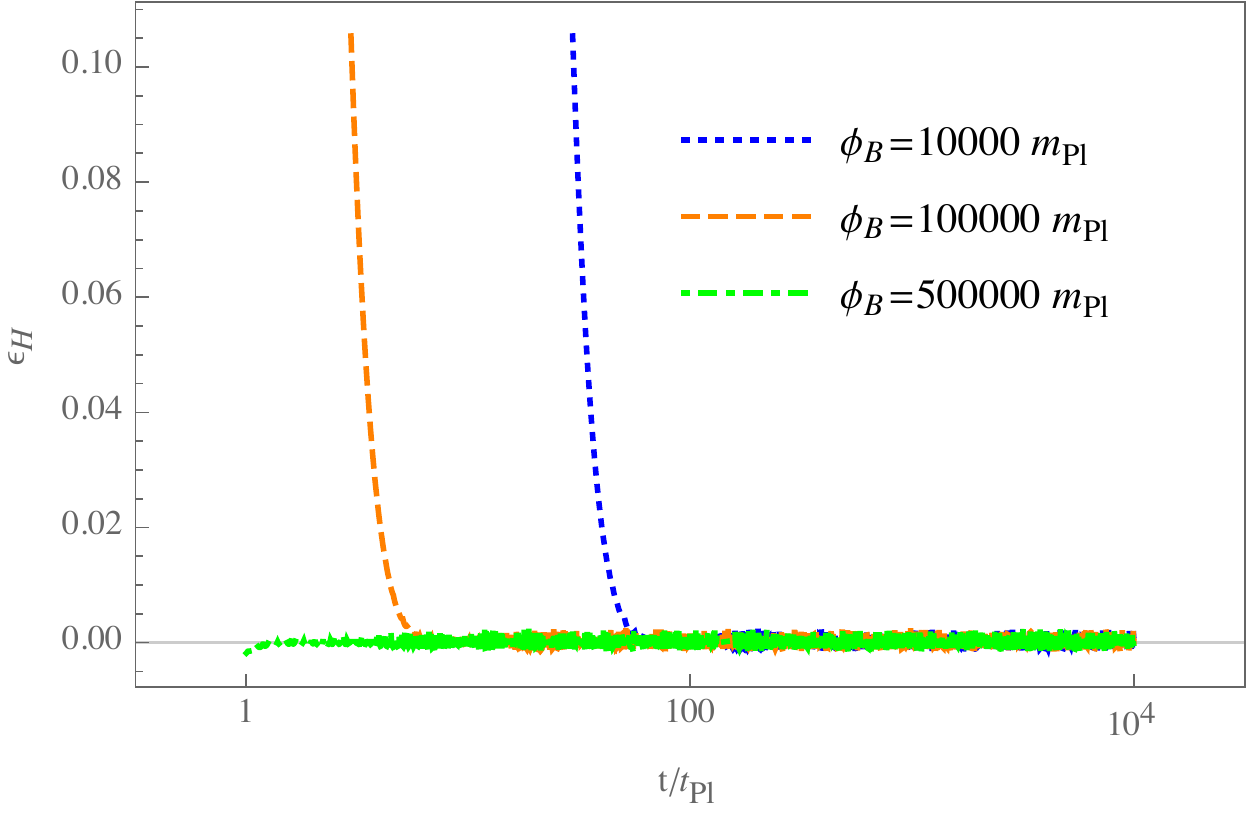}}
\caption{Numerical solution for the quadratic potential with the initial condition that the potential energy dominates the evolution of the universe at the bounce for $\dot\phi_\text{B}>0$.
Top panel: the evolution of the scale factor $a(t)$. The analytical solution given by Eq.~(\ref{scalar_analytical}) is also illustrated. Middle panel: the equation of state $w(\phi)$ for the same set of initial conditions
 as those given  in Top panel. Bottom panel: the slow-roll parameter $\epsilon_\text{H}$.}
 \label{quadratic_Vdominated}
\end{figure}

When the potential energy $V(\phi)$ of the inflaton dominates  at the quantum bounce,   the background evolution of the universe is illustrated in Fig.~\ref{quadratic_Vdominated}, from which we can see that the universality of the evolution of $a(t)$
disappears. In fact,  the bouncing phase does not exist any more, although the slow-roll inflationary phase $w(\phi) \simeq -1$ can be still achieved.

\subsection{Power-law potential with $n=1/2$}

We continue considering the power-law potential in this subsection  but now focus on $n=1/2$, for which the value of the scalar field $\phi$ must be  positive in order for the potential to be real. Similar to the
quadratic potential case, we further divide the initial conditions into two subclasses, the kinetic energy dominated  and the potential energy dominated.

\begin{figure}
{\label{scalar_n12_1}
\includegraphics[width=8.1cm]{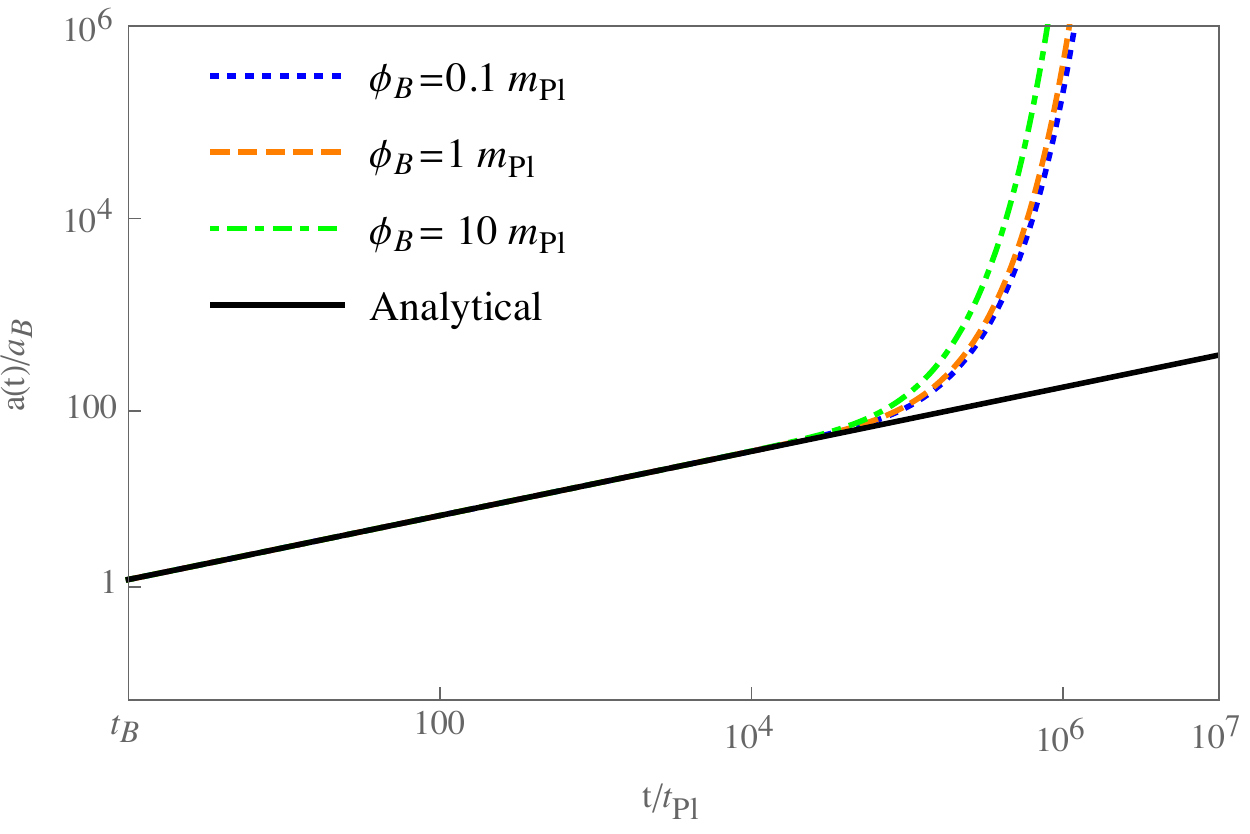}}\\
{\label{wphi_n12_1}
\includegraphics[width=8.1cm]{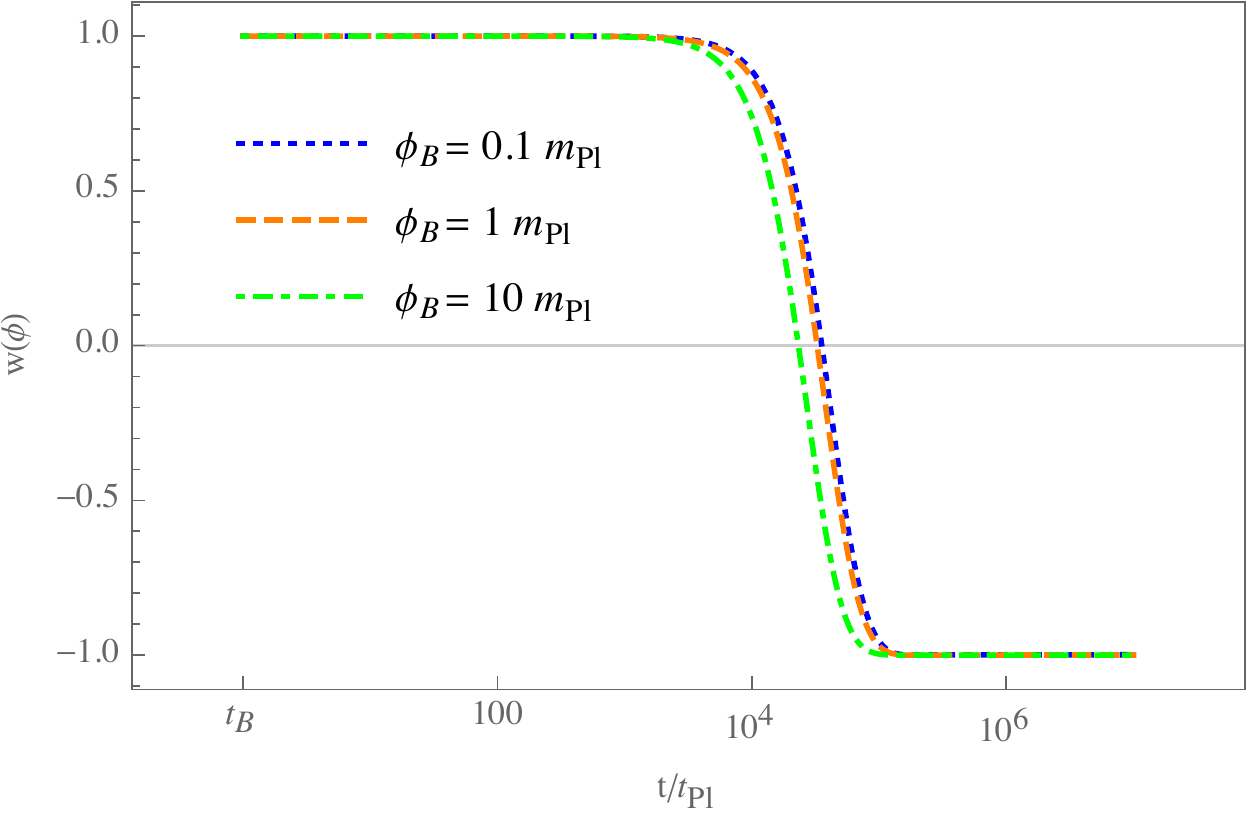}}\\
{\label{eH_n12}
\includegraphics[width=8.1cm]{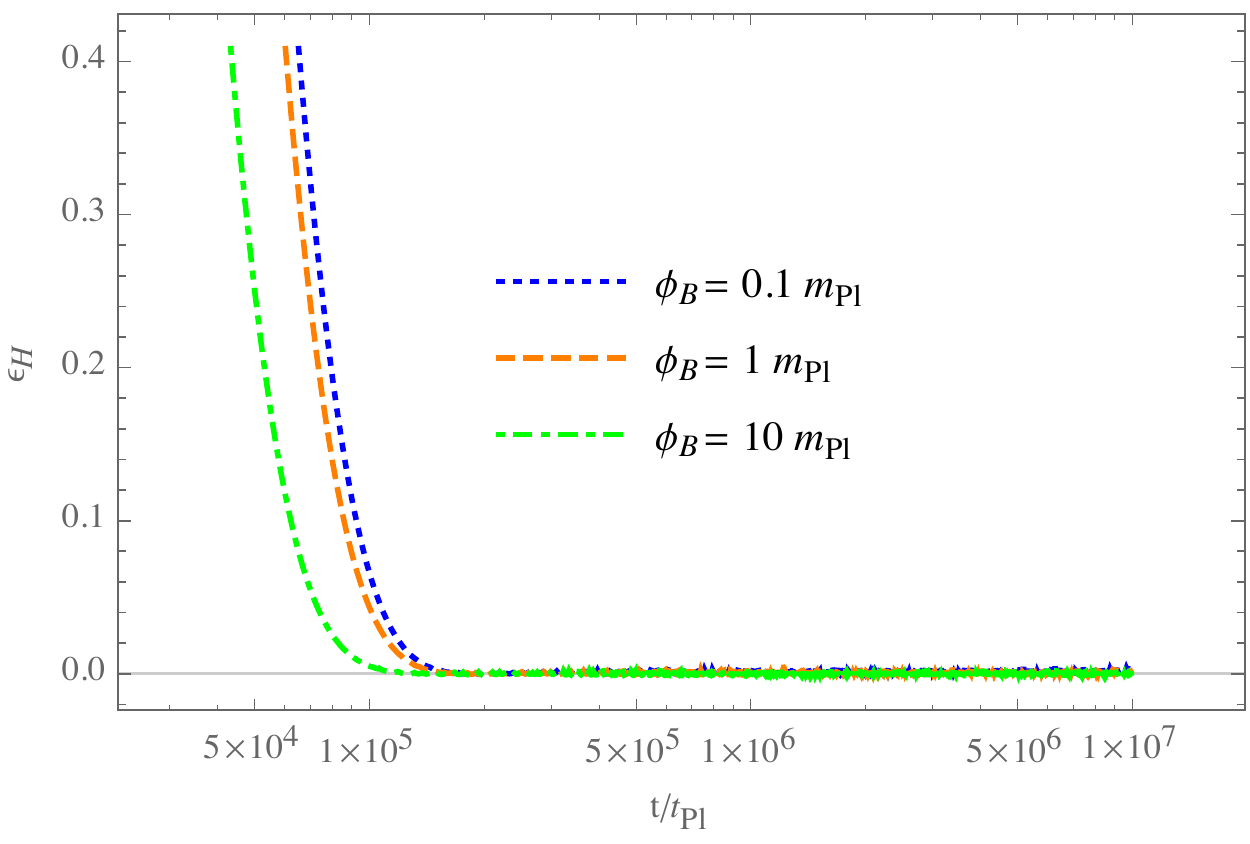}}\\
\caption{Numerical solution for the power-law potential with $n=1/2$ and the kinetic energy dominates the evolution at the bounce with $\dot\phi_\text{B}>0$. Top panel: the evolution of the scale factor $a(t)$.
The  analytical solution given in Eq.~(\ref{scalar_analytical}) is also illustracted. Middle panel: the equation of state $w(\phi)$ with the same set of the initial conditions as those given  in Top panel.
Bottom panel: the slow-roll parameter $\epsilon_\text{H}$.}
\label{n12_kinetic_1}
\end{figure}

\begin{figure}
{\label{scalar_n12_2}
\includegraphics[width=8.1cm]{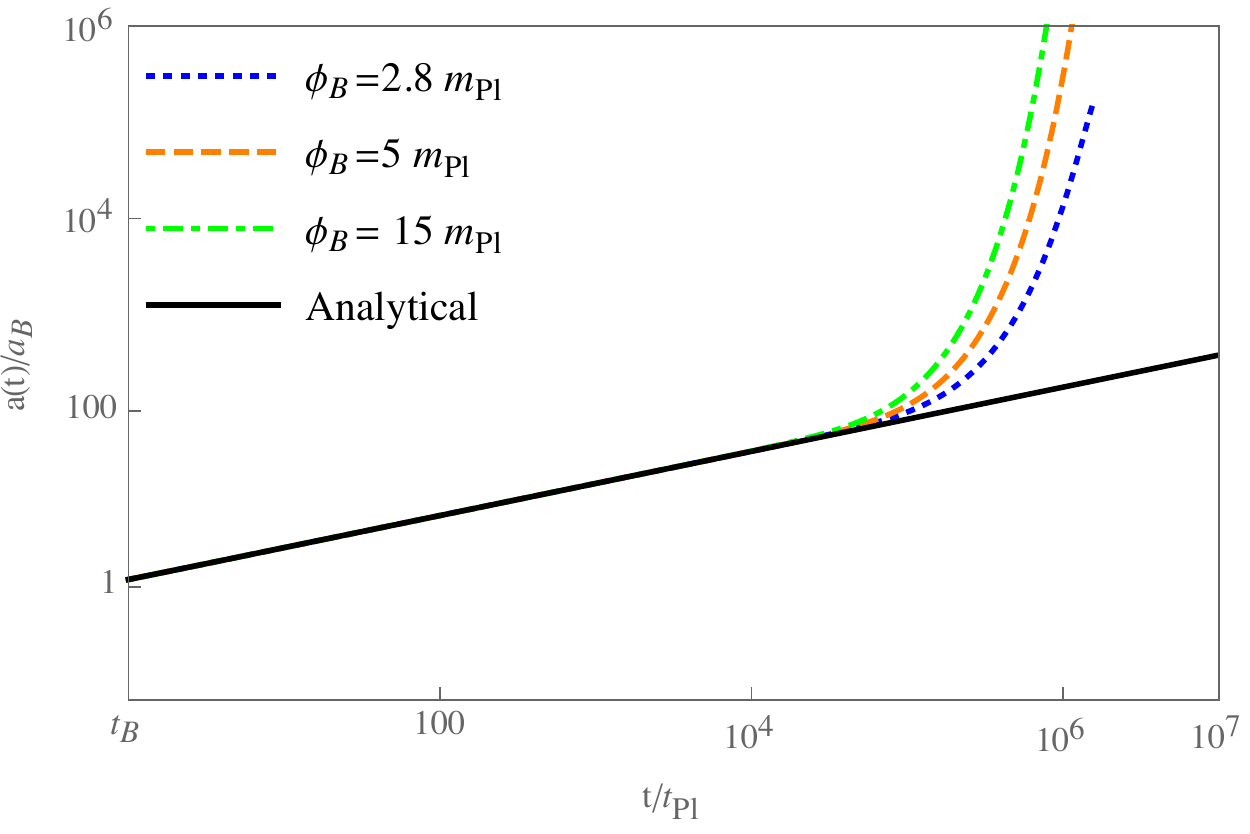}}\\
{\label{wphi_n12_2}
\includegraphics[width=8.1cm]{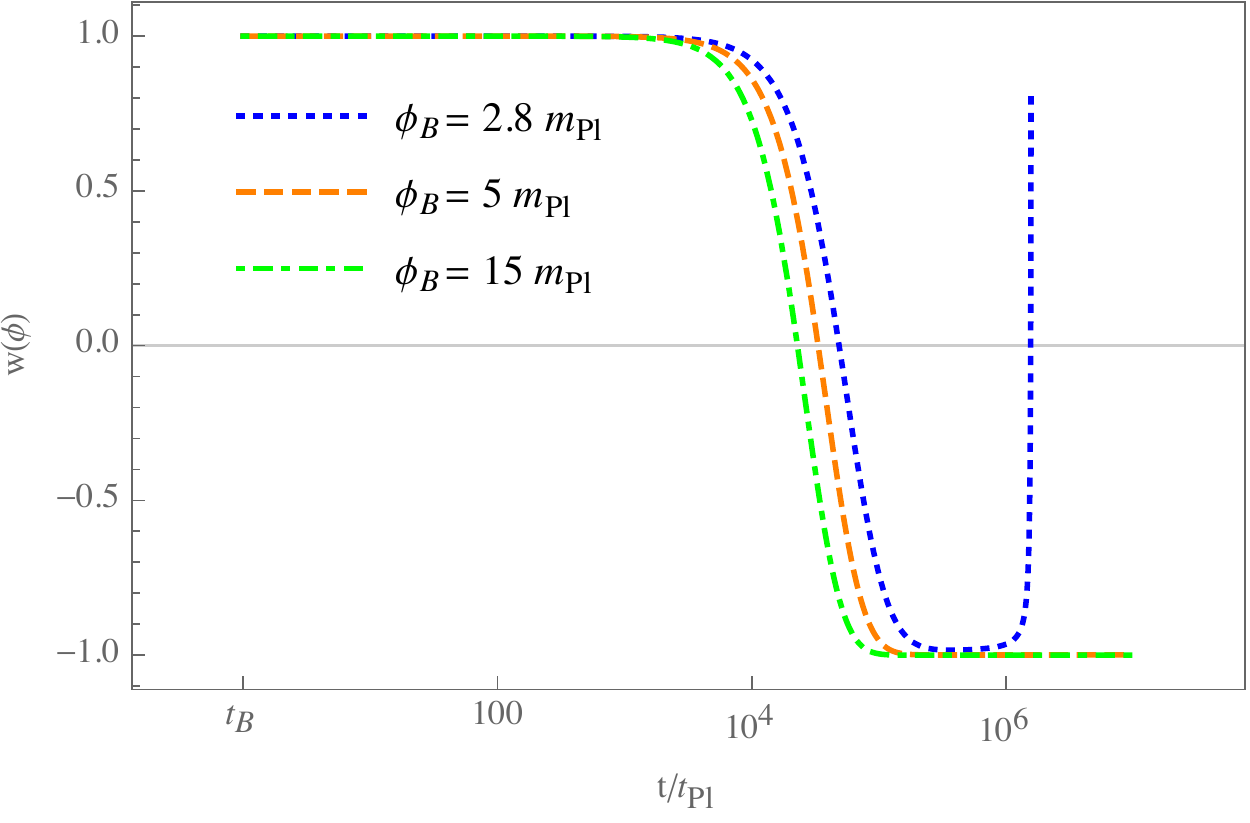}}\\
{\label{eH_n12_2}
\includegraphics[width=8.1cm]{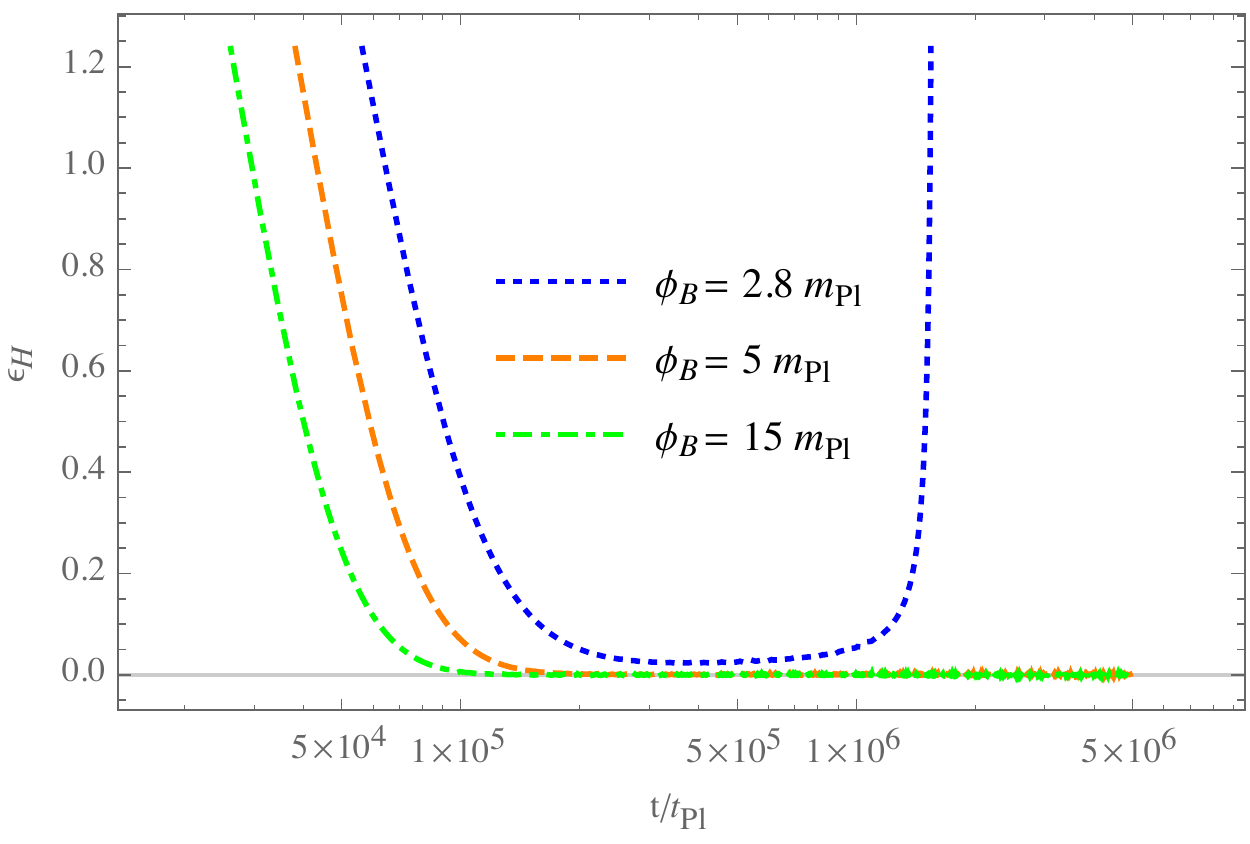}}\\
\caption{Numerical solution for the power-law potential with $n=1/2$ and the kinetic energy dominates the evolution at the bounce with $\dot\phi_\text{B}<0$. Top panel: the evolution of the scale factor $a(t)$. The
 analytical solution given in Eq.~(\ref{scalar_analytical}) is also illustrated. Middle panel: the equation of state $w(\phi)$ for the same set of initial conditions as those in Top panel. Bottom panel: the slow-roll parameter
 $\epsilon_\text{H}$.}
\label{n12_kinetic_2}
\end{figure}

\begin{figure}
{\label{VK_n12_1}
\includegraphics[width=8.1cm]{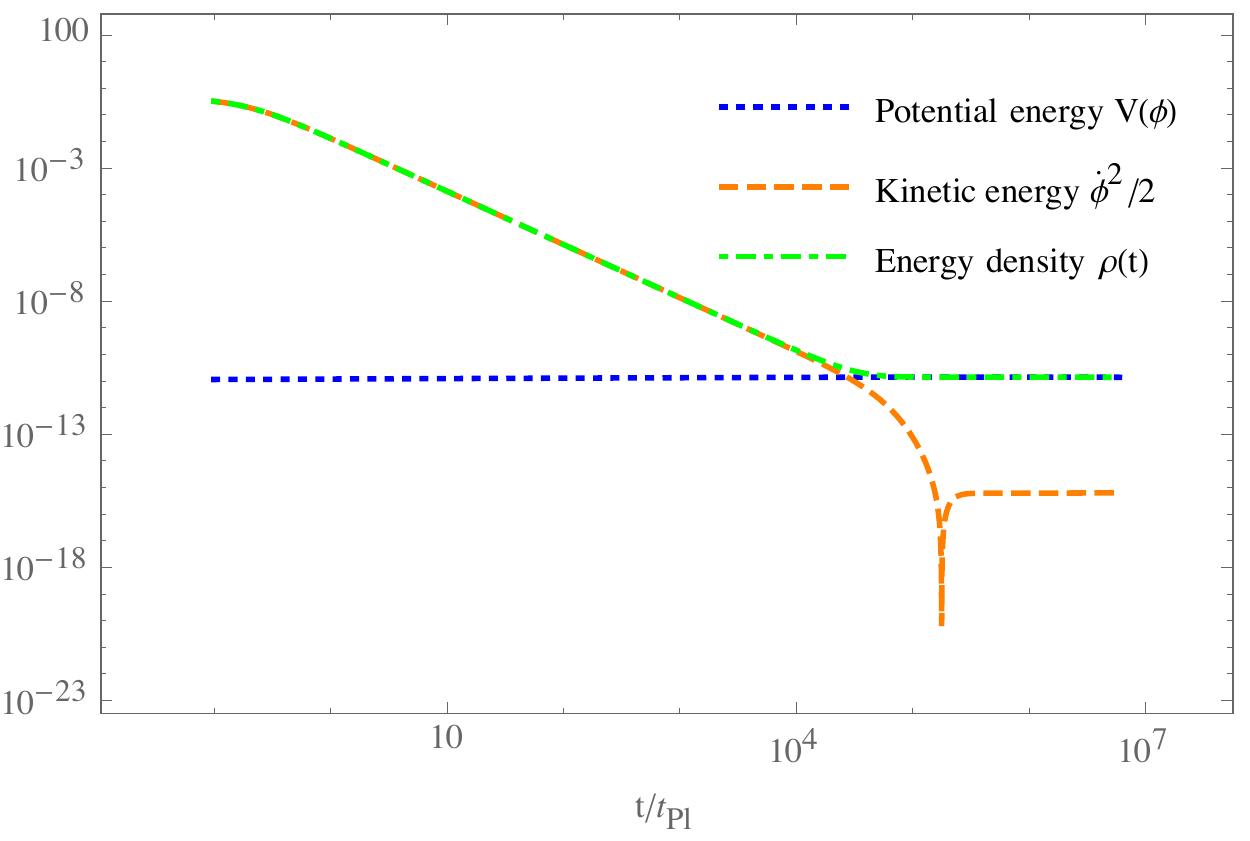}}\\
{\label{VK_n12_1}
\includegraphics[width=8.1cm]{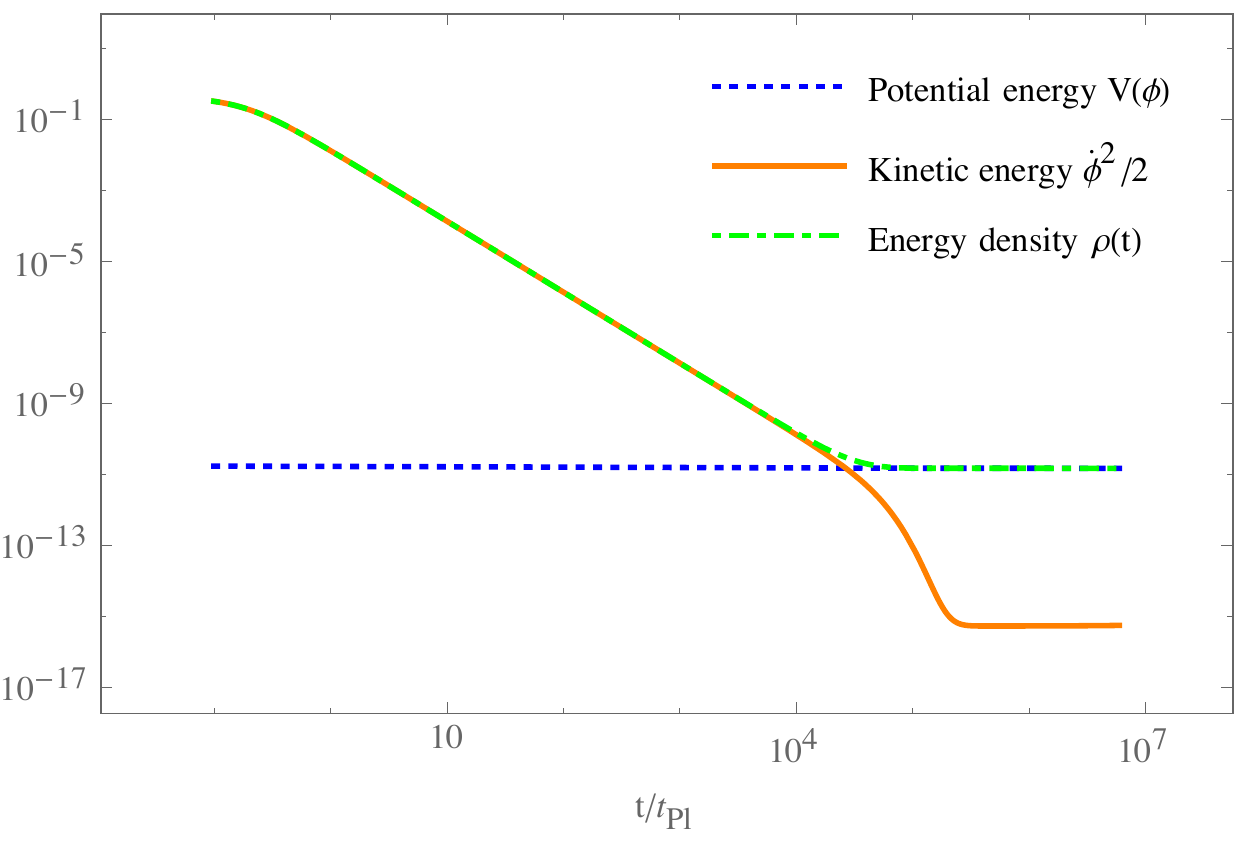}}
\caption{Comparison between the potential energy $V(\phi)$ and the kinetic energy $\dot \phi^2 /2$. The energy density $\rho = \dot \phi ^2 /2 + V(\phi)$ for the  power law potential with $n=1/2$ is also illustrated.
Top panel: for the initial condition $\phi_\text{B}=4 m_\text{Pl}$ and $\dot \phi_\text{B}>0$. Bottom panel: for the initial condition $\phi_\text{B}=9 m_\text{Pl}$ and $\dot \phi_\text{B} <0$. } \label{VK_n12}
\end{figure}

Let us first consider the case in which the evolution is dominated at the bounce by the kinetic energy of the inflaton. Then, the background evolutions for a set of  initial conditions  with $\dot \phi_\text{B}>0$ and $\dot \phi_\text{B}<0$ are
illustrated, respectively,  in Fig.~\ref{n12_kinetic_1} and Fig.~\ref{n12_kinetic_2}, in which the scale factor $a(t)$, the equation of state $w(\phi)$, and the slow-roll parameters $\epsilon_H$ are all obtained numerically.
To see the universality of the evolution of the scale factor $a(t)$,  its analytical solution of Eq.~(\ref{scalar_analytical}) is also illustrated.   From these figures, we find that: (a) similar to the quadratic potential case, the evolution of the
universe can be divided into three different phases, the bouncing, transition and slow-roll inflation. (b) During the bouncing phase, the evolution of the scale factor $a(t)$ is independent of not only the initial conditions of
$\phi_B$ and $\dot\phi_B$, but also the potential. It is well described by the analytical solution given by Eq.~(\ref{scalar_analytical}) for the quadratic potential ($n = 2$) as well as for the power-low potential with $n = 1/2$.
In fact, as we shall show below, this is also true for the Starobinsky potential. This is mainly because the amplitude of the potential during the bouncing phase is very small in comparison with the kinetic one, and its effects on the evolution
of the background during the bouncing phase are negligible. This can be seen clearly from Fig. \ref{VK_quadratic} for the quadratic potential and Fig. \ref{VK_n12} for the power-low potential with $n = 1/2$.

\begin{figure}
{\label{Ninf_n12_1}
\includegraphics[width=8.1cm]{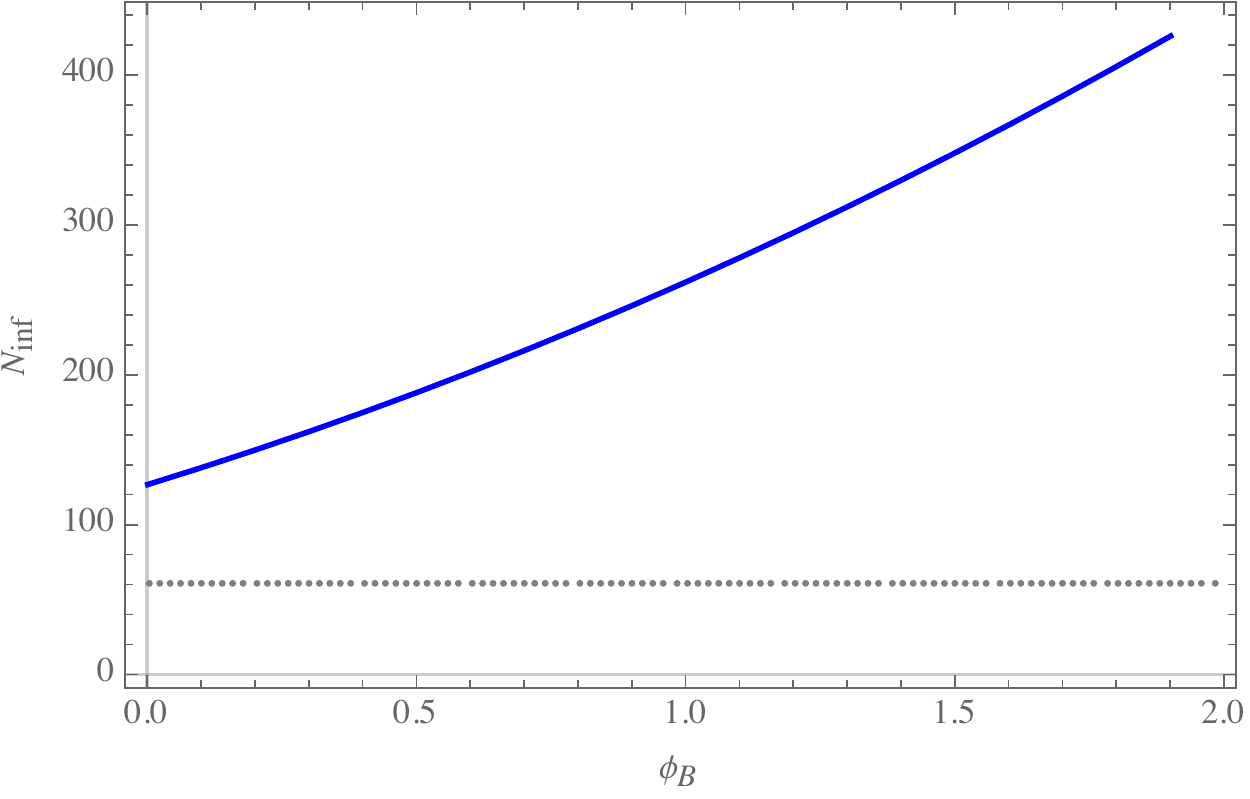}}\\
{\label{Ninf_n12_2}
\includegraphics[width=8.1cm]{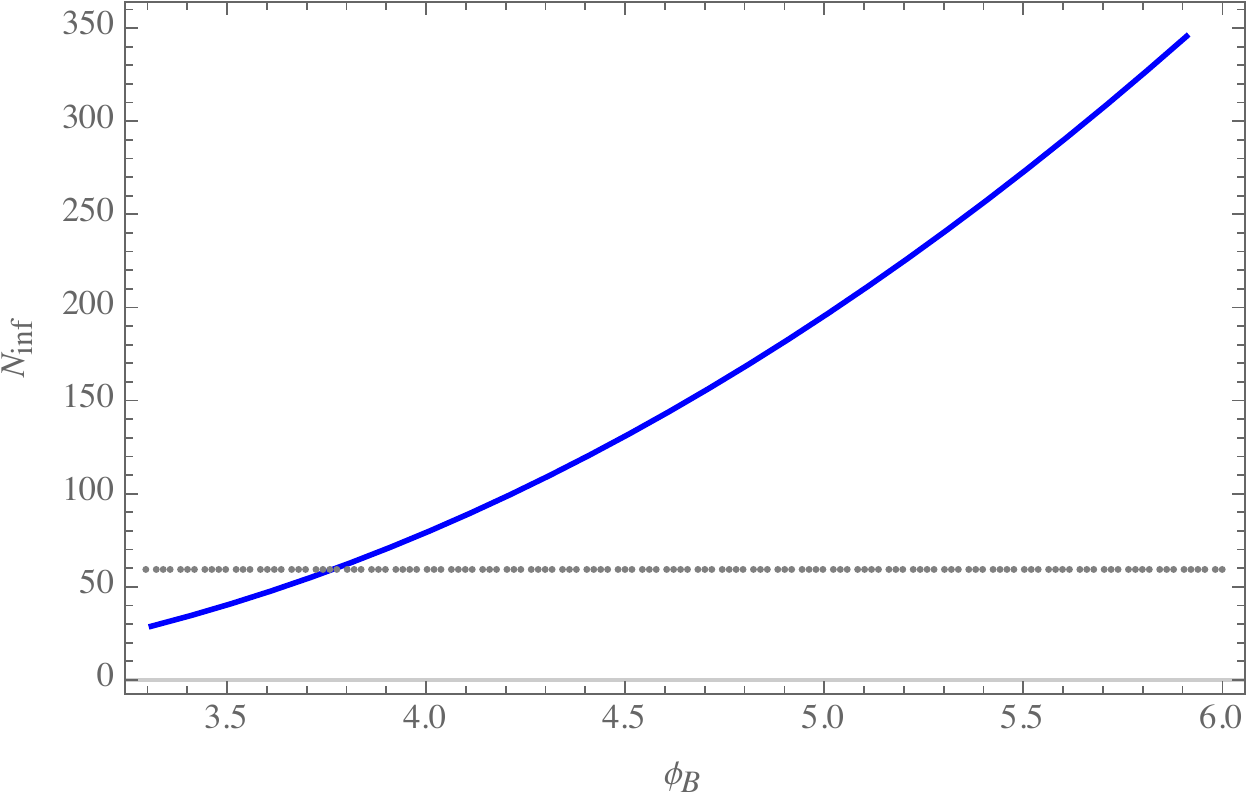}}\\
\caption{The e-folds $N_\text{inf}$ during the slow-roll inflation as a function of $\phi_\text{B}$  for the power-low potential with $n = 1/2$. Top panel: for $\dot \phi_\text{B}>0$. Bottom panel: for $\dot \phi_\text{B}<0$.} \label{Ninf_n12}
\end{figure}

 The corresponding  e-folds $N_\text{inf}$ during the slow-roll inflation as a function of $\phi_\text{B}$ is illustrated in Fig.~\ref{Ninf_n12}. For the initial conditions with $\dot \phi_\text{B}>0$, the top panel of Fig.~\ref{Ninf_n12}
 shows that the desired slow-roll inflation can be produced for any value of $\phi_\text{B}$ in the range of $(0,\phi_\text{max})$. Here $\phi_\text{max} = \frac{4 \rho_\text{c}^2}{m^{7}}$. Moreover, the slow-roll inflation can
 last long enough to produce more than $60$ e-folds. For the initial conditions with $\dot \phi_\text{B}<0$, in order to produce at least $60$ e-folds during the slow-roll inflationary phase, the values of $\phi_\text{B}$ have to be restricted to
\bqn
\phi_\text{B} \in [3.7 m_\text{Pl}, \phi_\text{max}].
\eqn
Similar to the quadratic potential, the e-folds $N_\text{inf}$ increases when the value of $\phi_\text{B}$ is increasing.

\begin{figure}
{\label{n12_Vdominated_1}
\includegraphics[width=8.1cm]{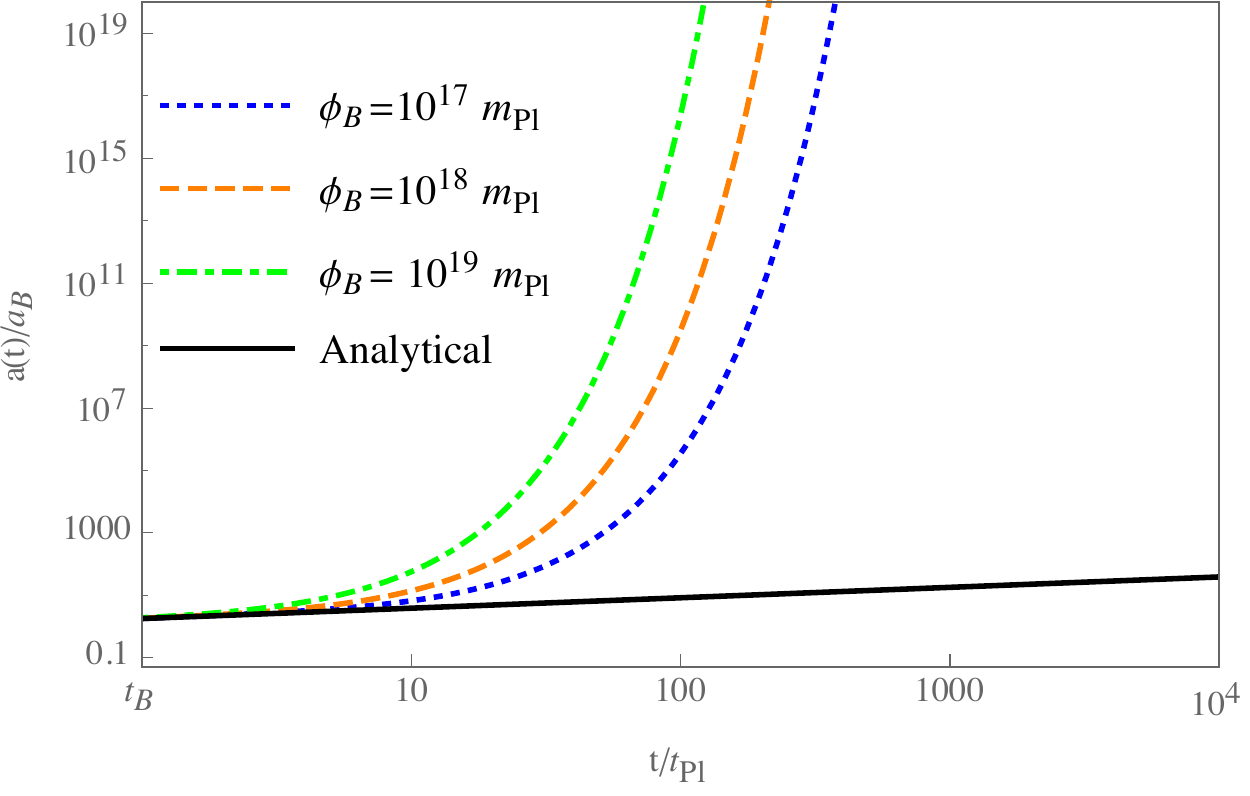}}
{\label{wphi_quadratic_Vdominated}
\includegraphics[width=8.1cm]{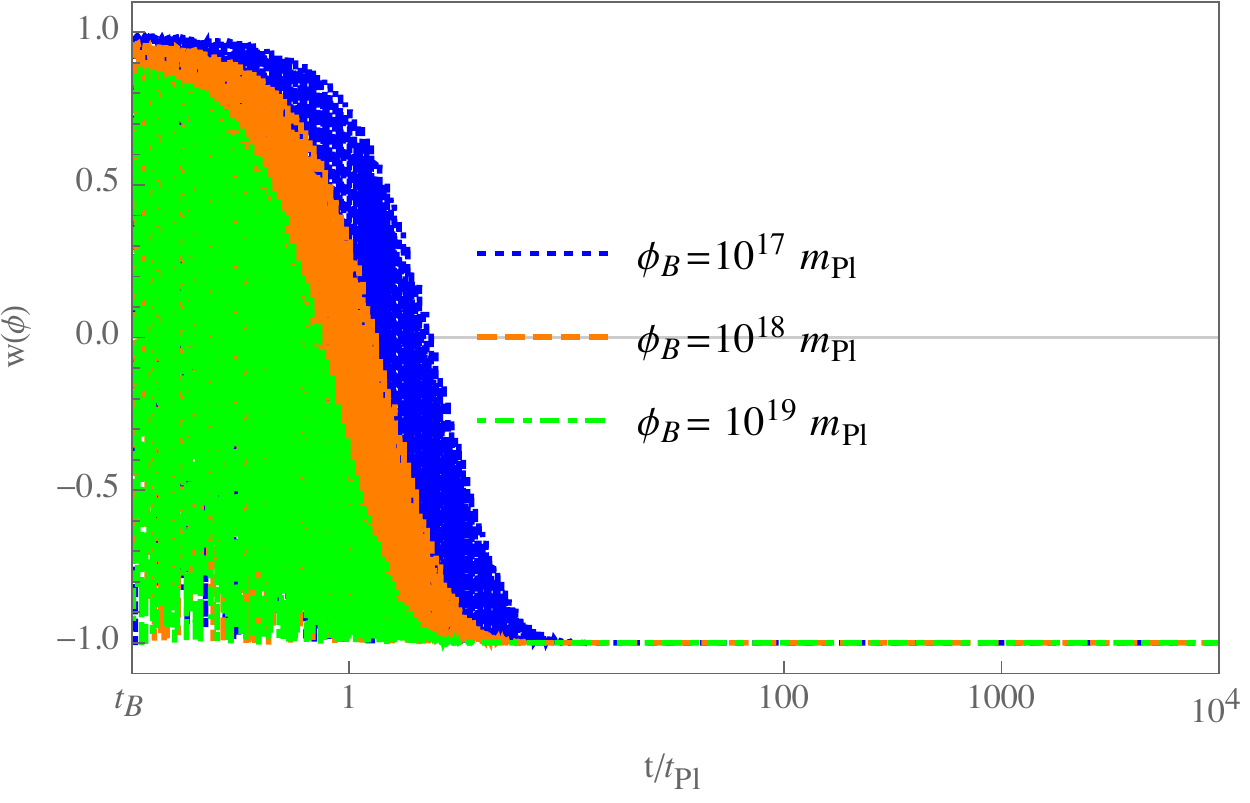}}\\
{\label{eH_n12_Vdominated}
\includegraphics[width=8.1cm]{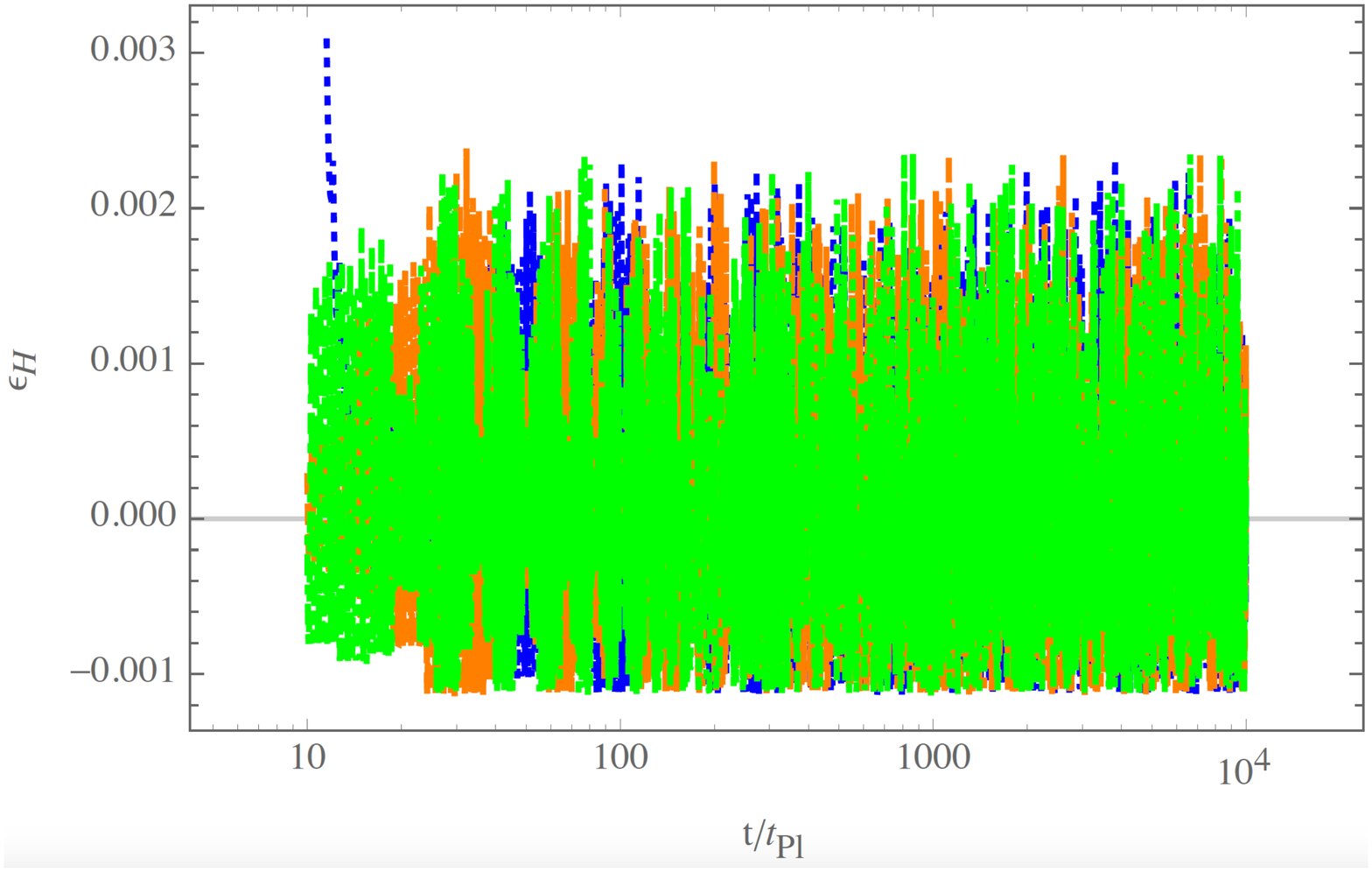}}\\
\caption{Numerical solution  for the potential energy dominated initial conditions for  the power law potential with $n=1/2$  and $\dot\phi_\text{B}>0$. Top panel: the evolution of the scale factor $a(t)$
including the analytical solution given by Eq.~(\ref{scalar_analytical}). Middle panel: the equation of state $w(\phi)$. Bottom panel: the slow-roll parameter $\epsilon_H$.} \label{n12_Vdominated}
\end{figure}

For the evolution in which the potential energy dominates at the quantum bounce,  the background evolution is presented in Fig.~\ref{n12_Vdominated}. Again, in this case the evolution of the scale 
factor sensitively depends on the choice
of the initial conditions of $\phi_B$ and $\dot\phi_B$. But, the slow-roll inflation is still achieved, and similar to the quadratic potential, the potential energy dominated initial conditions can lead to a large number of e-folds $N_\text{inf}$
during the slow-roll inflation.

\subsection{Starobinsky potential}

The behavior of the background evolution with the Starobinsky potential has been studied in detail in \cite{bonga_inflation_2016, bonga_phenomenological_2016}.
Here we again summarize some of their main results, by paying particular attention on the universal properties.

The potential energy of the scalar field cannot exceed the maximum value $\rho_\text{c}$, so that the value of $\phi_\text{B}$ can only lie in the range of $(\phi_\text{min}, +\infty)$, where
\bqn
\phi_\text{min}=-\frac{\sqrt{3}m_\text{Pl}}{4\sqrt{\pi}} \ln \left(1+\frac{\sqrt{32 \pi \rho_\text{c}}}{\sqrt{3} M m_\text{Pl}}\right).
\eqn
For the value of $M$ given by Eq.~(\ref{MV}) 
(see Ref. \cite{bonga_phenomenological_2016} for details), we have $\phi_\text{min}=-3.47 m_\text{Pl}$, and  the potential energy can dominate the evolution  only in a very narrow range
 $\phi \in [\phi_\text{min}, -3 m_\text{Pl})$. In order to identify what kind of initial conditions can lead to sufficiently long slow-roll inflation, we search all the parameter space of $\phi$  numerically.

 \begin{figure}
{\label{scalar_starobinsky_1}
\includegraphics[width=8.1cm]{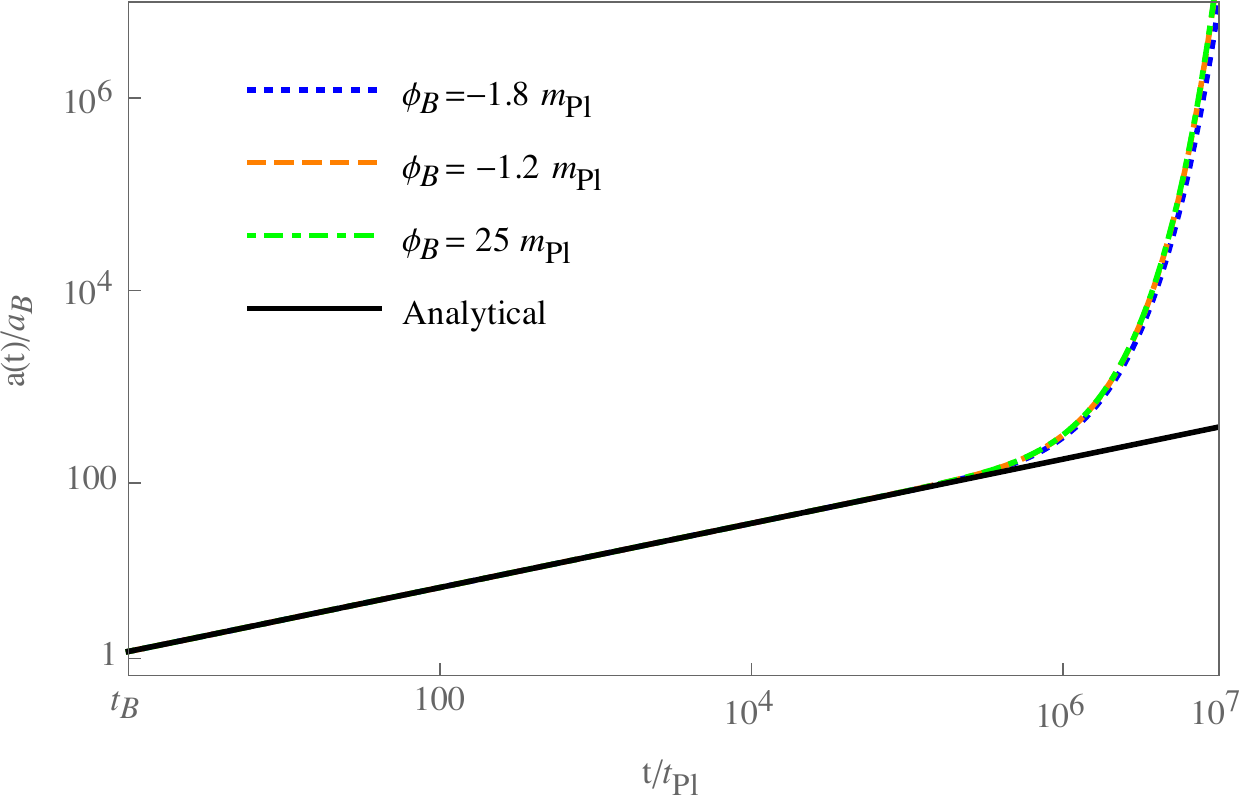}}\\
{\label{wphi_starobinsky_1}
\includegraphics[width=8.1cm]{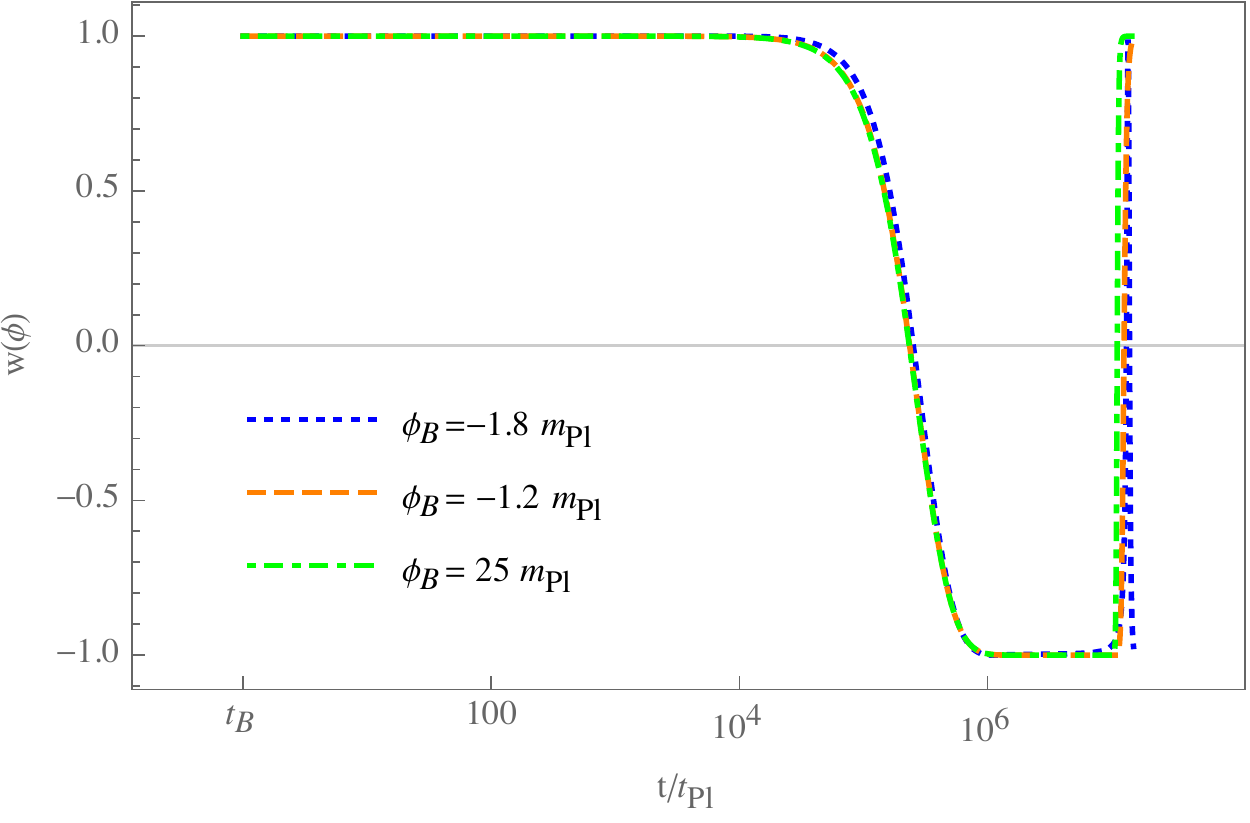}}\\
{\label{eH_starobinsky_1}
\includegraphics[width=8.1cm]{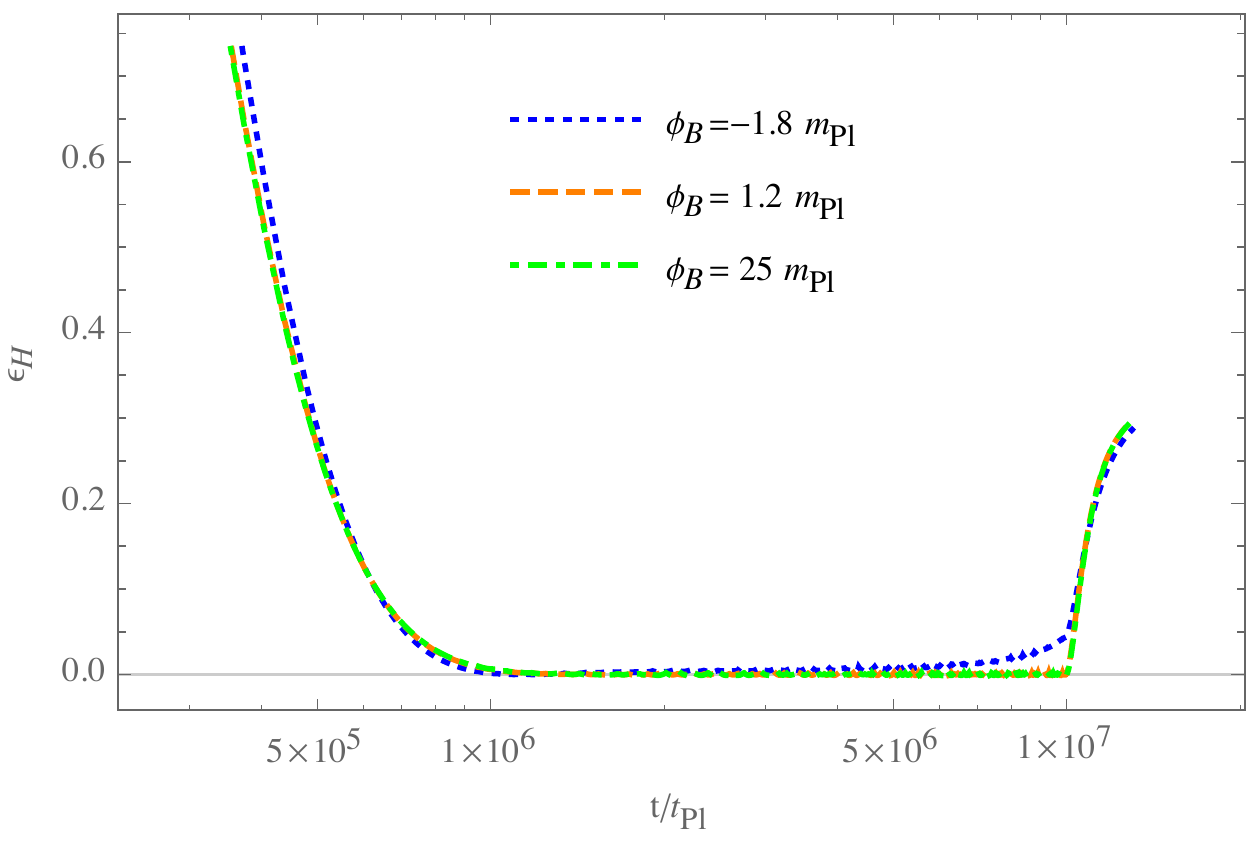}}\\
\caption{Numerical solution for the Starobinsky potential with $\dot\phi_\text{B}>0$. Top panel: the evolution of the scale factor $a(t)$ and the analytical solutions given by Eq.~(\ref{scalar_analytical}).
Middle panel: the equation of state $w(\phi)$. Bottom panel: the slow-roll parameter $\epsilon_\text{H}$.} \label{starobinsky_1}
\end{figure}

\begin{figure}
{\label{scalar_starobinsky_2}
\includegraphics[width=8.0cm]{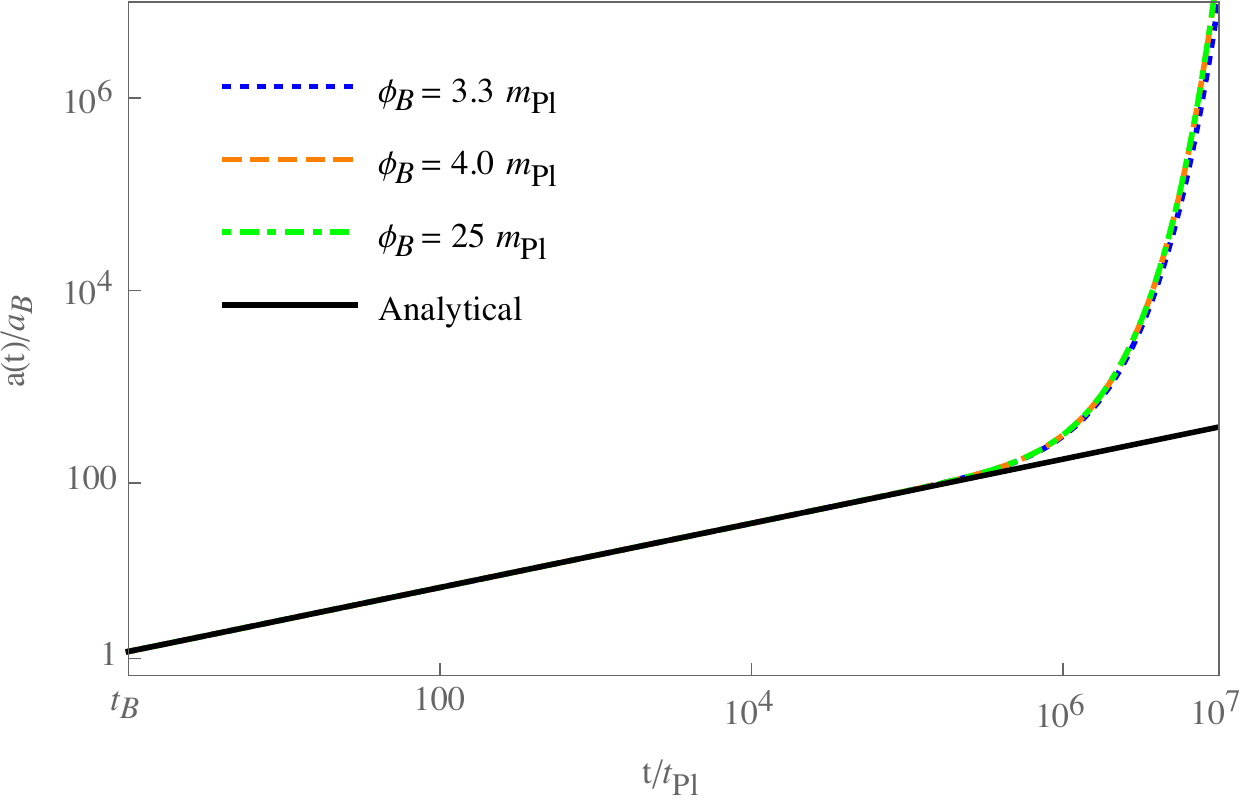}}\\
{\label{wphi_starobinsky_2}
\includegraphics[width=8.0cm]{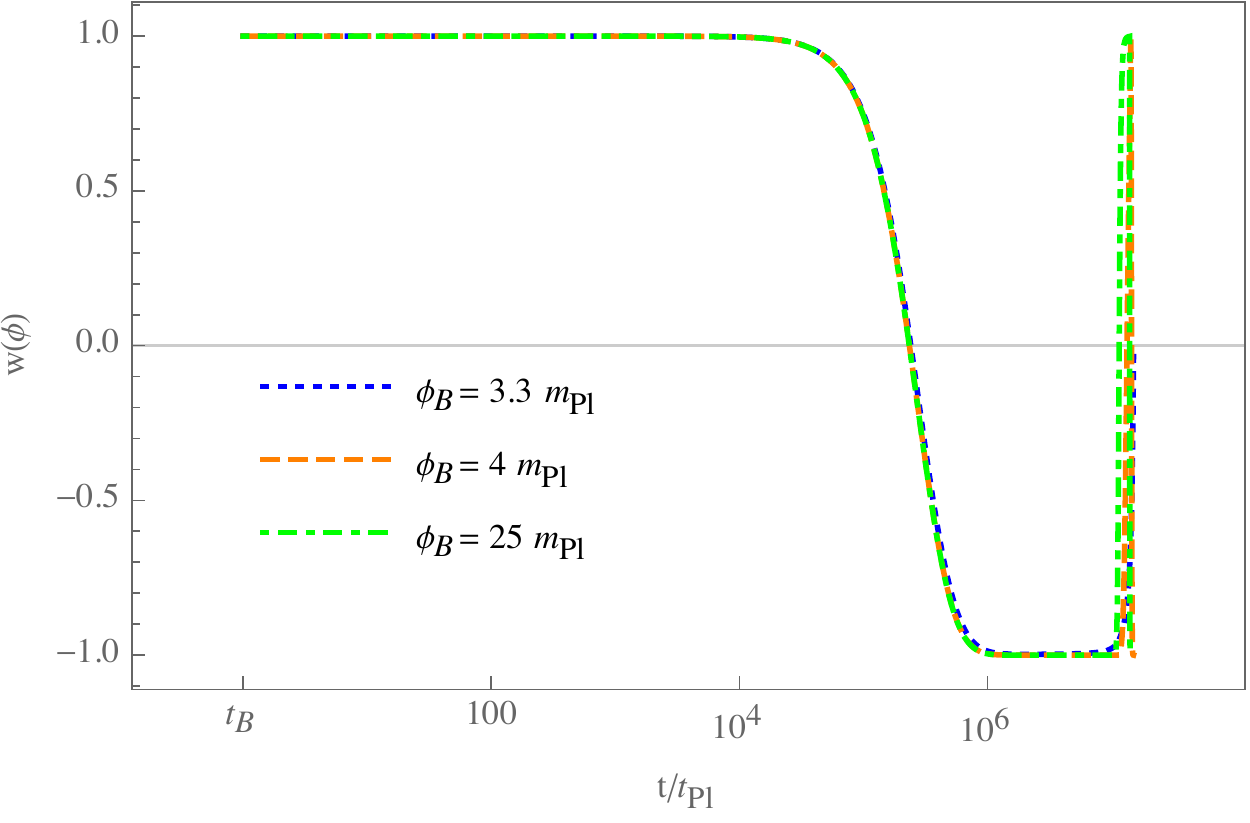}}\\
{\label{eH_starobinsky_2}
\includegraphics[width=8.0cm]{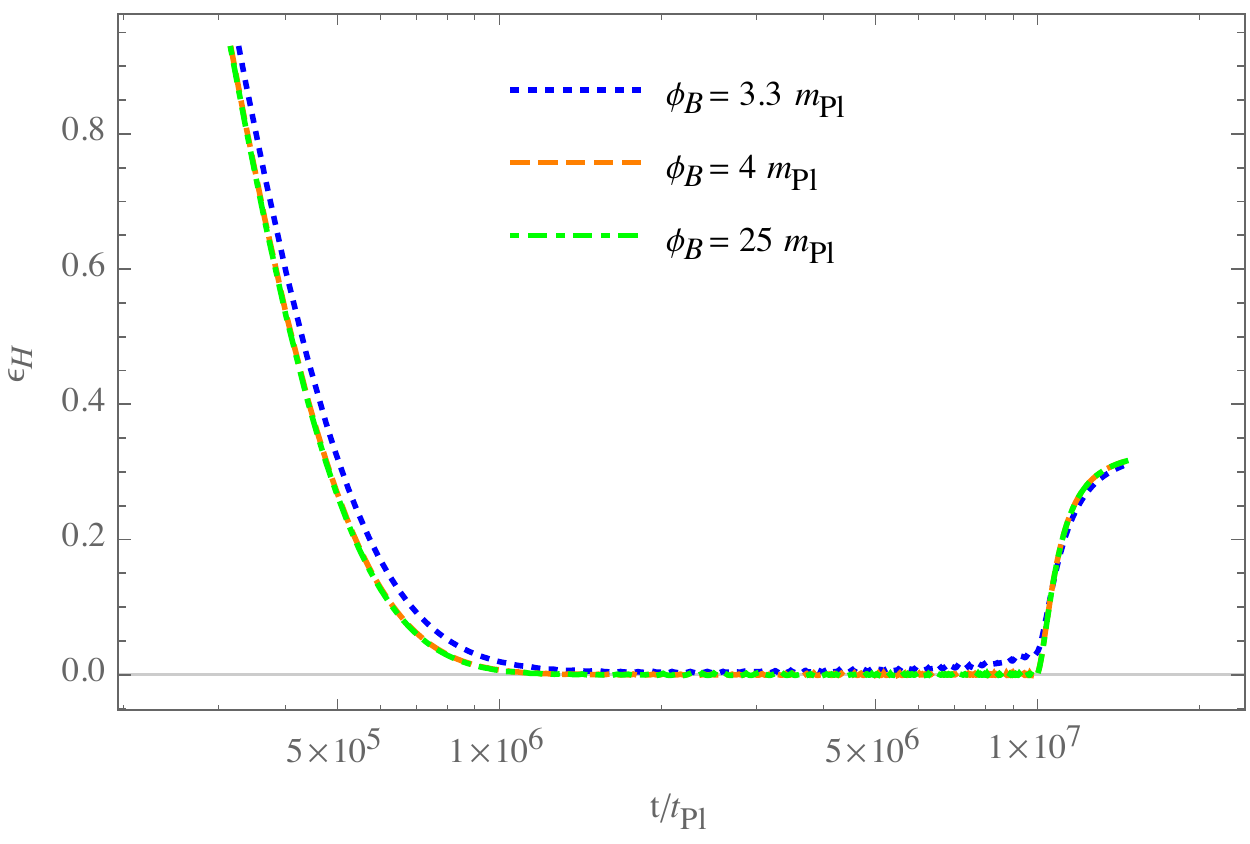}}\\
\caption{Numerical solution for the theory of Starobinsky potential with $\dot\phi_\text{B}<0$. Top panel: the evolution of the scale factor $a(t)$ from both the numerical and analytical solutions. Middle panel: the equation of state $w(\phi)$ for the same set of initial conditions. Bottom panel: the slow-roll parameter $\epsilon_\text{H}$ during the slow-roll inflation.} \label{starobinsky_2}
\end{figure}

 The  results of the background  evolutions for the kinetic  energy dominated initial conditions are illustrated in Fig.~\ref{starobinsky_1} and Fig.~\ref{starobinsky_2}, with $\dot \phi_\text{B}>0$ and $\dot \phi_\text{B}<0$, respectively.  In both
 figures,  the scale factor $a(t)$, the equation of state $w(\phi)$, and the slow-roll parameters $\epsilon_H$ are all obtained  numerically  for the same set of initial values of $\phi_\text{B}$ but with different signs of  $\dot \phi_\text{B}$.
 Again, similar to the last two cases, the evolutions of the background can be divided into three different phases, the bouncing, transition and slow-roll inflation, as one can see clearly from the behavior of the equation of state $w(\phi)$.
 Moreover, during the bouncing phase the evolution of the scale factor $a(t)$ is universal: {\em it does not depend on the initial values of $\phi_\text{B}$ and  $\dot \phi_\text{B}$, neither on the form of potentials}. As long as the evolution is
 dominated by the kinetic energy of the inflaton, the evolution of $a(t)$ is well described by the analytical solution (\ref{scalar_analytical}) during the bouncing phase,  no matter whether the inflationary  potential is the Starobinsky one
 or the  power-low one!

 \begin{figure}
{\label{VK_Starobinsky_1}
\includegraphics[width=8.1cm]{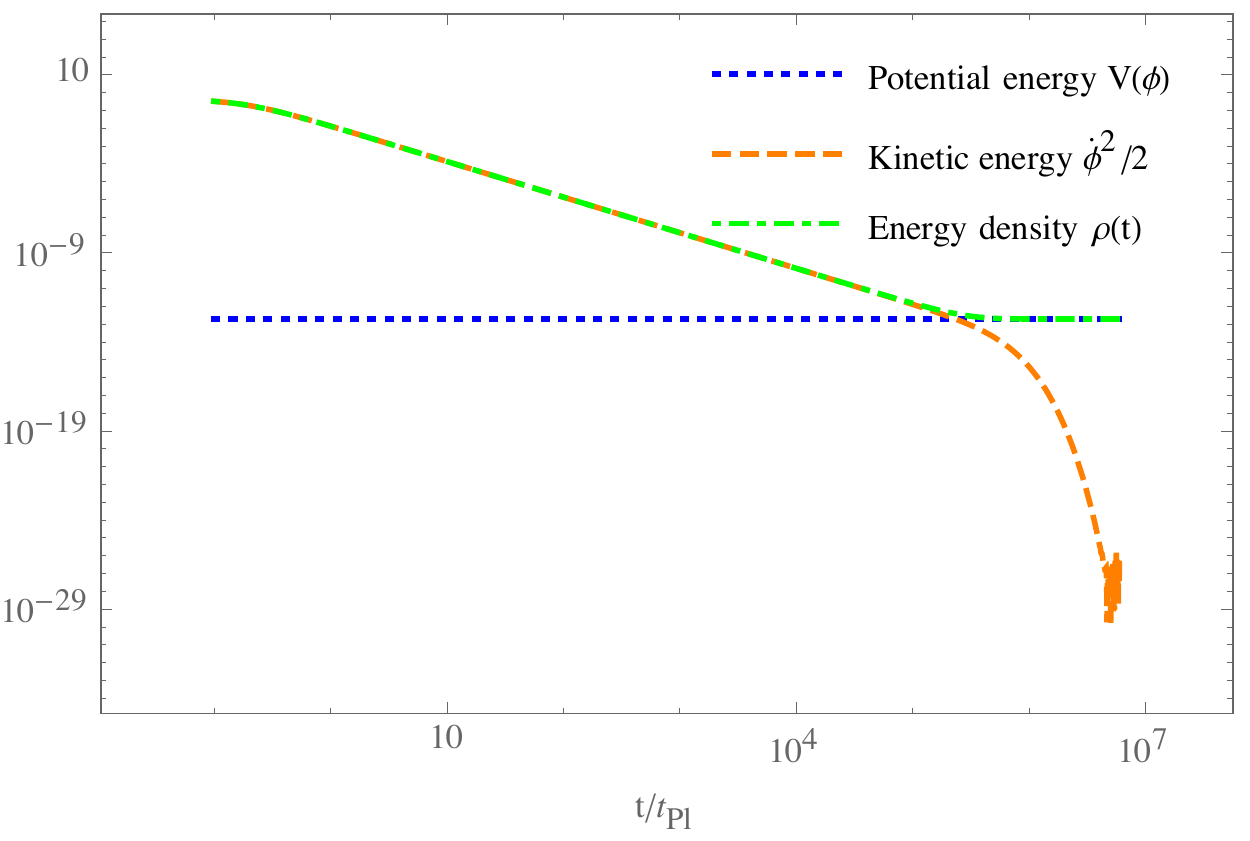}}\\
{\label{VK_Starobinsky_1}
\includegraphics[width=8.1cm]{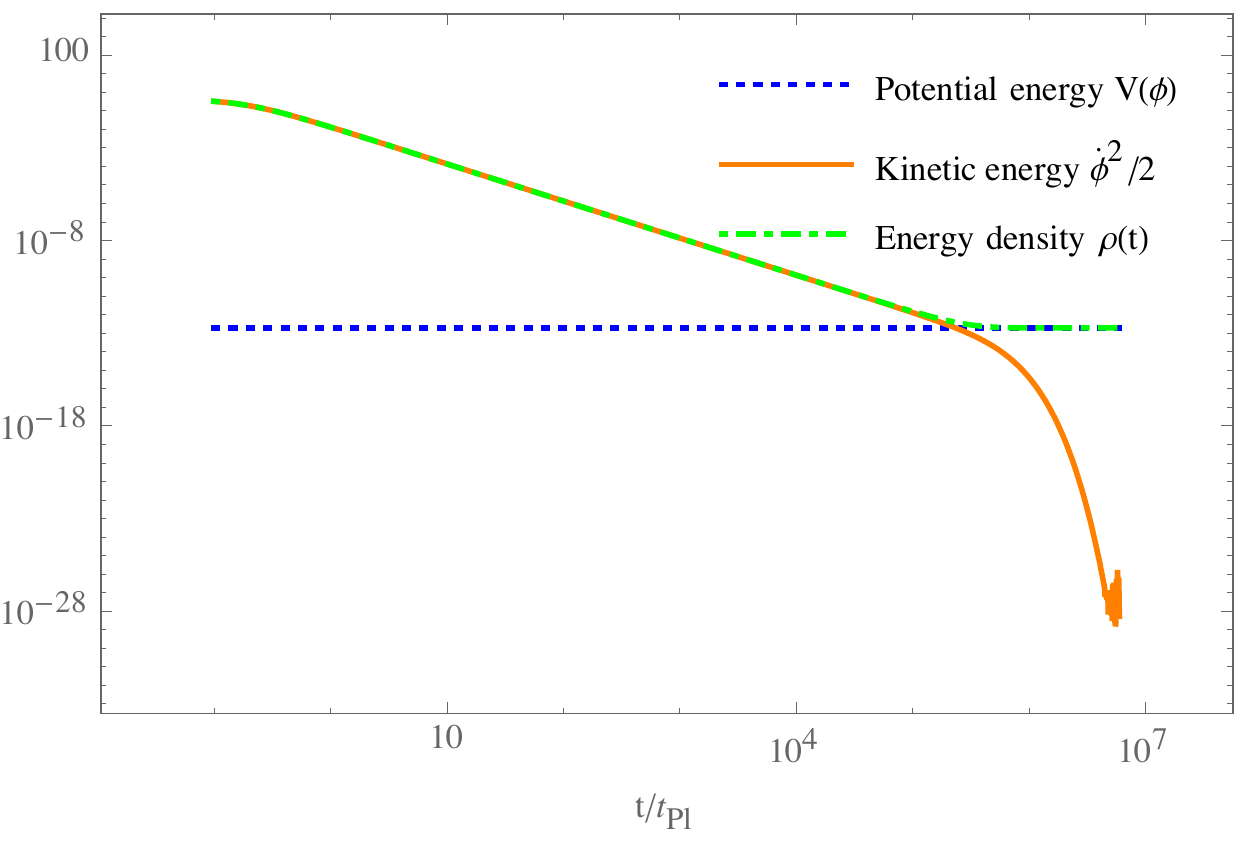}}
\caption{Comparison between the potential energy $V(\phi)$ and the kinetic energy $\dot \phi^2 /2$ for the theory of Starobinsky potential. Top panel: for the initial condition $\phi_\text{B}=4 m_\text{Pl}$ and $\dot \phi_\text{B}>0$. Bottom panel: for the initial condition $\phi_\text{B}=9 m_\text{Pl}$ and $\dot \phi_\text{B} <0$.} \label{VK_Starobinbsky}
\end{figure}

\begin{figure}
{\label{Ninf_starobinsky_1}
\includegraphics[width=8.1cm]{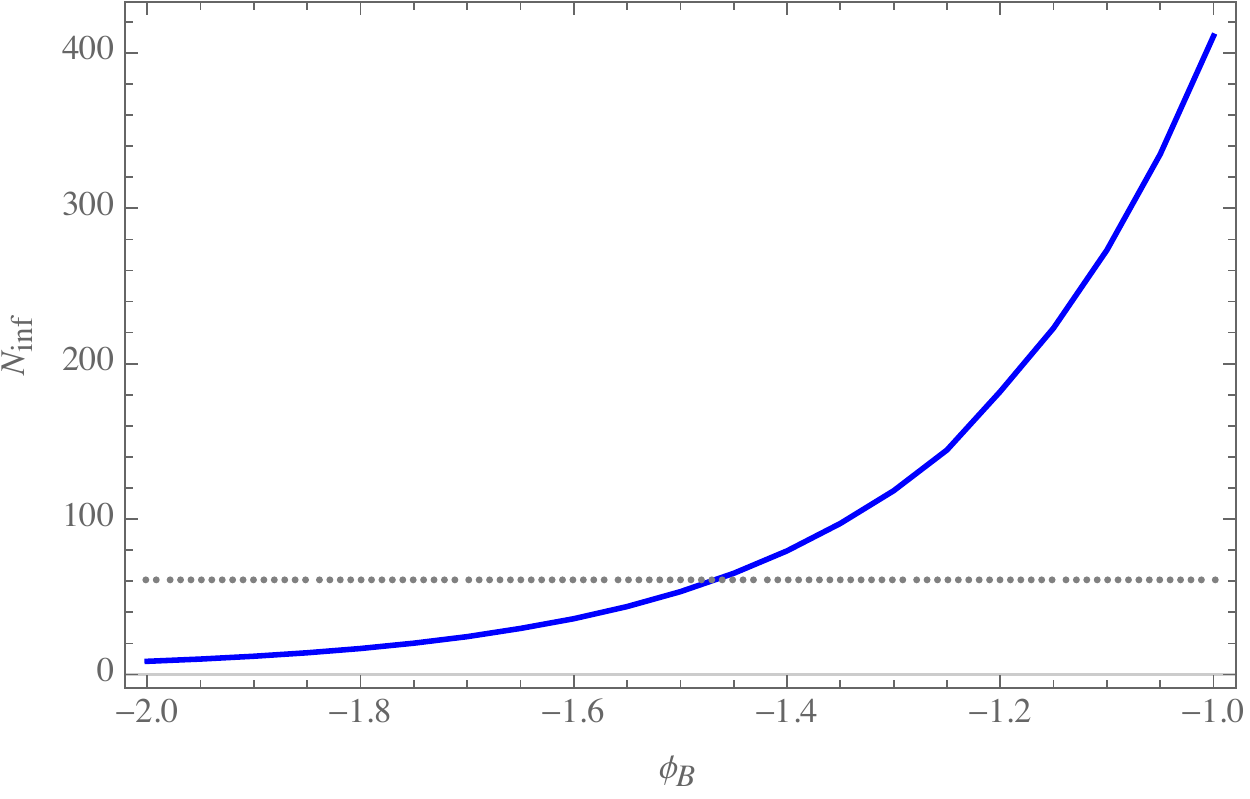}}\\
{\label{Ninf_starobinsky_2}
\includegraphics[width=8.1cm]{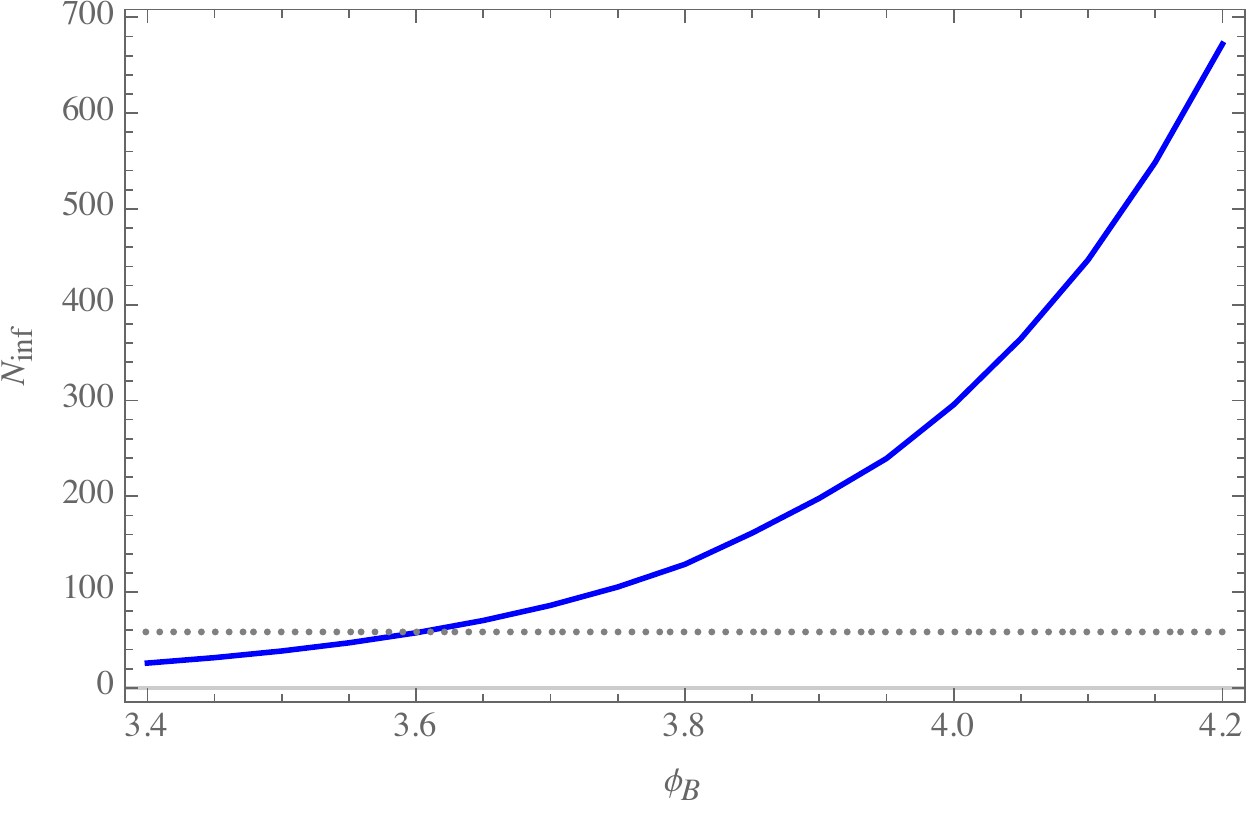}}
\caption{The e-folds $N_\text{inf}$ during the slow-roll inflation as a function of $\phi_\text{B}$. Top panel: for $\dot \phi_\text{B}>0$. Bottom panel: for $\dot \phi_\text{B}<0$.} \label{Ninf_starobinsky_1}
\end{figure}

 From Fig. \ref{VK_Starobinbsky} it can be seen that the universality is closely related to the fact that the kinetic energy of the inflaton dominates the evolution of the background during the whole bouncing phase, once it dominates at the
 quantum bounce. Then, the kinetic energy suddenly decreases at $t/t_\text{Pl} \simeq 10^{5}$, and the potential energy starts to take over ($w(\phi) \simeq -1$), whereafter the slow-roll inflation starts, as one can see clearly from
 Figs. \ref{starobinsky_1}  and \ref{starobinsky_2}.

The initial conditions that  lead to sufficiently long slow-roll inflation are shown in Fig.~\ref{Ninf_starobinsky_1}, from which we can see  that,  in order to produce at least $60$ e-folds during the slow-roll inflation,
the values of $\phi_\text{B}$ have to be in the range of
\bqn
\phi_\text{B} \in (-1.47 m_\text{Pl}, +\infty)
\eqn
for $\dot \phi_\text{B}>0$,  and
\bqn
\phi_\text{B} \in ( 3.61 m_\text{Pl}, +\infty)
\eqn
for $\dot \phi_\text{B}<0$. Within the above ranges, Fig.~\ref{Ninf_starobinsky_1} shows that the e-folds  $N_\text{inf}$ increases when the value of $\phi_\text{B}$ increases.

\begin{figure}
{\label{Starobinsky_Vdominated_1}
\includegraphics[width=8.1cm]{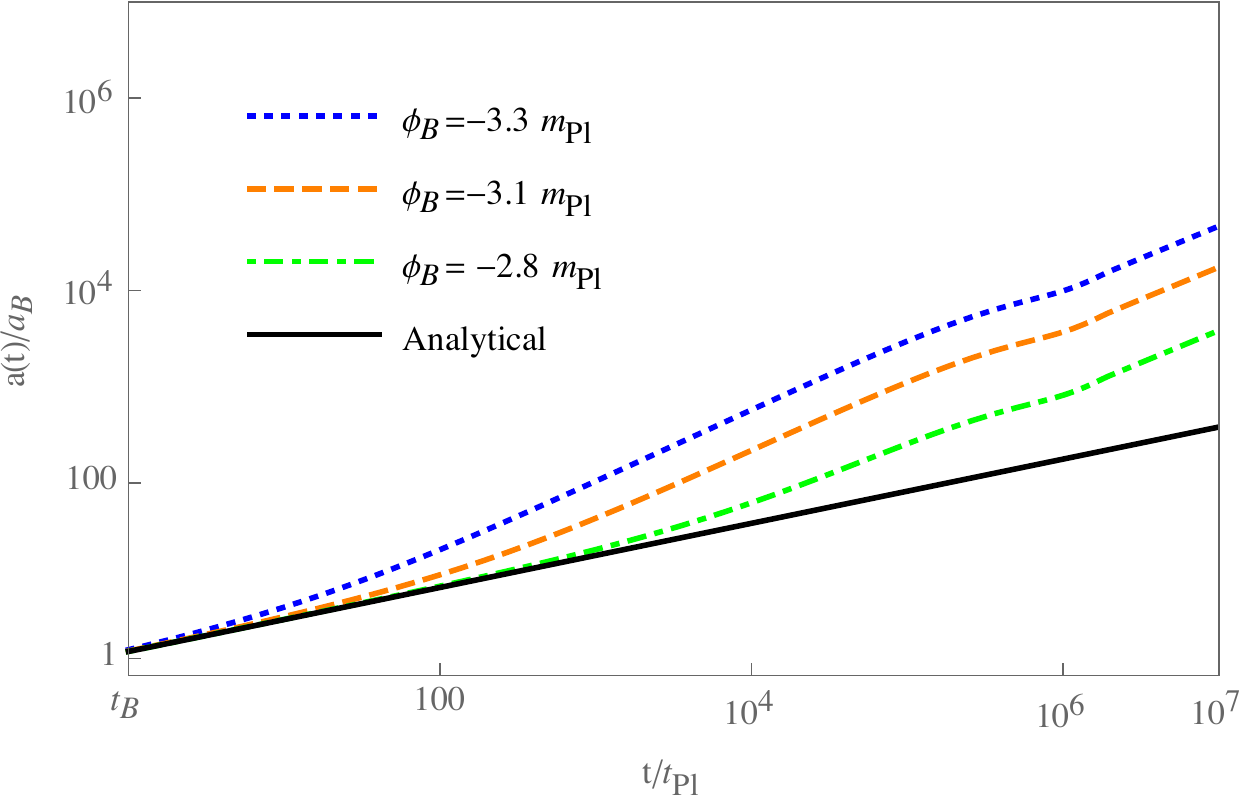}}
{\label{wphi_Starobinsky_Vdominated}
\includegraphics[width=8.1cm]{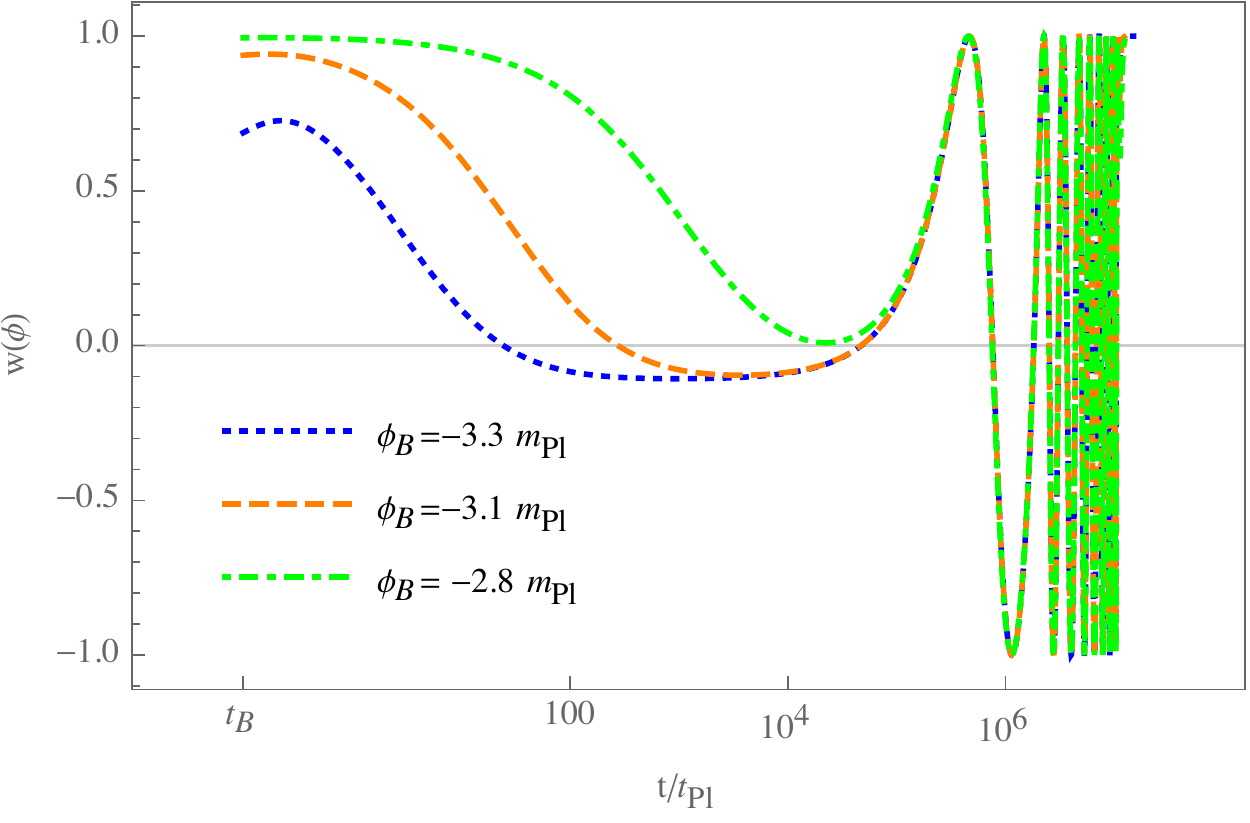}}\\
{\label{eH_Starobinbsky_Vdominated}
\includegraphics[width=8.1cm]{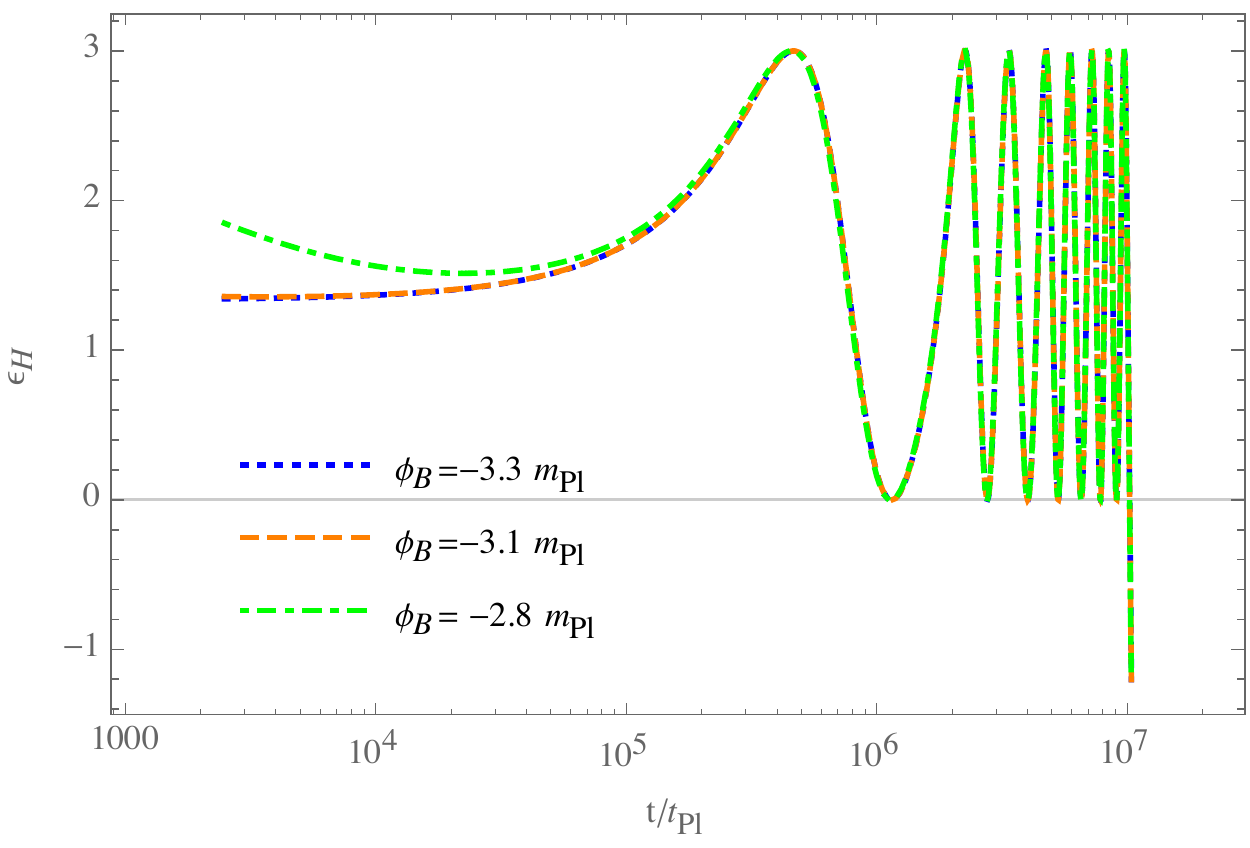}}\\
\caption{Numerical solution for the Starobinsky potential with the potential energy dominated initial conditions. For the sake of simplicity, we show here only for $\dot\phi_\text{B}>0$.
Top panel: the evolution of the scale factor $a(t)$  and the analytical solution of Eq.~(\ref{scalar_analytical}).
Middle panel: the equation of state $w(\phi)$. Bottom panel: the slow-roll parameter $\epsilon_H$.} \label{Starobinsky_Vdominated}
\end{figure}

 However, in contrast to the power-low potential cases, now the potential energy dominated initial conditions cannot lead to a slow-roll inflation, as can be seen clearly from Fig. \ref{Starobinsky_Vdominated}.
 This is consistent with what was obtained in \cite{bonga_inflation_2016, bonga_phenomenological_2016}.

\subsection{Summary of the background evolution of the FLRW universe}

Before proceeding further,  let us   summarize the main results obtained so far on the background evolutions of the flat FLRW universe in the framework of the dressed metric approach in LQC.
The evolutions can be divided into two classes, one is  dominated by the kinetic energy of the inflaton at the quantum bounce, and the other is dominated by its potential  energy  at the bounce.

In the case where the evolution of the universe is dominated by the potential energy of the inflaton, a slow-roll inflation phase may or may not be possible, depending on
the inflationary models. In particular, for the power-law potential, it is always the case, but for the Starobinsky potential the evolution never leads to a slow-roll inflationary phase
\cite{bonga_inflation_2016, bonga_phenomenological_2016}.

\begin{figure}
\includegraphics[width=8.5cm]{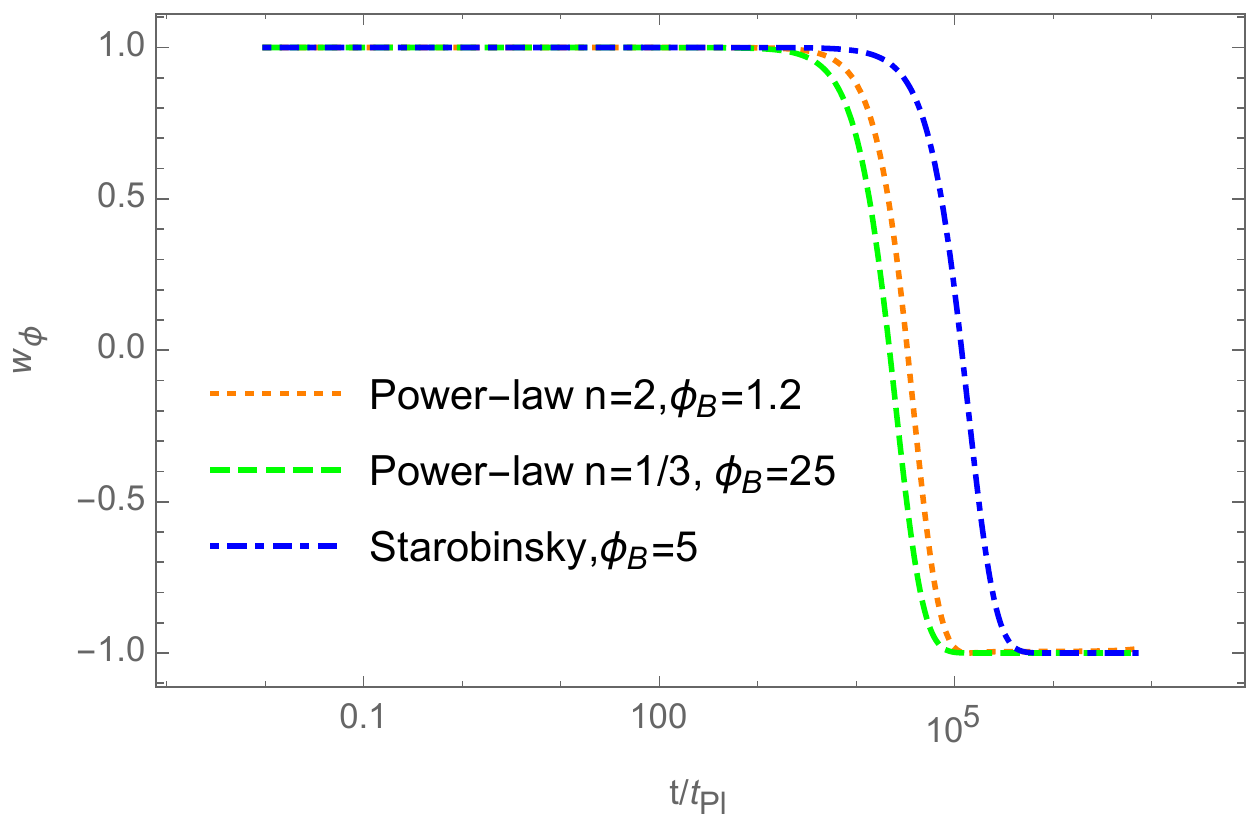}
\caption{The equation of state  $w_{\phi}$ for the power-law and Starobinsky  potentials.
We choose $ m=1.3\times 10^{-6}$ for $n=2$, $m=1.1\times 10^{-3}$ for $n=1/3$, and $M=2.5\times 10^{-6}$ for the Starobinsky potential. In all the cases we set $m_\text{Pl} =1$.} \label{EoS}
\end{figure}

In contrast, a slow-roll inflationary phase is always achieved in the case where the evolution of the universe is dominated initially by  the kinetic energy of the inflaton at the quantum bounce.
In this case, the evolution of the universe prior to preheating always experiences three different phases, {\em the bouncing, transition and slow-roll inflation}, as it can be seen clearly from the equation of
state $w(\phi)$ shown in Figs. \ref{quadratic_kinetic},  \ref{n12_kinetic_1},  \ref{n12_kinetic_2},  \ref{starobinsky_1} and  \ref{starobinsky_2}. To see this more clearly, we collect these results together
in Fig. \ref{EoS}. Note that, instead of plotting the case of the power-law potential with $n = 1/2$, we plotted out the case with $n = 1/3$ in order to show further the universal properties of the evolution.

\begin{figure}
\includegraphics[width=8.5cm]{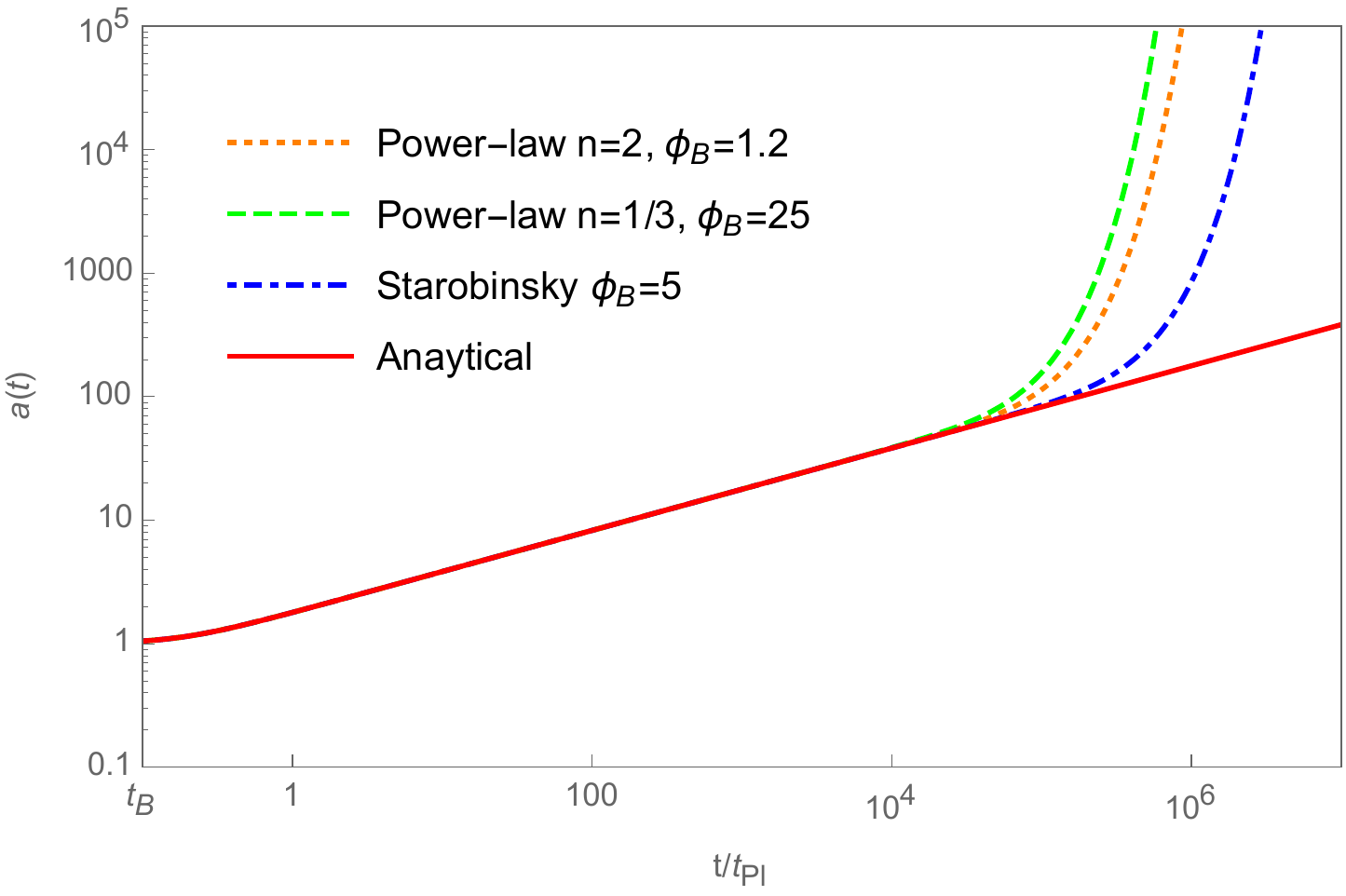}
\caption{The evolution of the scale factor $a(t)$ in the case where the evolution is dominated by the kinetic energy of the inflaton at the quantum bounce for the power-law and Starobinsky  potentials.
The parameters are chosen as the same as those given in Fig. \ref{EoS}.} \label{scalar_factor}
\end{figure}

It is also remarkable to note  that during the bouncing phase  the evolution of the scale factor $a(t)$ with the kinetic energy dominated initial conditions is universal: it is not only independent of the initial conditions
$\phi_B$ and $\dot\phi_B$, but also independent of the inflationary potentials, as shown explicitly in Figs. \ref{quadratic_kinetic},  \ref{n12_kinetic_1},  \ref{n12_kinetic_2},  \ref{starobinsky_1} and
\ref{starobinsky_2}. For the sake of comparison, we collect these curves into a single figure, Fig. \ref{scalar_factor}. The main reason is that
 the potential energy  $V(\phi)$ remains very small and the kinetic energy is completely dominant during this whole phase. For example, for the potential $V(\phi) = V_0\phi^2$,  we find that
$V(\phi)/m_\text{Pl}^4 \in(2\times 10^{-11}, 4.5\times 10^{-11})$; for $n=1/3$,  $V(\phi)/m_\text{Pl}^4 \in(9 \times 10^{-12},
1.2\times 10^{-11})$; and for the Starobinsky potential,  we have $V(\phi)/m_\text{Pl}^4 \in (7\times 10^{-13}, 7.3\times 10^{-13})$. Clearly, in this whole phase, we can safely ignore the effects of
the potential and simply set it to zero, $V(\phi) = 0$.
This explains why the evolution  of $a(t)$  is independent of inflationary models during this bouncing phase.

During the bouncing phase, the universe expands about 4 e-folds, and the exact number depends on the choices of  initial conditions of $\phi_B$ and $\dot\phi_B$, as well as on the  inflationary
models (for details, see the analysis to be carried out in the next section). Afterwards,  the kinetic energy drops dramatically (about 12 orders from its initial Planck scale), so the equation of state
$w(\phi)$ suddenly changes from $w(\phi) \simeq +1$ to $w(\phi) \simeq -1$,
whereby the slow-roll inflationary phase starts. The transition phase is normally very short in comparison to the other two phases.

\section{Background Evolution: Analytical}
\renewcommand{\theequation}{3.\arabic{equation}} \setcounter{equation}{0}
\lb{analytical}

As shown numerically in the last section, a slow-roll inflationary phase can be achieved for any potential considered so far only for the case where the kinetic energy
of the inflaton dominates the evolution of the universe at the quantum bounce. Therefore, in the rest of this paper we shall mainly focus on this case. Then,
 the evolution of the background can be divided into three phases, the bouncing, transition, and slow-roll inflation. In the following we shall first find analytical solutions
 of the scale factor in each of these three phases, and then connect them smoothly across the boundaries of these phases.

\subsection{The Bouncing Phase}

Since  the kinetic energy of the inflaton is dominating during this phase, as shown in the last section, we can safely  ignore the effects of the potential term in the equations of motion, and find
\bqn
\lb{fri}H^2 =\frac{8\pi}{3m_{\text{Pl}}^2} \frac{1}{2}\dot \phi^2 \left(1-\frac{\dot \phi^2}{2\rho_{\text{c}}}\right),\\
\lb{kg}\ddot \phi+3 H \dot \phi =0.
\eqn
The above set of equations can be solved analytically. In particular, from the Klein-Gordon equation (\ref{kg}) we obtain
\bqn\lb{phi_dot}
\dot \phi(t)= \pm \sqrt{2\rho_{\text{c}}} \left(\frac{a_\text{B}}{a(t)}\right)^{3}.
\eqn
Substituting this into Eq. (\ref{fri}), we find
\bqn\lb{scalar_analytical}
a(t)=a_\text{B}\left(1+\gamma_\text{B} \frac{t^2}{t_{\text{Pl}}^2}\right)^{1/6},
\eqn
where $\gamma_\text{B}\equiv \frac{24\pi \rho_{\text{c}}}{m_{\text{Pl}}^4} \simeq 30.9$ is a dimensionless constant. The relation between the conformal time $\eta$ and the cosmic time $t$ is then given by
\bqn
\eta(t)-\eta_B= \,_2F_1\left(\frac{1}{6},\frac{1}{2},\frac{3}{2}; -\gamma_\text{B}  \frac{t^2}{t_{\text{Pl}}^2}\right)\; t,
\eqn
which is a monotonically increasing function of $t$, 
where $_2F_1(a,b,c; z)$ is the hypergeometric function. With the analytical solution for $a(t)$, from Eq. (\ref{phi_dot}) one finds
\bqn \lb{phi_sol}
\phi(t)=\phi_{\text{B}} \pm \frac{m_{\text{Pl}}}{2\sqrt{3\pi}}\text{arcsinh}{\left(\sqrt{\gamma_\text{B}}\frac{t}{t_{\text{Pl}}}\right)},
\eqn
and
\bqn \lb{phidot_sol}
\dot \phi(t)= \pm \frac{\sqrt{2\rho_\text{c}}}{(1+\gamma_\text{B} t^2/t_{\text{Pl}}^2)^{1/2}}.
\eqn
In the top panels of Figs.~\ref{quadratic_kinetic}, \ref{n12_kinetic_1}, \ref{n12_kinetic_2}, \ref{starobinsky_1}, and \ref{starobinsky_2}, we compared the analytical solution of the scale factor given  by
Eq.~(\ref{scalar_analytical}) with the numerical (exact) one, obtained by various initial conditions and potentials, and found that all the numerical solutions are universal, and can be described well by the
analytical solution.

Note that, in contrast to the kinetic energy dominated bouncing phase,  for the potential dominated bouncing, as shown in Figs.~\ref{quadratic_Vdominated}, \ref{n12_Vdominated}
and \ref{Starobinsky_Vdominated}, the universality of the evolution of $a(t)$ is lost, and it sensitively depends on the initial conditions specified by $\phi_B$ and $\dot\phi_B$, as well as the inflationary
 potential $V(\phi)$.

\subsection{The Transition Phase}

After the bouncing phase, the universe  enters  the transition phase. During this period, the kinetic energy of the scalar field decreases dramatically, and the potential energy  soon becomes dominant.
A special point during this process is the time $t_c$, when the potential energy is equal to the kinetic energy, i.e.,  $w(\phi(t_c)) = 0$. As shown by the numerical results,   the variations of both the $e$-folds
$\delta N$ and the scalar field $\phi$ during the transition phase are almost negligible, in comparison to those obtained during the bouncing and slow-roll inflationary phases. This implies that the analytical
solutions of the scale factor $a(t)$ and scalar field $\phi(t)$ given by, respectively, Eqs.~(\ref{scalar_analytical}) and (\ref{phi_sol}) for the bouncing phase can extend their validity until $t=t_c$. In particular, 
we have 
\bqn \lb{phidot_sol_B}
\dot \phi_c\simeq \pm \frac{\sqrt{2\rho_\text{c}}}{(1+\gamma_\text{B} t_c^2/t_{\text{Pl}}^2)^{1/2}},
\eqn
where $\phi_c \equiv \phi(t_c)$. On the other hand, since  $w(\phi(t_c)) = 0$, we have 
\bqn\lb{eqaution_eq}
 \dot \phi_c= \pm \sqrt{2V(\phi_c)}.
\eqn
Then, combining the above two equations we can  find $\phi_c$ in terms of $t_c$.  To show this, let us first note 
the   examples in the second figures of Figs. \ref{quadratic_kinetic}, \ref{n12_kinetic_1}, \ref{n12_kinetic_2}, \ref{starobinsky_1} and \ref{starobinsky_2}, from which we find that 
$t_c/t_\text{Pl} \simeq 10^4 - 10^5$ (locating at points when $w_{\phi}\sim 0$ at these figures), that is, in general we have  $t_{Pl}/t_{c} \ll 1$. Then, to the leading order of  $t_{Pl}/t_{c}$, we find that,
\bqn
\lb{acphic}
\phi_c&=&\phi_\text{B} \pm \frac{m_\text{Pl}}{2\sqrt{3\pi}} \text{arcsinh}\left(\sqrt{\gamma_\text{B} \frac{t_c}{t_\text{Pl}}}\right) \nb\\
&\simeq& \phi_\text{B} \pm \frac{m_\text{Pl}}{2\sqrt{3\pi}} \ln\left(2\sqrt{\gamma_\text{B} \frac{t_c}{t_\text{Pl}}}\right),\\
\dot \phi_c &=&\pm \sqrt{\frac{\gamma_\text{B} m_\text{Pl}^4}{12 \pi (1+\gamma_\text{B} t_c^2/t_\text{Pl}^2)}} \nb\\
& \simeq& \pm \frac{m_\text{Pl}^2}{\sqrt{12 \pi}} \frac{t_\text{Pl}}{t_c}.
\eqn
With these expressions,  $t_c$ can be found explicitly for the quadratic and Starobinsky potentials.
In particular, for the quadratic potential from Eq.~(\ref{eqaution_eq}) we obtain
\bqn\lb{teq_quadratic}
t_c =
\begin{cases}
\frac{1}{m W\left(2 \sqrt{\gamma_ \text{B} }\frac{  m_\text{Pl} }{m}e^{2 \sqrt{3 \pi }\phi_\text{B}/m_\text{Pl}}\right)}, & \dot\phi_c >0,\\
 \frac{- 1}{m W_{-1}\left(- 2 \sqrt{\gamma_ \text{B} }\frac{  m_\text{Pl} }{m}e^{-2 \sqrt{3 \pi }\phi_\text{B}/m_\text{Pl}}\right)}, & \dot\phi_c >0,
\end{cases} ~~~~~
\eqn
where $W(x)$ and $W_{-1}(x)$ are the Lambert $W_0$ and $W_{-1}$ functions, respectively. For the Starobinsky potential, we find
\begin{widetext}
\bqn
t_c =
\begin{cases}
\frac{2}{3M}+\frac{e^{-\frac{2 \sqrt{3 \pi } \phi _B}{m_{\text{Pl}}}} \left[\left(\sqrt{36 \gamma_B m_{\text{Pl}}^2 e^{\frac{4 \sqrt{3 \pi } \phi _B}{m_{\text{Pl}}}}-3 M^2}+6
\sqrt{\gamma _B} m_{\text{Pl}} e^{\frac{2 \sqrt{3 \pi } \phi _B}{m_{\text{Pl}}}}\right)^{2/3}+(3M^2)^{1/3}\right]}{2 (9M)^{1/3} \sqrt{\gamma _B}
m_{\text{Pl}} \left({\sqrt{36 \gamma _B m_{\text{Pl}}^2 e^{\frac{4 \sqrt{3 \pi } \phi _B}{m_{\text{Pl}}}}-3 M^2}+6 \sqrt{\gamma _B} m_{\text{Pl}}
e^{\frac{2 \sqrt{3 \pi } \phi _B}{m_{\text{Pl}}}}}\right)^{1/3}}, & \dot\phi_c >0,\\
\frac{2}{3M} \left[1+\left(\frac{4\sqrt{\gamma_\text{B}}m_\text{Pl}}{3M}\right)^{2/3} e^{-\sqrt{16\pi/3} \phi_\text{B}/m_\text{Pl}} \right], & \dot\phi_c <0.
\end{cases}
\eqn
\end{widetext}
The detailed derivation of $t_c$ for the Starobinsky potential is presented in Appendix~\ref{teqST}.

\begin{table*}
\caption{Numerical and analytical results with different values of $\phi_\text{B}$ for a quadratic potential and $\dot \phi_\text{B}>0$. We use units $m_{\rm Pl}=1$ and $a_{\rm B}=1$.}
\lb{t_c_values}
\begin{ruledtabular}
\begin{tabular}{|c|cccccccc|}
 $\phi_\text{B}$ & $1.2  $  & $3  $   & $6 $  & $8 $ & $10 $ & $15 $ & $30 $ & $ 100  $\\
\hline
$\phi_c$ (num.) &  $3.27857$ & $5.01001$ & $7.93546$ & $9.89956$ & $11.8701$ & $16.8135$ & $31.7103$ & $ 101.52 $\\
$\phi_c$ in Eq.~(\ref{phic_2a}) &  $3.30919$ & $5.04065$ & $7.96612$ & $9.93022$ & $11.9007$ & $16.8442$ & $31.741$ & $101.552$\\
relative errors (\%) &  $0.933958$ & $ 0.611686$ & $ 0.386329$ & $ 0.309709$ & $ 0.258305$ & $ 0.182371$ & $ 0.0966972$ & $ 0.00315$\\
\hline
$10^6 \dot \phi_c$ (num.) &  $4.26214$ & $6.513$ & $10.3161$ & $12.8694$ & $15.4311$ & $21.8575$ & $41.2234$ & $131.98$\\
$10^6 \dot \phi_c$ in Eq.~(\ref{phidotc_2a}) & $4.30195$ & $6.55285$ & $10.356$ & $12.9093$ & $15.471$ & $21.8974$ & $41.2633$ & $132.017$\\
relative errors (\%)&  $0.933974$ & $ 0.611895$ & $ 0.386329$ & $ 0.309711$ & $ 0.258344$ & $ 0.18245$ & $ 0.0967077$ & $ 0.02803$\\
\hline
$N_c$ (num.) &  $4.08222$ & $3.9431$ & $3.79139$ & $3.71822$ & $3.65808$ & $3.54258$ & $3.33171$ & $ 2.94444$\\
$N_c$ in Eq.~(\ref{Nc}) &  $4.08574$ & $3.94546$ & $3.79291$ & $3.71945$ & $3.65911$ & $3.54331$ & $3.3321$ & $2.94445$\\
relative error (\%) &  $0.0862255$ & $ 0.0598019$ & $ 0.0399598$ & $ 0.0328818$ & $ 0.0280036$ & $ 0.0205946$ & $ \
0.0117816$ & $0.00034$\\
\hline
\multicolumn{9}{|c|}{$\;$}\\
  \hline
   $\phi_i$ (num.) & $3.31229$ & $5.04457$ & $7.97064$ & $9.93496$ & $11.9056$ & $16.8493$ & $31.7463$ & $101.557$\\
  $\phi_i$ in Eq.~(\ref{phi_i}) & $3.41687$ & $ 5.15005$ & $ 8.07676$ & $ 10.0413$ & $ 12.0121$ & $ 16.956$ & $ 31.8533$ & $ 101.664$\\
  relative errors (\%) & $3.15746$ & $ 2.09092$ & $ 1.33138$ & $ 1.07037$ & $ 0.89445$ & $ 0.633335$ & $ 0.336934$ & $ 0.105518$\\
\hline
$N_i$ (num.) &  $4.18948$ & $4.05307$ & $3.90334$ & $3.83086$ & $3.77119$ & $3.65638$ & $3.44632$ & $3.05956$\\
$N_i$ in Eq.~(\ref{N_i}) &  $4.2849$ & $ 4.1475$ & $ 3.99703$ & $ 3.9243$ & $ 3.86445$ & $ 3.74938$ & $ 3.53901$ & $ 3.15202$\\
relative error (\%) &  $2.27761$ & $ 2.32974$ & $ 2.40028$ & $ 2.43898$ & $ 2.47292$ & $ 2.54341$ & $ 2.6897$ & $ 3.02191$\\
\end{tabular}
\end{ruledtabular}
\end{table*}

\begin{table*}
\caption{Numerical and analytical results with different values of $\phi_\text{B}$ for a quadratic potential and $\dot \phi_\text{B}<0$. We use units $m_{\rm Pl}=1$ and $a_{\rm B}=1$.}
\lb{t_c_values_2}
\begin{ruledtabular}
\begin{tabular}{|c|cccccccc|}
 $\phi_\text{B}$ & $5.3  $  & $6  $   & $8 $  & $10 $& $12 $ & $15 $ & $30 $ & $ 40  $\\
\hline
$\phi_c$ (num.) & $3.21517$ & $3.94896$ & $6.01803$ & $8.06595$ & $10.1027$ & $13.1457$ & $28.2707$ & $38.3202$\\
$\phi_c$ in Eq.~(\ref{phic_2a})& $3.18455$ & $3.91832$ & $5.98738$ & $8.03529$ & $10.0721$ & $13.1151$ & $28.24$ & $38.2896$\\
relative errors (\%)& $0.952133$ & $ 0.775671$ & $ 0.509318$ & $ 0.380079$ & $ 0.303482$ & $ 0.233247$ & $ 0.108462$ & $ 0.080017$\\
\hline
$10^6 \dot \phi_c$ (num.) &  $-4.17972$ & $-5.13364$ & $-7.82344$ & $-10.4857$ & $-13.1336$ & $-17.0895$ & $-36.7519$ & $-49.8163$\\
$10^6 \dot \phi_c$ in Eq.~(\ref{phidotc_2a}) & $-4.13992$ & $-5.09382$ & $-7.78359$ & $-10.4459$ & $-13.0937$ & $-17.0496$ & $-36.712$ & $-49.7765$\\
relative errors (\%) & $0.961286$ & $ 0.781726$ & $ 0.511909$ & $ 0.381523$ & $ 0.304391$ & $ 0.233919$ & $ 0.108579$ & $ 0.0800772$\\
\hline
$ N_c$ (num.) &  $4.1026$ & $4.03269$ & $3.89019$ & $3.79158$ & $3.71595$ & $3.62766$ & $3.37149$ & $3.2699$\\
$N_c$ in Eq.~(\ref{phidotc_2a}) &  $4.09854$ & $4.02942$ & $3.88809$ & $3.79003$ & $3.71472$ & $3.62672$ & $3.37106$ & $3.26958$\\
relative errors (\%)&  $0.0992386$ & $ 0.0811183$ & $ 0.0540453$ & $ 0.0409425$ & $ 0.0331148$ & $ 0.0258796$ & $ \
0.0127129$ & $ 0.00960974$\\
\hline
\multicolumn{9}{|c|}{$\;$}\\
\hline
$\phi_i$ (num.) & $3.17574$ & $3.91015$ & $5.98013$ & $8.02847$ & $10.0655$ & $13.1087$ & $28.234$ & $38.2837$\\
$\phi_i$ in Eq.~(\ref{phi_i})& $3.12504$ & $3.85941$ & $5.92935$ & $7.97768$ & $10.0147$ & $13.0579$ & $28.1832$ & $38.2329$\\
relative errors (\%)& $1.59655$ & $ 1.29764$ & $ 0.849147$ & $ 0.632656$ & $ 0.504679$ & $ 0.387547$ & $ 0.179945$ & $ \
0.132708$\\
\hline
$N_i$ (num.) &  $4.22812$ & $4.1562$ & $4.01083$ & $3.91087$ & $3.83446$ & $3.74546$ & $3.48806$ & $3.38619$\\
$N_i$ in Eq.~(\ref{N_i}) &  $4.21348$ & $4.14326$ & $4.00032$ & $3.9015$ & $3.82575$ & $3.73735$ & $3.48099$ & $3.37935$\\
relative error (\%) &  $0.346113$ & $ 0.311432$ & $ 0.261958$ & $ 0.239621$ & $ 0.227182$ & $ 0.216663$ & $ 0.202858$ & $ \
0.202082$\\
\end{tabular}
\end{ruledtabular}
\end{table*}

\begin{table*}
\caption{Numerical and analytical results with different values of $\phi_\text{B}$ for the Starobinsky potential and $\dot \phi_\text{B}>0$. We use units $m_{\rm Pl}=1$ and $a_{\rm B}=1$.}
\lb{starobinsky}
\begin{ruledtabular}
\begin{tabular}{|c|cccccccc|}
$\phi_\text{B}$ & $-1.47$ & $-1.30$ & $ -1.1$ & $ -0.4$ & $0.1$ & $0.2$ & $1$ & $4$\\
\hline
$\phi_c$ (num.) & $0.928828$ & $1.09732$ & $1.29648$ & $1.99585$ & $2.49582$ & $2.59582$ & $3.39582$ & $6.39582$\\
$\phi_c$ in Eq.~(\ref{phic_2a})& $0.959718$ & $1.1281$ & $1.3272$ & $2.02652$ & $2.52649$ & $2.62649$ & $3.42648$ & $6.42648$\\
relative errors (\%)& $3.32573$ & $ 2.80515$ & $ 2.36924$ & $ 1.53662$ & $ 1.22872$ & $ 1.18138$ & $ 0.903066$ & $ \
0.47948$\\
\hline
$10^6 \dot \phi_c$ (num.) &  $0.599505$ & $0.606327$ & $0.610156$ & $0.613022$ & $0.613174$ & $0.613181$ & $0.613196$ & $0.613196$\\
$10^6 \dot \phi_c$ in Eq.~(\ref{phidotc_2a}) & $0.601132$ & $0.60714$ & $0.610515$ & $0.613043$ & $0.613176$ & $0.613183$ & $0.613196$ & $0.613196$\\
relative errors (\%) & $0.271265$ & $ 0.134118$ & $ 0.0588624$ & $ 0.00335137$ & $ 0.000448314$ & $ 0.000301966$ & $ \
0.0000209013$ & $ 4.22436*10^-6$\\
\hline
$ N_c$ (num.) &  $4.73968$ & $4.73738$ & $4.73611$ & $4.73516$ & $4.73511$ & $4.73511$ & $4.73511$ & $4.7351$\\
$N_c$ in Eq.~(\ref{phidotc_2a}) &  $4.74174$ & $4.73843$ & $4.73658$ & $4.7352$ & $4.73513$ & $4.73513$ & $4.73512$ & $4.73512$\\
relative errors (\%)&  $0.0434831$ & $ 0.0220978$ & $ 0.00990369$ & $ 0.000787333$ & $ 0.000327793$ & $ \
0.000309995$ & $ 0.000270085$ & $ 0.000286131$\\
\hline
\multicolumn{9}{|c|}{$\;$}\\
\hline
$\phi_i$ (num.) & $0.964351$ & $1.13323$ & $1.33261$ & $2.03215$ & $2.53212$ & $2.63212$ & $3.43212$ & $6.43212$\\
$\phi_i$ in Eq.~(\ref{phi_i})& $1.01542$ & $1.18417$ & $1.38348$ & $2.08296$ & $2.58293$ & $2.68293$ & $3.48293$ & $6.48293$\\
relative errors (\%)& $5.29573$ & $ 4.4954$ & $ 3.8172$ & $ 2.50043$ & $ 2.00661$ & $ 1.93038$ & $ 1.48042$ & $ 0.789942$\\
\hline
$N_i$ (num.) &  $4.85272$ & $4.85166$ & $4.85108$ & $4.85066$ & $4.85063$ & $4.85063$ & $4.85063$ & $4.85063$\\
$N_i$ in Eq.~(\ref{N_i}) &  $4.8497$ & $4.84707$ & $4.8456$ & $4.84451$ & $4.84445$ & $4.84445$ & $4.84444$ & $4.84444$\\
relative error (\%) &  $0.0622625$ & $ 0.094637$ & $ 0.113005$ & $ 0.126749$ & $ 0.127458$ & $ 0.127488$ & $ \
0.127546$ & $ 0.127531$\\
\end{tabular}
\end{ruledtabular}
\end{table*}

\begin{table*}
\caption{Numerical and analytical results with different values of $\phi_\text{B}$ for the Starobinsky potential and $\dot \phi_\text{B}<0$. We use units $m_{\rm Pl}=1$ and $a_{\rm B}=1$.}
\lb{starobinsky_b}
\begin{ruledtabular}
\begin{tabular}{|c|cccccccc|}
 $\phi_\text{B}$ & $3.61$ & $3.7$ & $3.8$ & $ 3.9$ & $4.1$ & $ 4.3$ & $ 5$ & $8$\\
\hline
$\phi_c$ (num.) & $1.21293$ & $1.30332$ & $1.40361$ & $1.50381$ & $1.70402$ & $1.90411$ & $2.60418$ & $5.60418$\\
$\phi_c$ in Eq.~(\ref{phic_2a})& $1.18224$ & $1.27263$ & $1.37293$ & $1.47313$ & $1.67335$ & $1.87344$ & $2.57351$ & $5.57352$\\
relative errors (\%)& $2.53051$ & $ 2.35464$ & $ 2.18599$ & $ 2.04003$ & $ 1.79998$ & $ 1.61067$ & $ 1.1776$ & $ 0.547214$\\
\hline
$10^6 \dot \phi_c$ (num.) &  $-0.608918$ & $-0.61024$ & $-0.611236$ & $-0.611895$ & $-0.612623$ & $-0.612943$ & $-0.613182$ & $-0.613195$\\
$10^6 \dot \phi_c$ in Eq.~(\ref{phidotc_2a}) & $-0.608407$ & $-0.609875$ & $-0.610986$ & $-0.611727$ & $-0.612547$ & $-0.61291$ & $-0.61318$ & $-0.613196$\\
relative errors (\%) & $0.0838578$ & $ 0.0599164$ & $ 0.0407656$ & $ 0.0274843$ & $ 0.0123245$ & $ 0.00546677$ & $ \
0.000304264$ & $ 0.000112727$\\
\hline
$ N_c$ (num.) &  $4.73794$ & $4.73706$ & $4.7364$ & $4.73596$ & $4.73548$ & $4.73527$ & $4.73511$ & $4.7351$\\
$N_c$ in Eq.~(\ref{phidotc_2a}) &  $4.73773$ & $4.73693$ & $4.73632$ & $4.73592$ & $4.73547$ & $4.73527$ & $4.73513$ & $4.73512$\\
relative errors (\%)&  $0.00437058$ & $ 0.0027701$ & $ 0.00166291$ & $ 0.000973656$ & $ 0.000256096$ & $ \
0.0000465156$ & $ 0.00026842$ & $ 0.000294261$\\
\hline
\multicolumn{9}{|c|}{$\;$}\\
\hline
$\phi_i$ (num.) & $1.17634$ & $1.26682$ & $1.36718$ & $1.46742$ & $1.66768$ & $1.86779$ & $2.56788$ & $5.56788$\\
$\phi_i$ in Eq.~(\ref{phi_i})& $1.12548$ & $1.21597$ & $1.31634$ & $1.41659$ & $1.61686$ & $1.81698$ & $2.51707$ & $5.51707$\\
relative errors (\%)& $4.32418$ & $ 4.01394$ & $ 3.71831$ & $ 3.46372$ & $ 3.0472$ & $ 2.72049$ & $ 1.97868$ & $ 0.912557$\\
\hline
$N_i$ (num.) &  $4.85439$ & $4.85322$ & $4.85234$ & $4.85177$ & $4.85113$ & $4.85085$ & $4.85064$ & $4.85063$\\
$N_i$ in Eq.~(\ref{N_i}) &  $4.84764$ & $4.84665$ & $4.8459$ & $4.84541$ & $4.84487$ & $4.84463$ & $4.84445$ & $4.84444$\\
relative error (\%) &  $0.138954$ & $ 0.135372$ & $ 0.13271$ & $ 0.130961$ & $ 0.129044$ & $ 0.128204$ & $ 0.127572$ & $ \
0.12753$\\
\end{tabular}
\end{ruledtabular}
\end{table*}


Given $t_c$, we are now able to calculate $a(t_c)$, $\phi(t_c)$ and $\dot \phi(t_c)$, which are given by
\bqn\lb{ac_2a}
a_c = a_\text{B} \left(1+\gamma_\text{B} \frac{t_c^2}{t_\text{Pl}^2}\right)^{1/6},\\
\lb{phic_2a}
\phi_c = \phi_\text{B} + \frac{m_\text{Pl}}{2 \sqrt{3 \pi }} \ln \left( 2 \sqrt{\gamma_\text{B}} \frac{t_c}{t_\text{Pl}}\right),
\eqn
and
\bqn\lb{phidotc_2a}
\dot \phi_c = \frac{m m_{\text{Pl}}}{2 \sqrt{3 \pi }} W\left(2 \sqrt{\gamma_\text{B}}\frac{ m_\text{Pl}}{m} e^{2 \sqrt{3 \pi } \phi_\text{B}/m_\text{Pl}}\right).
\eqn
Then, we obtain,
\bqn
\lb{Nc}
N_c &\equiv & \ln \left(\frac{a_c}{a_\text{B}}\right)\nb\\
&=& \frac{1}{6} \ln \left[1+\frac{\gamma_\text{B} m_\text{Pl}^2}{m^2 W\left(2 \sqrt{\gamma_\text{B}}\frac{ m_\text{Pl}}{m} e^{2 \sqrt{3 \pi } \phi_\text{B}/m_\text{Pl}}\right)^2}\right].\nb\\
\eqn
The numerical values of $N_c$, $\phi_c$, and $\dot \phi_c$ derived from the above expressions are presented in Table.~\ref{t_c_values}-\ref{starobinsky_b} for different potentials and signs of $\dot \phi_c$.
From these tables it can be seen that the upper bounds of errors of these quantities between their numerical (exact) and analytical values are less than one percent.

Right after the moment $t=t_c$, the kinetic energy of the scalar field is continuously decreasing, and the universe remains decelerating until $w=-1/3$, at which point we have
\bq
\lb{is}
\ddot a(t)=0,\;\;\;\; \dot \phi^2=V(\phi), \;\;\; (w(\phi) = -1/3).
\eq
Denoting this moment as  $t_i$, we can see that after tis moment  the universe enters  an accelerating  phase, $\ddot{a}(t) > 0$. Right after $t_i$, the slow-roll parameter $|\epsilon_H|$ is still  large and oscillating around
its zero point, but soon becomes very small, $|\epsilon_H| \ll 1$, and then the slow-roll inflationary phase starts.  Therefore, practically we can consider the moment  $t=t_i$ as the beginning of the slow-roll inflation.
To estimate the values of $\phi(t_i)$ and $a(t_i)$, we expand  $\phi(t)$ and   $a(t)$ at $t=t_c$ as
\bqn\lb{phieq_expansion}
\phi(t) &=& \phi_c+ t_c \dot \phi_c \ln\frac{t}{t_c},\\
a(t) &=& a_c \left(1+ t_c H_c \ln \frac{t}{t_c}\right),
\eqn
where $H_c$ can be calculated by using the Friedmann equation,
\bqn
H_c = \sqrt{\frac{8\pi G}{3} \left(\frac{\dot \phi^2_c}{2}+V(\phi_c)\right)}.
\eqn
Note that in writing the above equation we had ignored the term $\rho(t_c)/\rho_c$ in Eq.~(\ref{friedmann}), which is of the order $\rho(t_c)/\rho_c \simeq 10^{-12}$, as shown in the last section Then,  at $t=t_i$ we have
\bqn
\phi_i = \phi_c + \dot \phi_c t_c \ln \frac{t_i}{t_c},\\
\dot \phi_i = \frac{t_c }{t_i} \;  \dot \phi_c.
\eqn
Given $w=-1/3$ at $t=t_i$, we also have
\bqn\lb{equation_tac}
\dot \phi_i^2 = V(\phi_i).
\eqn
In general it is very difficult  to solve above equations to get $t_i$ for any given  potential $V(\phi)$. However, we can always   expand $V(\phi_i)$ at $\phi_c$ as
\bqn
V(\phi_i) = V(\phi_c)+ V_{, \phi}(\phi_c) t_c \dot \phi_c \ln\frac{t_i}{t_c},
\eqn
and Eq.~(\ref{equation_tac}) yields
\bqn
 \dot \phi_c   =\pm\frac{t_i}{t_c}\left(\sqrt{V(\phi_c)} + \frac{t_c \dot \phi_c}{2} \frac{V_\phi(\phi_c)}{\sqrt{V(\phi_c)}} \ln \frac{t_i}{t_c}\right),
\eqn
where $``\pm"$ correspond to $\dot \phi_c>0$ and $\dot \phi_c <0$, respectively. Solving the above equations we find
\begin{widetext}
\bqn
t_i =
\begin{cases}
\frac{2 \sqrt{V(\phi_c)}}{V_\phi(\phi_c) W\left[\frac{2 \sqrt{V(\phi_c)}}{V_\phi(\phi_c) t_c}\exp\left(\frac{2 V(\phi_c)}{t_c \dot \phi_c V_\phi(\phi_c)}\right)\right]}, & \dot \phi_c >0, \\
- \frac{2 \sqrt{V(\phi_c)}}{ V_\phi(\phi_c) W_{-1}\left[- \frac{2 \sqrt{V(\phi_c)}}{V_\phi(\phi_c) t_c}\exp\left(\frac{2 V(\phi_c)}{t_c \dot \phi_c V_\phi(\phi_c)}\right)\right]}, & \dot \phi_c <0.
\end{cases}
\eqn
\end{widetext}
Once  $t_i$ is given, we can then calculate $a_i$ and $\phi_i$, which are given by
\bqn
a_i &=& a_c \left(1+ t_c H_c \ln \frac{t_i}{t_c}\right),\\
\phi_i &=& \phi_c + t_c \dot \phi_c \ln \frac{t_i}{t_c}. \lb{phi_i}
\eqn
Then, we find that
\bqn\lb{N_i}
N_i &\equiv& \ln\left(\frac{a_i}{a_B}\right)\nb\\
&=& N_c + \ln\left(1+ t_c H_c \ln \frac{t_i}{t_c}\right).
\eqn
The numerical values of $\phi_i$ and $N_i$ obtained  from the above expressions and the numerical ones are presented in Table.~\ref{t_c_values}-\ref{starobinsky_b} for different  potentials (quadratic and Starobinsky potentials)
and signs of $\dot \phi_c$, from which it can be seen that the upper bounds of errors of our analytical estimations for $\phi_i$ are less than $5.3\%$, while  for   $N_i$ they are less than $3.1\%$.

\subsection{The Slow-Roll Inflationary Phase}

After $t=t_i$, the universe soon enters   the slow-roll inflationary phase. During it, the potential energy of the scalar field is dominating.
To ensure the slow-roll evolution of the background, we also need to impose two additional conditions,
\bq
\lb{SCs}
(i) \; \frac{1}{2}\dot \phi^2 \ll V(\phi), \;\;\;\;\;  (ii) \; |\ddot \phi| \ll |H\dot \phi|.
\eq
Then, the Friedmann and Klein-Gordon equations can be approximated by
\bqn
H^2 \simeq \frac{8\pi}{3m_\text{Pl}^2} V(\phi),\\
3 H \dot \phi +\frac{dV(\phi)}{d\phi} \simeq 0.
\eqn
From the Friedmann equation for a slowly varying $V(\phi)$, we obtain,
\bqn\lb{scalar_slow}
a(t)  \propto e^{H_{\text{inf}} t },
\eqn
where $H_\text{inf}$ denotes the Hubble parameter during the slow-roll inflation.
Thus, the e-folds $N_\text{inf}$ can be calculated via the relation
\bqn\lb{Ninf_general}
N_\text{inf} &\equiv& \ln\left(\frac{a_\text{end}}{a_i}\right) = \int_{t_i}^{t_\text{end}} H(t) dt \nb\\
&=&  \int_{\phi_i}^{\phi_\text{end}} \frac{H}{\dot \phi} d\phi \simeq \frac{8\pi}{m_{\rm Pl}^2}\int_{\phi_\text{end}}^{\phi_i} \frac{V}{V_\phi} d\phi.
\eqn

\section{Primordial scalar and tensor perturbations}
\renewcommand{\theequation}{4.\arabic{equation}} \setcounter{equation}{1}
\lb{general_solution}

Let us now turn to consider the linear perturbations of the background of the universe presented in the last two sections. In general, there are mainly two different approaches to implement cosmological perturbations
in the framework of LQC, {\em the  dressed metric} \cite{agullo_quantum_2012, agullo_extension_2013, agullo_pre-inflationary_2013} {\em and  deformed algebra approaches} \cite{barrau_conceptual_2016, bounce}.
 In both approaches, the primordial perturbations with quantum gravitational effects have been studied (see, for example,
 Refs. \cite{schander_primordial_2016,bolliet_observational_2016,bolliet_comparison_2015,agullo_pre-inflationary_2013,agullo_detailed_2015,bonga_inflation_2016,bonga_phenomenological_2016,ashtekar_quantum_2017,%
 mielczarek_observational_2010} and references therein). In particular, the deformed algebra approach, with some (reasonable) assumptions, seems already in conflict with current observations \cite{bolliet_observational_2016,Grain16}.
 Therefore, in this paper we shall focus ourselves only on  {\em the dressed metric approach}.

 \begin{figure}
{
\includegraphics[width=8.1cm]{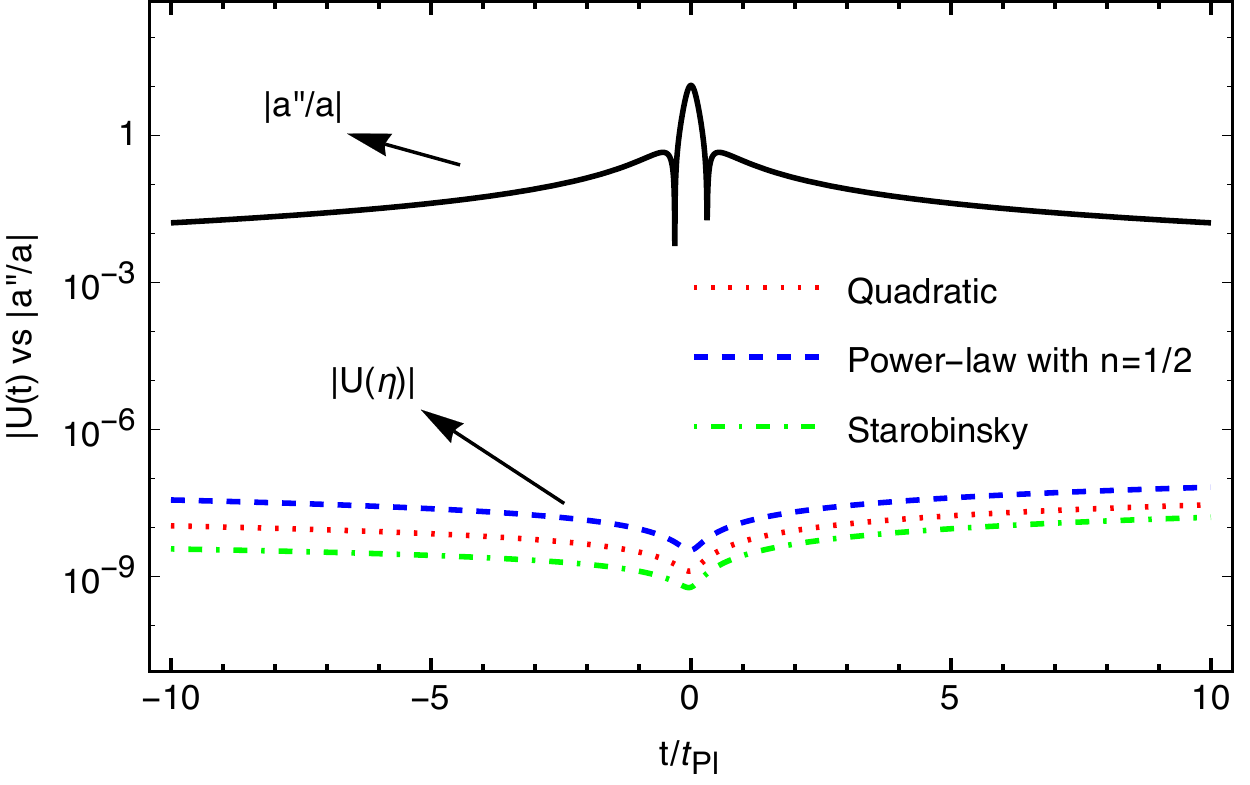}}
\caption{Comparison between $|U(\eta)|$ and $|a''/a|$ for different inflationary potentials during the bouncing phase for the case where  kinetic energy of the inflaton  dominates the evolutiion of the background at the quantum  bounce.
 In plotting these curves,  we assumed $\dot \phi_\text{B}>0$,  and set $\phi_\text{B}=3 m_\text{Pl}, \; 5 m_\text{Pl}, \;2 m_\text{Pl}$, respectively,  for the quadratic potential, the power-law potential with $n=1/2$, and the Starobinsky potential.}
\label{UvsA}
\end{figure}

\subsection{Cosmological perturbations in the quantum FRLW background spacetime}

In this subsection, we present a brief introduction of the dressed metric approach to the cosmological scalar and tensor perturbations in LQC. In the standard inflationary framework, both the cosmological scalar and tensor perturbations are treated as quantum fields in the classical FRLW background spacetime. In this treatment, as the energy density of the background is well below the Planck energy, the quantum gravitational effects of the background spacetime is negligible.  However, the above formalism breaks down when one extends the standard inflationary phase to the pre-inflationary era (near the bounce), in which case the quantum gravitational effects on the background has to be taken into account. For this reason, one has to study the quantum cosmological perturbations on a quantum FRLW background spacetime.

The quantum theory of cosmological perturbations on a quantum FRLW spacetime was developed in the context of LQC in \cite{agullo_quantum_2012, agullo_extension_2013, agullo_pre-inflationary_2013} based on the framework developed in \cite{AKL09}. In this picture, the quantum cosmological perturbation fields propagate on a quantum geometry described by the quantum state $\Psi_0(a, \phi)$, which is free of singularities. Moreover, as shown in \cite{agullo_extension_2013}, one can treat the cosmological perturbation field as test fields if their back-reaction on the FLRW spacetime is small. In this case, the mathematical treatments can be very much simplified.

In the test field approximation, the dynamics of the cosmological perturbation fields on the quantum geometry $\Psi_0(a,\phi)$ is equivalent to that of quantum fields on a quantum modified effective geometry described by a dressed metric $\tilde{g}_{ab}$ \cite{agullo_quantum_2012, agullo_extension_2013, agullo_pre-inflationary_2013},
\bqn
\tilde{g}_{ab}dx^a dx^b = \tilde{a} (-d \tilde \eta^2 + dx_id x^i),
\eqn
where the dressed scale factor $\tilde a$ and the dressed conformal time $\tilde \eta$ are
\bqn
\tilde a &=& \left(\frac{\langle \hat H_0^{-1/2} \hat a^4 \hat H_0^{-1/4} \rangle}{\langle \hat H_0^{-1} \rangle}\right)^{1/4},\\
d \tilde \eta &=& \langle \hat H_0^{-1/2} \rangle  ( \langle \hat H_0^{-1/2} \hat a^4 \hat H_0^{-1/2}\rangle)^{1/2} d\phi.
\eqn
Here $\hat H_0$ is the background Hamiltonian and the expectation values are taken with respect to the background quantum geometry state given by $\Psi_0(a,\phi)$. In this dressed metric, the equations of motion of the operators representing scalar and tensor perturbations are formally the same as the equations appearing in classical spacetimes, which in Fourier space are
\bqn\lb{dressed_scalar}
\mu_k^{(s)}(\tilde \eta)''+\left(k^2-\frac{\tilde a''}{\tilde a}+\tilde U(\tilde \eta)\right)\mu_k^{(s)}(\tilde \eta)=0,
\eqn
and
\bqn\lb{dressed_tensor}
\mu_k^{(t)}(\tilde \eta)''+\left(k^2-\frac{\tilde a''}{\tilde a}\right)\mu_k^{(t)}(\tilde \eta)=0,
\eqn
where $\mu_k^{(s)}(\tilde \eta)=z_s \mathcal{R}_k$ with $\mathcal{R}_k$ denotes the cosmological comoving curvature perturbation and $z_s(\tilde \eta)=\tilde a\dot \phi/H$, $\mu_k^{(t)}(\tilde \eta) = \tilde a h_k$ denotes tensor perturbation, and 
\bqn
\lb{up}
\tilde U(\tilde \eta) &=& \frac{\langle \hat H_0^{-1/2} \hat a^2 \hat U(\phi) \hat a^2 \hat H_0^{-1/2} \rangle}{\langle  \hat H_0^{-1/2} \hat a^4 \hat H_0^{-1/2}\rangle}.
\eqn

It should be note that although the effective equation of motions (\ref{dressed_scalar}) and (\ref{dressed_tensor}) takes the same form as their classical version, the background quantities in these equations, namely $\tilde a$, $\tilde U$, and $\tilde \eta$ are different from their classical counterparts. In the opposite, these quantities are now representing the quantum expectation values in the background state $\Psi_0(a,\phi)$. However, it is pointed out in \cite{agullo_pre-inflationary_2013} that for sharply peaked background states $\Psi_0(a,\phi)$, the dressed effective quantities (like $\tilde a$, $\tilde \eta$, and $\tilde U(\tilde U)$) are very well approximated by their peaked values ($a$, $\eta$, and $U(\phi)$) from the deep Planck era up to the entire expanding phase, where
\bqn
U(\phi) = a^2 \left(\mathfrak{f}^2 V(\phi)+2 \mathfrak{f} V_{,\phi}(\phi)+V_{,\phi\phi}(\phi)\right), \lb{upB}
\eqn
with $\mathfrak{f}\equiv \sqrt{24 \pi G}\dot \phi/\sqrt{\rho}$. 
With the above approximation, the equation of motion of the scalar perturbation becomes
\bqn\lb{eom_scalar}
\mu_k^{(s)}(\eta)''+\left(k^2-\frac{a''}{a}+U(\eta)\right)\mu_k^{(s)}(\eta)=0.
\eqn
During the bouncing phase, if the energy density of the scalar field is dominated by its kinetic energy at the quantum bounce, then it will dominate the evolution of the background during the whole bouncing phase, as shown in the last sections. In this case, it can be shown that $ U(\eta)$ is negligible in comparison with $a''/a$. Fig.~\ref{UvsA} shows  the absolute values of $U(\eta)$ and $a''/a$ for different potentials, from which we can see clearly that $|U(\eta)/(a''/a)| \ll 1$ for any given potentials. Thus,    the $U(\eta)$ term in Eq.~(\ref{eom_scalar}) can be safely ignored during the bouncing phase. 

In the slow-roll inflationary  phase, during which the energy density is dropped down to about $10^{-12} \rho_\text{c}$, the equation of motion reduces to the classical one obtained in general relativity,
\bqn
\mu_k^{(s)}(\eta)''+\left(k^2-\frac{z_s''}{z_s}\right)\mu_k^{(s)}(\eta)=0.
\eqn
For the tensor perturbations,  similar to the scalar one, the equation of motion becomes,
\bqn
\lb{eomt}
\mu_k^{(t)}(\eta)''+\left(k^2-\frac{a''}{a}\right)\mu_k^{(t)}(\eta)=0,
\eqn
which takes exactly the same form as that in classical general relativity.

\subsection{The characteristic length during the bouncing phase}

\begin{figure}
{\includegraphics[width=8.1cm]{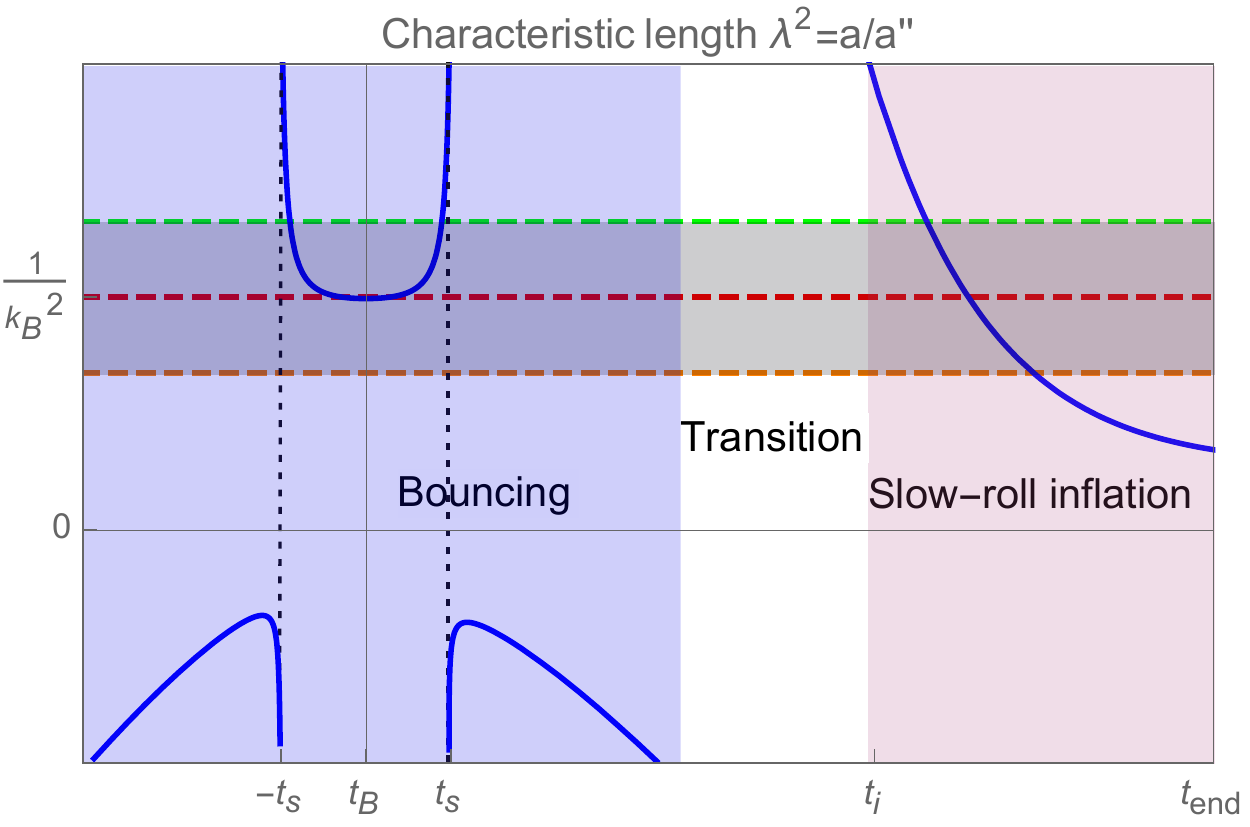}}
\caption{Schematic plot of $a/a''$, where $a''/a|_{t=t_{s}}=0$ with $t_s \sim 0.2 t_\text{Pl}$, $\ddot a(t_i)=0$ with $t_i$ being the starting time of the inflationary phase,
and during the slow-roll inflation, $a/a''=L_\text{H}^2/2$. In drawing this figure, we analytically extend the solution of $a(t)$  to a contracting phase $t<t_\text{B}$.} \label{length}
\end{figure}

The evolutions of the scalar and tensor perturbations depend on both the background and the wave-number $k$ of the perturbations. As we consider only the kinetic energy dominated case, both
scalar and tensor perturbations follow the same equation of motion during the bouncing phase ($t/t_\text{Pl} \le 10^{4}$). In this case, the term $a''/a$ in Eq.~(\ref{eom_scalar}) defines a characteristic radius $\lambda$ as
\bqn
\lambda^2=\frac{a}{a''},
\eqn
for $a'' > 0$, which plays the same role as that of the comoving Hubble radius $L_H$ defined as $L_H=(aH)^{-1}$. However, for a better understanding, we find that, instead of $L_H$,
 it is more proper to use $\lambda^2$, as shown schematically in Fig.~\ref{length}. For example, when the modes are inside this radius (i.e., $1/k^2< \lambda^2 $)
the solution of Eq.~(\ref{eom_scalar}) is of the form,
\bqn
\mu^{s,t}_k(\eta) \sim e^{\pm i \int \sqrt{k^2-a''/a} d\eta}.
\eqn
 When the modes are outside of the radius (i.e., $1/k^2> \lambda^2$), we have growing/decaying  solutions,
\bqn
\mu^{s,t}_k(\eta) \sim e^{\pm \int \sqrt{a''/a-k^2}d\eta }.
\eqn
Here we would like to note that $\lambda^2$ is only defined when $a''>0$, and right after the bounce $a''$ changes its sign from positive to negative. 
This defines two special points $t=\pm t_s$ as shown in Fig.~\ref{length}, at which we have
\bqn
a''(t_s)=0.
\eqn
Another specific time is the transition point $t=t_i$ which divides the decelerating and accelerating expansions of the universe, i.e., at $t_i$ we have $\ddot a=0$. This point is also considered as the starting point of the slow-roll
 inflationary phase in this paper.

The term $a''/a$ has its maximum at the bounce, $a''/a|_{t=t_B}= a_{\text{B}}^2 \gamma_{\rm B} m_{\text{Pl}}^2/3$, which defines a characteristic energy scale,
\bq
\lb{kB}
k_\text{B}\equiv \left. \sqrt{\frac{a''}{a}}\right|_{t=t_{\rm B}} = \sqrt{\frac{\gamma_{\rm B}}{3}}\;  a_{\text{B}} m_{\text{Pl}},
\eq
the blue solid curve shown in Fig.~\ref{length}, so that we can use it to classify different modes. Some modes with large values of $k^2\gg k_\text{B}^2$ (the region below the low (orange) dashed line in Fig.~\ref{length}) are
 inside the radius all the time until it exits the Hubble horizon during the slow-roll inflation. Some of the modes with smaller  $k^2\ll k_\text{B}^2$ (the region above the upper (green) dashed line  in Fig.~\ref{length}) exit and
 re-enter the radius during the bouncing phase, and will finally re-exit the Hubble radius during the slow-roll inflation. Since the modes with $k\gg k_\text{B}$ are inside the radius during the whole pre-inflationary phase, they
 will have the same power-law spectra as those given in GR. We are interested in the modes with $k\simeq k_\text{B}$ (the shaded region in Fig.~\ref{length}) as they are modes whose physical energy during the bouncing
 phase are of the Planck scale $k_{\text{phy}}=k/a_{\text{B}} \simeq m_{\text{Pl}}$. However, the perturbations for these modes have different behavior when they are inside and outside the radius, which makes Eq.~(\ref{eom_scalar})
very difficult to solve analytically.

\subsection{Perturbations During the Bouncing Phase}

In general, we can consider the equation of motion of primordial perturbations as a specific type of the Schr\"odinger equation,
\bqn\lb{sch}
\mu_k''(\eta)+\left[k^2-\mathscr{V}(\eta)\right]\mu_k(\eta)=0,
\eqn
in which $\mathscr{V}(\eta)$ serves as an effective potential, and for the scalar and tensor perturbations we have
\bqn\lb{adda}
\mathscr{V}(\eta)=
\begin{cases}
{a''}/{a}-U(\eta), & \;\;\; \text{scalar}, \\
{a''}/{a}, &\;\;\; \text{tensor}.
\end{cases}
\eqn
However,   as we mentioned above, we can safely ignore the $U(\eta)$ term in the scalar perturbation equation during
 the bouncing phase. As a result, both scalar and tensor perturbations obey the same equation with the same effective potential  $a''/a$. So, in this subsection we only need to consider one of them   during the bouncing phase.

\begin{figure}
\includegraphics[width=8.1cm]{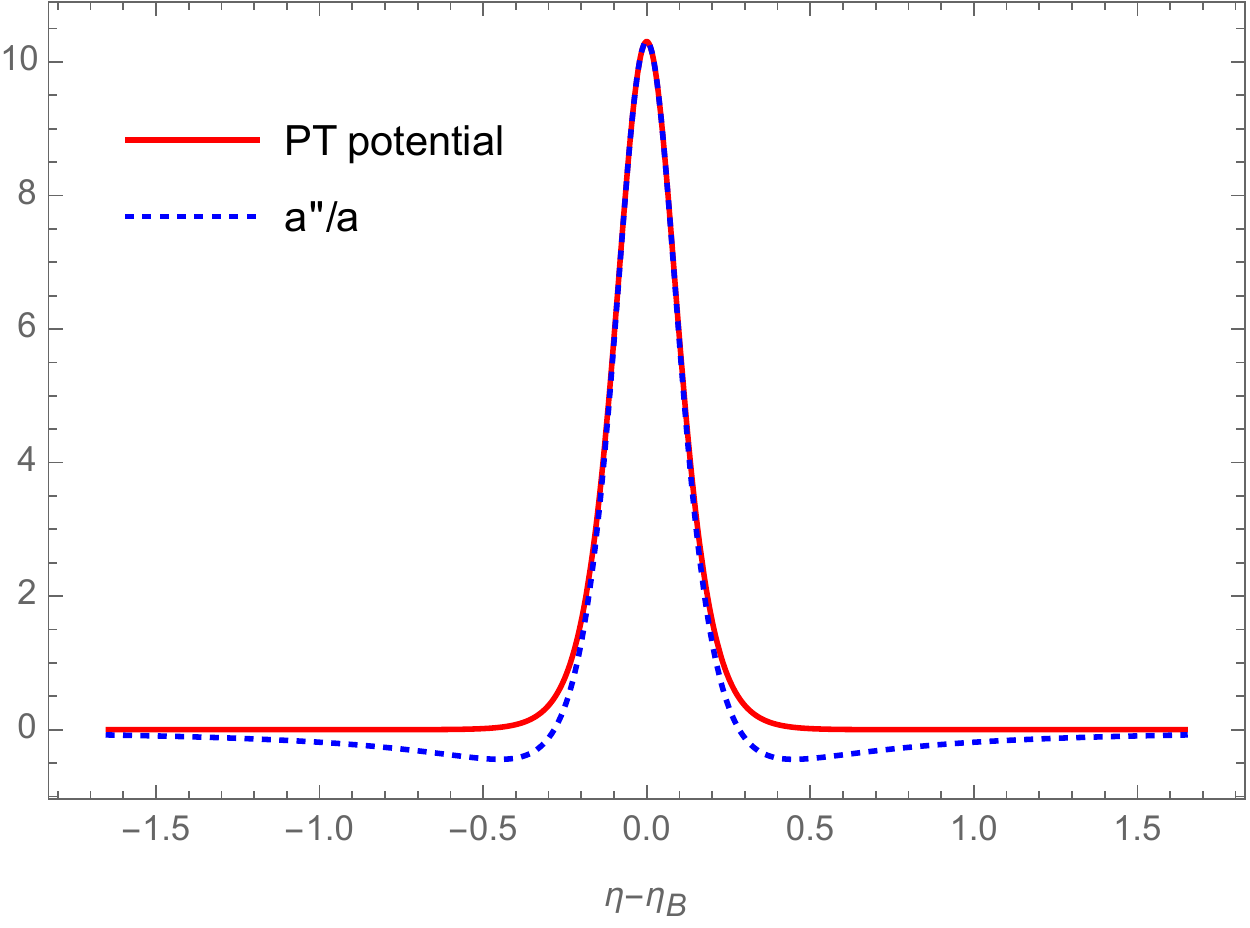}
\caption{Comparison between $a''/a$  and the PT potential given in Eq. (\ref{PT}). We set $a_\text{B}=1$ and $m_\text{Pl}=1$.} \label{compar}
\end{figure}

In order to solve the perturbation equation, let us first consider the analytical expression of $a''/a$. From the analytical equation of $a(t)$ and the relation $dt/a(t)=d\eta$, we find
\bqn\lb{potential}
\mathscr{V}(\eta)\equiv \frac{a''}{a}=a_\text{B}^2\frac{\gamma_\text{B} m_{\text{Pl}}^2(3- \gamma_\text{B} t^2/t_{\text{Pl}}^2)}{9(1+\gamma_\text{B} t^2 /t_{\text{Pl}}^2)^{5/3}}.
\eqn
  If we consider Eq.~(\ref{sch}) as the Schr\"odinger equation, the term $\mathscr{V}(\eta)$ serves as an effective barrier during the bouncing phase. This potential can be approximated
  by a PT potential for which we know the analytical solution,
\bqn\lb{PT}
\mathscr{V}_{\text{PT}}(\eta) = \frac{\mathscr{V}_0}{\cosh^2{\alpha (\eta-\eta_\text{B})}}.
\eqn
Here $\mathscr{V}_0$ is the height of the effective potential and $-2 \mathscr{V}_0 \alpha^2$ is the curvature of the potential at its maximum. Specifically for $\mathscr{V}(\eta)$ given by Eq. (\ref{potential}), one has
\bqn
\lb{cc}
\mathscr{V}_0=\frac{a_\text{B}^2 \gamma_\text{B} m_{\text{Pl}}^2}{3} =  k_{\rm B}^2 = \frac{\alpha^2}{6}.
\eqn

In Fig.~\ref{compar}, we plot the PT potential and the one given by Eq. (\ref{potential}), from which  we can see that $\mathscr{V}_{\text{PT}}(\eta)$ mimics $\mathscr{V}(\eta)$ very well.

To find the analytical solution of Eq.~(\ref{sch}) with the PT potential, we define two new  variables $x$ and $\mathcal{Y}(x)$ via the relations,
\bqn
x(\eta) &=& \frac{1}{1+e^{-2 \sqrt{6} k_{\rm B} (\eta-\eta_\text{B})}}, \\
\mathcal{Y}(x)&=&[x (1-x)]^{i k /(2\sqrt{6}k_{\rm B})} \mu_k(\eta),
\eqn
and rewrite Eq.~(\ref{sch}) as
\bqn\lb{Hyper_eq}
x (1-x) \frac{d^2 \mathcal{Y}}{dx^2}+[a_3-(a_1+a_2+1)x]\frac{d\mathcal{Y}}{dx}-a_1a_2 \mathcal{Y}=0,\nb\\
\eqn
where
\bqn
\lb{acs}
a_1 &\equiv& \frac{1}{2}\left(1+\frac{1}{\sqrt{3}}\right)  -\frac{i k}{\sqrt{6}\; k_{\rm B}},\nb\\
a_2 &\equiv & \frac{1}{2}\left(1-\frac{1}{\sqrt{3}}\right) -\frac{i k}{\sqrt{6}\; k_{\rm B}},\nb\\
a_3 &\equiv & 1-\frac{i k}{\sqrt{6}\; k_{\rm B}}.
\eqn
Eq.~(\ref{Hyper_eq}) is the standard hypergeometric equation, and has the general solution,
\bqn\lb{sol_PT}
\mu^{(\text{PT})}_k(\eta) &=& a_k x^{ik/(2\sqrt{6} k_{\rm B})} (1-x)^{-ik/(2 \sqrt{6} k_{\rm B})} \nb\\
&& \times \; _2F_1(a_1-a_3+1,a_2-a_3+1,2-a_3,x)\nb\\
&& +b_k [x (1-x)]^{-ik /(2 \sqrt{6} k_{\rm B})} \;_2F_1(a_1,a_2,a_3,x).\nb\\
\eqn
Here $a_k$ and $b_k$ are two independent integration constants, and $\alpha = \sqrt{6} k_B$ as given in Eq.~(\ref{cc}). They are uniquely determined by the
initial conditions.

\subsection{Perturbations During the Transition Phase}

After the bouncing phase, the kinetic energy of the scalar field $\phi$ keeps decreasing and the potential energy soon becomes dominant.
 As shown in Fig.~\ref{length}, during this transition phase we have
\bqn
k^2 \gg \left|\frac{a''}{a}-U(\eta) \right|, \;\;\;\; k^2 \gg \left|\frac{a''}{a}\right|.
\eqn
Thus,  the equation of motion for both scalar and tensor perturbations can be described by
\bqn
\mu_k''(\eta) + k^2 \mu_k(\eta)=0,
\eqn
which has the  general solutions,
\bqn\lb{transition_solution}
\mu_k(\eta) = \frac{1}{\sqrt{2k}} \left(\tilde \alpha_k e^{- i k \eta} +  \tilde \beta_k e^{i k \eta}\right),
\eqn
where $\tilde \alpha_k $ and $\tilde \beta_k $ are two constants.

It is remarkable to note that during both, the bouncing and transition phase, the scalar and tensor perturbations obey the same equation of motion.
As to be shown below, this is no longer the case during the slow-roll inflationary phase.

\subsection{Perturbations During the Slow-Roll Inflationary Phase}

After the transition phase, the energy density dropped down to about $10^{-12} \rho_\text{c}$ and the universe enters the slow-roll inflationary phase. During this period, the equations of motion
for both scalar and tensor perturbations are relativistic, and can be described by
\bqn\lb{eom_sl}
\mu_k''(\eta)+\left(k^2-\frac{\nu_{s,t}^2+1/4}{\eta^2}\right)\mu_k(\eta)=0,
\eqn
where $\nu_{s,t}(\eta)$ can be expressed in terms of the slow-roll parameters and  the subscripts ``$s,t$"  denote scalar and tensor perturbations, respectively.  For the scalar perturbations, we have
\bqn
\nu^2_{s} \simeq  \eta^2 \frac{z_s''(\eta)}{z_s(\eta)}+\frac{1}{4},
\eqn
with $z_s=a \dot \phi/H$, while   for the tensor perturbations, we have
\bqn
\nu^2_t \simeq  \eta^2 \frac{z_t''(\eta)}{z_t(\eta)}+\frac{1}{4},
\eqn
with $z_t=a(t)$.  As both $\nu_s$ and $\nu_t$ are the slow-roll quantities during the slow-roll inflation, we can  take  them as constant approximately. Then,
 the approximate solutions of Eq.~(\ref{eom_sl}) can be solved analytically and expressed as a linear combination of Hankel functions,
\bqn\lb{hankel}
\mu_k^{(s,t)}(\eta) \simeq \frac{\sqrt{-\pi \eta}}{2} \left[\alpha_k H^{(1)}_{\nu_{s,t}}(-k\eta) +\beta_k H^{(2)}_{\nu_{s,t}}(-k\eta)\right],\nb\\
\eqn
where we already ignored the irrelevant phase factor $e^{i(1+2 \nu_{s,t}) \pi/4}$.

\subsection{Matching the Solutions Together}

Now we need to determine the coefficients $\alpha_k$ and $\beta_k$ by connecting the solutions (\ref{hankel}), (\ref{transition_solution}), and (\ref{sol_PT}) together  in their intermediate regions. For this purpose,
we first consider the limit $\eta-\eta_\text{B} \gg 0$ for the solution (\ref{sol_PT}), from which we find
\bqn
x \sim 1- e^{-2 \sqrt{6} k_{\rm B} (\eta-\eta_\text{B})} \to 1.
\eqn
Thus, we obtain
\bqn
1-x \sim e^{-2 \sqrt{6} k_{\rm B} (\eta-\eta_\text{B})}.
\eqn
Using the   relations,
\begin{widetext}
\bqn
&&_2F_1(a_1-a_3+1,a_2-a_3+1,2-a_3,x)\nb\\
&&~~~=(1-x)^{a_3-a_1-a_2} \frac{\Gamma(2-a_3)\Gamma(a_1+a_2-a_3)}{\Gamma(a_1-a_3+1)\Gamma(a_2-a_3+1)}\; _2F_1(1-a_1,1-a_2, 1+a_3-a_1-a_2,1-x)\nb\\
&&~~~~~+\frac{\Gamma(2-a_3)\Gamma(a_3-a_1-a_2)}{\Gamma(1-a_1)\Gamma(1-a_2)}\; _2F_1(a_1-a_3+1, a_2-a_3+1, a_1+a_2-a_3+1,1-x),
\eqn
and
\bqn
_2F_1(a_1,a_2,a_3,x)&=&(1-x)^{a_3-a_1-a_2} \frac{\Gamma(a_3)\Gamma(a_1+a_2-a_3)}{\Gamma(a_1)\Gamma(a_2)}\;_2F_1 (a_3-a_1,a_3-a_2,a_3-a_1-a_2+1,1-x)\nb\\
&&+\frac{\Gamma(a_3)\Gamma(a_3-a_1-a_2)}{\Gamma(a_3-a_1)\Gamma(a_3-a_1)}\;_2F_1(a_1,a_2,a_1+a_2+1-a_3,1-x),
\eqn
we find
\bqn\lb{solPT_2}
\mu_k^{\text{PT}}(\eta) &=& \left[ a_k\frac{\Gamma(2-a_3)\Gamma(a_1+a_2-a_3)}{\Gamma(a_1-a_3+1)\Gamma(a_2-a_3+1)}
+b_k \frac{\Gamma(a_3)\Gamma(a_1+a_2-a_3)}{\Gamma(a_1)\Gamma(a_2)} \right] e^{-ik(\eta-\eta_\text{B})}\nb\\
&&+\left[a_k \frac{\Gamma(2-a_3)\Gamma(a_3-a_1-a_2)}{\Gamma(1-a_1)\Gamma(1-a_2)}+ b_k\frac{\Gamma(a_3)
\Gamma(a_3-a_1-a_2)}{\Gamma(a_3-a_1)\Gamma(a_3-a_2)}  \right]e^{i k (\eta-\eta_\text{B})}.
\eqn
\end{widetext}
Comparing Eqs.~(\ref{solPT_2}) and (\ref{transition_solution}), we obtain,
\bqn\lb{alphak}
\frac{\tilde \alpha_k}{\sqrt{2k}}&=&\Bigg[a_k \frac{\Gamma(2-a_3)\Gamma(a_1+a_2-a_3)}{\Gamma(a_1-a_3+1)\Gamma(a_2-a_3+1)}\nb\\
&&~~~~+ b_k \frac{\Gamma(a_3)\Gamma(a_1+a_2-a_3)}{\Gamma(a_1)\Gamma(a_2)} \Bigg]e^{ i k\eta_\text{B}},\\
\lb{betak}
\frac{\tilde \beta_k}{\sqrt{2k}} &=&\Bigg[ a_k\frac{\Gamma(2-a_3)\Gamma(a_3-a_1-a_2)}{\Gamma(1-a_1)\Gamma(1-a_2)}  \nb\\
&&~ +b_k \frac{\Gamma(a_3)\Gamma(a_3-a_1-a_2)}{\Gamma(a_3-a_1)\Gamma(a_3-a_2)} \Bigg] e^{-i k\eta_\text{B}}.
\eqn

Then,   considering  the solution (\ref{hankel}) in the limit $- k \eta \to \infty$,  we obtain
\bqn\lb{hankel_2}
\mu_k(\eta)=\frac{\alpha_k}{\sqrt{2k}}e^{-ik\eta}+\frac{\beta_k}{\sqrt{2k}} e^{i k\eta}.
\eqn
Comparing it with Eq.~(\ref{transition_solution}), we find
\bqn
\lb{alphak}
\frac{\alpha_k}{\sqrt{2k}}&=& \frac{\tilde \alpha_k}{\sqrt{2k}}\nb\\
&=&\Bigg[a_k \frac{\Gamma(2-a_3)\Gamma(a_1+a_2-a_3)}{\Gamma(a_1-a_3+1)\Gamma(a_2-a_3+1)}\nb\\
&&~~~~+ b_k \frac{\Gamma(a_3)\Gamma(a_1+a_2-a_3)}{\Gamma(a_1)\Gamma(a_2)} \Bigg]e^{ i k\eta_\text{B}},\\
\lb{betak}
\frac{\beta_k}{\sqrt{2k}} &=& \frac{\tilde \beta_k}{\sqrt{2k}}\nb \\
&=&\Bigg[ a_k\frac{\Gamma(2-a_3)\Gamma(a_3-a_1-a_2)}{\Gamma(1-a_1)\Gamma(1-a_2)}  \nb\\
&&~ +b_k \frac{\Gamma(a_3)\Gamma(a_3-a_1-a_2)}{\Gamma(a_3-a_1)\Gamma(a_3-a_2)} \Bigg] e^{-i k\eta_\text{B}}.
\eqn
Now  several comments are  in order:

\begin{itemize}

\item Since $a_i$'s depend on $k/k_{\rm B}$ via Eq.~(\ref{acs}), we can see that $\alpha_k$ and $\beta_k$ in
general also depend on $k/k_B$.

\item In general relativity, the BD vacuum is normally imposed \cite{baumann_tasi_2009},
\bq
\lb{BDc}
\alpha_k^{\rm GR} = 1, \;\;\;\; \beta_k^{\rm GR} = 0,
\eq
 whenever the mode is inside the Hubble  horizon. When the effects of the pre-inflationary dynamics are taken into account, in general this is no longer the case.
 In fact, both of them now depend on the constants $a_k$ and $b_k$, which are closely related to the pre-inflationary dynamics during the bouncing phase.

\item The quantity  $|\beta_k|^2$ represents the rate of particle creation due to the expansion of the universe.  In general $\beta_k$ does not vanish,  that is,
particles  are generically created due to the expansion of the universe during the bouncing and transition phases.

\item Eqs.~(\ref{alphak}) and (\ref{betak}) are valid for both scalar and tensor perturbations. Thus, if the same initial conditions are chosen for these two types of perturbations, the
effects of pre-inflationary dynamics are also the same. This is {\em a unique characteristic of  the  dressed metric approach}.

\end{itemize}

Now with the analytical solution of Eq.~(\ref{hankel}) and $\alpha_k$ and $\beta_k$ given by Eqs.~(\ref{alphak}) and (\ref{betak}),
 let us turn to compute the power spectra for both  scalar and tensor perturbations.
For the scalar perturbations, the primordial power spectrum
 can be calculated in the limit $-k \eta \to 0^+$, and is given by,
\bqn
\mathcal{P}_{\mathcal{R}}(k)\equiv \frac{k^3}{4\pi^2} \left|\mathcal{R}_k(\eta)\right|^2=\frac{k^3}{4\pi^2 }\left|\frac{\mu^{(s)}_k(\eta)}{z_s(\eta)}\right|^2.
\eqn
Using the asymptotic form of the Hankel functions,
\bqn\lb{hankel_limit}
&&\lim_{-k \eta \to 0^+} H_{\nu}^{(1,2)}(-k\eta)\simeq \mp \frac{i}{\pi}\Gamma(\nu)\left(\frac{- k\eta}{2}\right)^{-\nu},
\eqn
where $+,\;-$ correspond to $H_\nu^{(2)}(-k\eta)$ and $H_\nu^{(1)}(-k \eta)$ respectively, we find
\bqn
\mu_k^{(s)}(\eta) \to i \sqrt{\frac{-\eta}{2\pi}}(\alpha_k+\beta_k)\Gamma(\nu_s)\left(\frac{-k\eta}{2}\right)^{-\nu_{s}}.  ~~~~
\eqn
Then, the power spectrum reads
\bqn\lb{scalar_pw}
\mathcal{P}_{\mathcal{R}}(k) &=& |\alpha_k+\beta_k|^2\mathcal{P}^{\text{GR}}_{\mathcal{R}}(k),
\eqn
where
\bqn
\mathcal{P}^{\text{GR}}_{\mathcal{R}}(k) \equiv \frac{k^2}{4\pi^3} \left(\frac{H}{a \dot \phi}\right)^2 \Gamma^2(\nu_s) \left(\frac{-k\eta}{2}\right)^{1-2\nu_s}, ~~~~~~~
\eqn
is the standard inflationary scalar spectrum for a single scalar field inflationary model in   GR, and $\alpha_k$ and $\beta_k$ are given by Eqs.~(\ref{alphak})
and (\ref{betak})  in terms of $a_k$ and $b_k$.  The latter are determined by the initial conditions.

Note that Eq.~(\ref{scalar_pw}) is {\em the most general expression for the scalar power spectrum in the framework of the dressed metric approach in LQC}.
Once the initial conditions are specified,  it will be uniquely determined.

In addition, as mentioned above, $\alpha_k$ and $\beta_k$ depend on the comoving wavenumber $k$ via the coefficients $a_i$'s through
Eq.~(\ref{acs}), so the power spectrum $\mathcal{P}_{\mathcal{R}}(k)$ is generically scale-dependent, due to the quantum gravitational effects. Clearly, to be
consistent with observations \cite{planck_collaboration_planck_2014-1,planck_collaboration_planck_2015-4}, which show that the  power spectrum is almost scale-invariant, the
dependence cannot be  strong. Otherwise, it will be inconsistent with observations. Such a dependence clearly is closely related to the initial conditions,
which will be considered in detail in the next section. For some given initial conditions, observations will impose strong constraints on the effects of pre-inflationary dynamics,
which will provide an excellent opportunity to test observationally the theory of LQC, or more precisely, the ideas of the dressed metric approach.

For  the tensor perturbation, the primordial power spectrum of   $h = 2 M_\text{Pl} \mu_k^{(t)}(\eta)/a(\eta)$ can be calculated in the limit $-k\eta \to 0^+$ via the expression,
\bqn
\mathcal{P}_h \equiv \frac{k^3}{4\pi^2} \left|h(k)\right|^2 = \frac{k^3}{4\pi^2} \left|\frac{2 \mu_k^{(t)}(\eta)}{M_\text{Pl}a(\eta)}\right|^2.
\eqn
Similar to the scalar case, using the asymptotic form of the Hankel functions given  by Eq.~(\ref{hankel}) in the limit $-k\eta \to 0^+$,  we find
\bqn
\mu_k^{(t)}(\eta) \to  i \sqrt{\frac{-\eta}{2\pi}}(\alpha_k+\beta_k)\Gamma(\nu_t)\left(\frac{-k\eta}{2}\right)^{-\nu_{t}}.  ~~~~~
\eqn
Then,  the tensor spectrum reads
\bqn\lb{tensor_pw}
\mathcal{P}_h(k) = \left|\alpha_k+\beta_k\right|^2 \mathcal{P}_h^{\text{GR}}(k),
\eqn
where
\bqn
\mathcal{P}_h^{\text{GR}}(k) \equiv \frac{k^2}{\pi^3M_\text{Pl}^2}\frac{1}{a^2} \Gamma^2(\nu_t) \left(\frac{-k \eta}{2}\right)^{1-2\nu_t},
\eqn
and $\alpha_k$ and $\beta_k$ are given by Eqs.~(\ref{alphak}) and (\ref{betak}), which are generically scale-dependent, as mentioned above. Once the initial conditions are given,
they are uniquely determined.  Eq.~(\ref{tensor_pw}) is {\em the most general expression for the tensor power spectrum in the framework of the dressed metric approach in LQC,
and in general depends on the comoving wavenumber $k$, due to the quantum gravitational effects during the pre-inflationary phases}.

It should  be noted that the corrections due to the quantum gravitational effects of the pre-inflationary dynamics are all proportional to the factor $\left|\alpha_k+\beta_k\right|^2$ for both scalar and
tensor perturbations, as one can see from Eqs.~(\ref{scalar_pw}) and (\ref{tensor_pw}). As mentioned above, if the same initial conditions for these two types of perturbations are imposed, for example,
all of therm are in the Bunch-Davies vacuum initially,  this factor will be the same. Then,  the ratio $r$ between the scalar and tensor perturbations remains the same as that given in GR!

\section{Initial conditions and Effects of Pre-inflationary Dynamics in Primordial Spectra}
\renewcommand{\theequation}{5.\arabic{equation}} \setcounter{equation}{0}
\lb{initial_conditions}

In the last section, the linear cosmological perturbations were calculated in each of the three phases prior to preheating, i.e., the bouncing, transition and slow-rill inflationary phases
[cf.  Figs. \ref{EoS} and \ref{length}], and analytical expressions of the mode functions for both scalar and tensor perturbations were found, whereby the corresponding power spectra
were computed and given explicitly by Eqs.~(\ref{scalar_pw}) and (\ref{tensor_pw}). The mode functions found in these three different phases were also matched together and
finally given by Eqs.~(\ref{alphak}) and (\ref{betak}), in which the parameters $\alpha_k$ and $\beta_k$ of the mode functions in the slow-roll inflationary phase are expressed as
functions of $a_k$ and $b_k$ of the mode functions in the bouncing phase. The latter will be determined by initial conditions. So, in this section, we shall first consider the initial conditions,
and then we study the effects of the quantum bounce and its subsequent pre-inflationary dynamics during the bouncing and transition phase on the power spectra.

\subsection{Initial Conditions of Primordial Perturbations}

In the framework of LQC, various sets of initial conditions  have been investigated \cite{ashtekar_loop_2010,ashtekar_quantum_2017,winitzki_cosmological_2005,GNS07,CK11,LB13,
agullo_unitarity_2015, agullo_preferred_2015,ashtekar_initial_2017,MBS17}. However,  this is a subtle issue, because in general there is not a preferred initial state for a quantum field in arbitrarily curved space-times.
If the universe is sufficiently spatially flat and evolves sufficiently slowly so that the characteristic scale for a perturbation mode is much larger than its wavelength, there is an approximate definition of the initial
state: the Bunch-Davies vacuum state. This is also the common initial state adopted in general relativity at the beginning of the slow-roll inflation where all the relevant perturbation modes are well inside the
Hubble horizon \cite{baumann_tasi_2009}.

However, in the pre-inflationary phases, especially near the bounce, the background geometry is far from the slow-roll inflationary phase. In particular, for the perturbations during the bouncing phase, as illustrated in
Fig.~\ref{length}, their wavelengths could be larger, equal, or smaller than the corresponding characteristic scale. Thus, it is in general impossible to assume that the universe is in the
 Bunch-Davies vacuum   at the bounce \cite{winitzki_cosmological_2005, agullo_unitarity_2015, agullo_preferred_2015}.
 In this subsection, we consider only  two of them that have been frequently used in the literature,  and show that they essentially lead to the same results.

 \begin{figure}
\includegraphics[width=8.1cm]{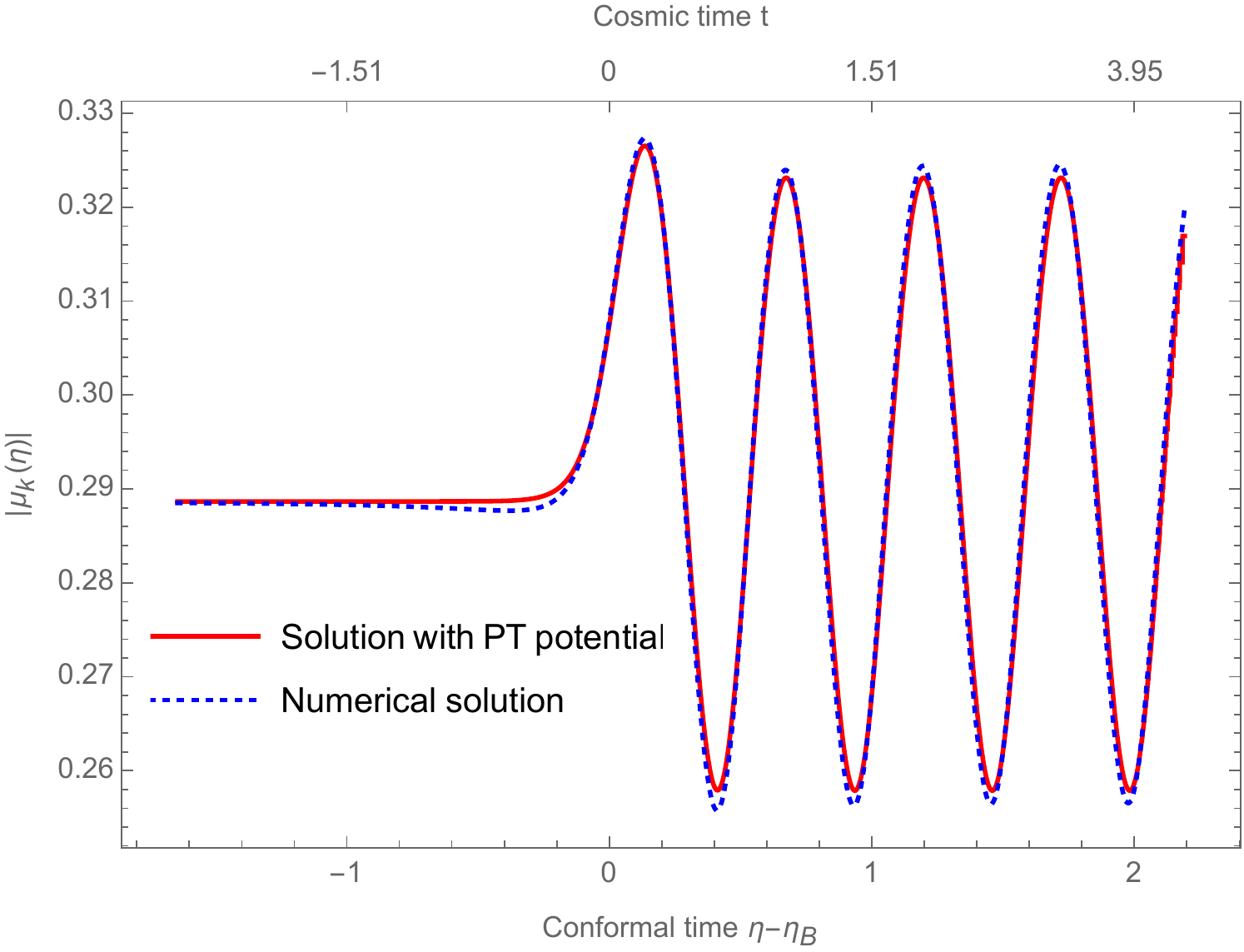}
\caption{Comparison between the analytical solution given by  Eq.~(\ref{sol_PT}) and the numerical solution with the initial condition (\ref{akbk}), $a_\text{B}=1$, and $m_\text{Pl}=1$.}
\label{compar_solution}
\end{figure}

\subsubsection{The BD Vacuum in Contracting Phase}

As illustrated in Fig.~\ref{length}, during the contracting phase right before the bounce, all the relevant perturbation modes are well inside the characteristic length $\lambda$. Then,
we  can naturally choose the  BD vacuum state as the initial conditions of the perturbations of both scalar and tensor \cite{LB13},
\bqn\lb{BD_vacuum}
\mu_k^{\text{initial}}(\eta) \sim \frac{1}{\sqrt{2 k}} e^{- i k \eta},
\eqn
 with which we can uniquely determine  the coefficients $a_k$ and $b_k$ appearing in Eqs.~(\ref{alphak}) and (\ref{betak}). Considering  that during the contracting phase, we have $\eta-\eta_\text{B} \ll 0$,  we
 find
\bqn
x \sim e^{2 \sqrt{6} k_{\rm B} (\eta-\eta_\text{B})} \to 0,
\eqn
and thus
\bqn
 x^{ik/(2\sqrt{6} k_{\rm B})} (1-x)^{-ik/(2 \sqrt{6} k_{\rm B})} &\sim& e^{i k (\eta-\eta_\text{B})},\nb\\
 (x (1-x))^{-ik/(2 \sqrt{6} k_{\rm B})} &\sim& e^{-i k (\eta-\eta_\text{B})}.
\eqn
Inserting the above expressions into Eq.~(\ref{sol_PT}), and then comparing it with the above initial condition, we find that,
\bqn\lb{akbk}
a_k =0 , \;\;\;\; b_k = \frac{e^{i k\eta_\text{B}}}{\sqrt{2k}}.
\eqn
Note that in the above calculations we used the fact $_2F_1(c_1,c_2,c_2,0)=1$. Fig. \ref{compar_solution} shows our analytical solution together with the numerical one, from which we can see that they match extremely
well during the bouncing phase.

Here we would like to note that most of the modes (with $k^2 \gg |a''/a|$) are well inside the characteristic length $\lambda$ during the contracting phase.
As a result, $\mu_k^{\rm PT} (\eta)$ of Eq.~(\ref{sol_PT}) with the initial condition of the  BD vacuum state imposed at the contracting phase are valid for all these modes,
 even if some of them (with $k<k_{\rm B}$) are outside the characteristic length $\lambda$ at the bounce.

\subsubsection{The Fourth-order Adiabatic Vacuum State at the Bounce}

Another moment  of the initial conditions  is right at the bounce \cite{ashtekar_loop_2010}. However, as we mentioned above, at this point the curvature of the background geometry has significant effects on
the perturbation modes. This is different from the case of the quasi-de-Sitter spacetime or the contracting phase right before the bounce. Therefore, it is not possible to impose the BD vacuum state
at the bounce.

Instead,  one can impose two conditions,  {\em the adiabatic regularization} and {\em maximal symmetry} of the perturbation modes, with which it was shown   \cite{agullo_quantum_2012, agullo_extension_2013, agullo_pre-inflationary_2013, agullo_unitarity_2015}  that the state of the perturbations at the bounce can be constructed as  {\em the fourth-order adiabatic vacuum}. According to \cite{agullo_pre-inflationary_2013},  the  fourth-order adiabatic vacuum of Eq. (\ref{sch})
with $V(\eta)$ being given by Eq.~(\ref{adda}) can be written in the form
\bqn\lb{fourth}
\mu_k(\eta)=\frac{1}{\sqrt{2 W_k^{(4)}(\eta)}}e^{-i \int^ \eta W_k^{(4)}(\tilde \eta)d\tilde \eta},
\eqn
where
\bqn
W_k^{(4)}(\eta)=\sum_{i=1}^{i=4} W_i ,
\eqn
with
\bqn
W_0&=&k,\\
W_1&=&0,\\
W_2&=&-\frac{1}{2k} \frac{a''}{a},\\
W_3&=&0,\\
W_4&=& \frac{1}{8 k^3} \left(\frac{2 a'' a'^2}{a^3}-\frac{2 a''^2}{a^2}-\frac{2 a' a'''}{a^2}+\frac{a''''}{a}\right).\nb\\
\eqn
The above state is constructed from the generalized WKB approximate solution of the fourth-order and can also be regarded as an expansion in the number of derivatives of the scale factor.
The leading order  in the expansion corresponds to the positive energy solution in the Minkowski space and the rest of the terms are higher-order contributions that vanish at different rates when
$a/k \to 0$ \cite{agullo_extension_2013, agullo_pre-inflationary_2013}.

Here we would like to note that, as also pointed out in \cite{agullo_extension_2013, agullo_pre-inflationary_2013}, 2.   This implies that the fourth-order adiabatic vacuum state constructed above can only apply to the modes with $k\geq k_\text{B}$,
here $k_\text{B}$ represents the maximum value of $\sqrt{a''/a}$ during the bouncing phase [c.f. Fig.~\ref{length}]. For the modes with $k<k_\text{B}$, as pointed out in \cite{agullo_extension_2013, agullo_pre-inflationary_2013},
ambiguity remains in constructing the vacuum state at the bounce.

Imposing  {\em the fourth-order adiabatic vacuum} at the bounce for modes with $k \geq k_\text{B}$, we find
\bqn
W_2 &=&-\frac{a_\text{B}^2 \gamma _\text{B}}{6 k t_{\text{Pl}}^2}=-\frac{k_\text{B}^2}{2 k},\\
W_4 &=& -\frac{13 a_\text{B}^4 \gamma _\text{B}^2}{72 k^3 t_{\text{Pl}}^4} = -\frac{13 k_\text{B}^4}{8 k^3}.
\eqn
Using these results and expanding the fourth-order adiabatic vacuum state up to the order of $O(k_\text{B}^4/k^4)$, we find
\bqn\lb{wkb_B}
\mu_k(\eta_\text{B}) &=&  \frac{1}{\sqrt{2k}} \left[1-\frac{1}{4}\frac{k_\text{B}^2}{k^2}- \frac{29}{32} \frac{k_\text{B}^4}{k^4}+\mathcal{O}\left(\frac{k_\text{B}^6}{k^6}\right)\right].\nb\\
\eqn
Now we are in the position  to determine the coefficients $a_k$ and $b_k$ appearing  in Eq.~(\ref{sol_PT}) by connecting the initial state with the analytical  solution (\ref{sol_PT}) at the bounce, at which
we find
\bqn\lb{sol_PT_bounce}
\mu^{(\text{PT})}(\eta_\text{B}) &=& a_k \;_2F_1\left(\frac{1-\sqrt{3}}{2}, \frac{1+\sqrt{3}}{2}, 1+\frac{i k}{\sqrt{6} k_\text{B}}, \frac{1}{2}\right)\nb\\
&&+b_k \sqrt{\pi }4^{\frac{ik}{2\sqrt{6}k_\text{B}}} \Gamma\left(1-\frac{i k}{\sqrt{6}k_\text{B}}\right)\nb\\
&&\;\;\; \times  \Gamma^{-1} \left(\frac{3}{4}-\frac{\sqrt{3}}{12}-\frac{ik}{2\sqrt{6}k_\text{B}}\right)\nb\\
&&\;\;\;\times  \Gamma^{-1} \left(\frac{3}{4}+\frac{\sqrt{3}}{12}-\frac{ik}{2\sqrt{6}k_\text{B}}\right).
\eqn
Then,  using the asymptotic expansions of the Gamma function appearing  in Eq.~(\ref{gamma}) of Appendix ~\ref{gamma_function}, we arrive at the same results of
Eq.~(\ref{akbk}). Note that in obtaining the above result, we had simply ignored an irrelevant phase difference between (\ref{sol_PT_bounce}) and (\ref{wkb_B}).

Thus, we conclude that   the solution of Eq.~(\ref{sol_PT}) starting with a BD vacuum state at the contracting phase will  reduce to the one obtained by imposing   {\em the fourth-order adiabatic vacuum state} at the bounce.
This is easy to understand. In fact, if one applies {\em the fourth-order adiabatic vacuum state} to Eq.~(\ref{fourth}) at the contracting phase, one can see  that the perturbation modes  also satisfy   the BD vacuum condition there.
However, it must be noted that the perturbations with the BD vacuum initial condition  and the fourth-order adiabatic vacuum one are the same only for the modes with $k\geq k_\text{B}$.
For other modes, {\em the fourth-order adiabatic vacuum state}  at the bounce is not applicable, while the BD vacuum initial state can be still applied to these modes but in the contracting phase.

\subsection{ Effects of the   Pre-inflationary Dynamics on Primordial Power Spectra}

\begin{figure}
\includegraphics[width=8.3cm]{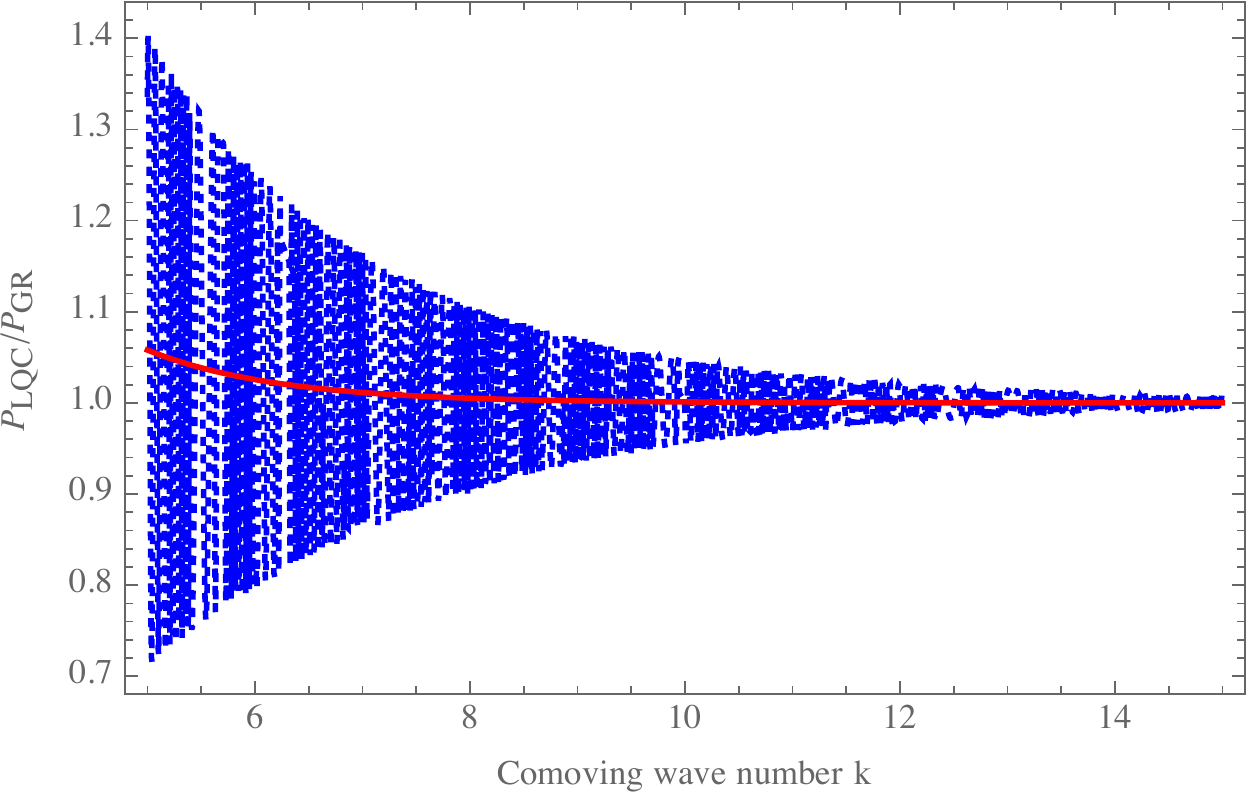}
\caption{Ratio between the power spectrum with the effects of the pre-inflationary dynamics  and the standard power-law spectrum obtained in GR.
The dotted blue curve denotes the analytical power spectrum given by Eq.~(\ref{pw}), which obviously oscillates rapidly with $k$. The solid red curve shows the average of the oscillating spectrum.}
\label{power_spectrum_PT}
\end{figure}

With the coefficients $a_k$ and $b_k$ being given by Eq.~(\ref{akbk}), the coefficients $\alpha_k$ and $\beta_k$ appearing  in Eqs.~(\ref{alphak}) and (\ref{betak}) can be casted in the form
\bqn
\lb{beta_k}
\alpha_k&=&  \frac{\Gamma(a_3)\Gamma(a_1+a_2-a_3)}{\Gamma(a_1)\Gamma(a_2)}e^{ 2 i k\eta_\text{B}},\nb\\
\beta_k&=&\frac{\Gamma(a_3)\Gamma(a_3-a_1-a_2)}{\Gamma(a_3-a_1)\Gamma(a_3-a_2)}.
\eqn
Then, we find that
\bqn\lb{pw}
&&|\alpha_k+\beta_k|^2 = 1+ \left[1+\cos \left(\frac{\pi }{\sqrt{3}}\right)\right] \text{csch}^2\left(\frac{\pi  k}{\sqrt{6} k_{\rm B} }\right)\nb\\
&&~~~~ +\sqrt{2} \sqrt{\cosh \left(\frac{2 \pi  k}{\sqrt{6} k_{\rm B} }\right)+\cos \left(\frac{\pi }{\sqrt{3}}\right)}\cos \left(\frac{\pi }{2 \sqrt{3}}\right) \nb\\
&&~~~~~~\times  \text{csch}^2\left(\frac{\pi  k}{\sqrt{6} k_{\rm B} }\right) \cos \left(2 k \eta_{\rm B}+\varphi _k\right),
\eqn
where
\bqn
\varphi_k \equiv \arctan{\left\{\frac{\text{Im}[\Gamma(a_1)\Gamma(a_2)\Gamma^2(a_3-a_1-a_2)]}{\text{Re}[\Gamma(a_1)\Gamma(a_2)\Gamma^2(a_3-a_1-a_2)]}\right\}}.\nb\\
\eqn
In Fig.~\ref{power_spectrum_PT} we display the ratio between the power spectrum with the bouncing effects and the standard one given  in GR,   as a function of the wave-number $k$.
We would like to note that Fig.~\ref{power_spectrum_PT} is consistent with the   one presented  in \cite{agullo_quantum_2012, agullo_pre-inflationary_2013}
[c.f. Fig.~1 in \cite{agullo_quantum_2012} and Fig.~5 in \cite{agullo_pre-inflationary_2013}]. While the results obtained in \cite{agullo_quantum_2012, agullo_pre-inflationary_2013} are purely numerical,
 here our results are derived directly from the analytical expression given by Eq.~(\ref{pw}).

It is remarkable to note that {\em the pre-inflationary dynamics leads to oscillations in the power spectra of both scalar and tensor perturbations, and the amplitudes of these oscillations are
independent of the slow-roll inflationary models, although they depend explicitly on $k$.} The amplitudes of these oscillations, which essentially depend on the parameter $\alpha [=\sqrt{6} k_{\rm B}]$, represent a characteristic feature of LQC. In Eq.~(\ref{pw}),
the last two terms, proportional to $\text{csch}^2(\pi k/(\sqrt{6} k_B))$, decrease exponentially as $k$ increases. In other words, the power spectra get reduced exponentially for  $k/k_B \gg 1$. However, as
$k/k_B \simeq 0$, they get enhanced  as $(k_B/k)^2$.
Hence,  the quantum gravitational  effects are important at the scales $k \lesssim k_\text{B}$. These modes, as we mentioned above, are essentially
the ones whose energies are of the Planck scale at the bounce. They are initially inside the radius defined by $\lambda=\sqrt{|a/a''|}$, and then leave and re-enter
it during the bouncing phase. The modes with $k\gg k_\text{B}$ are always inside the radius before they leave the Hubble horizon during the slow-roll inflationary phase,
 thus finally they lead to a standard power-law spectrum.

 Note that the solution with the PT potential is not a good approximation for the modes with a very small wavenumber (i.e.,  $k^2 \ll |a''/a|$   during the whole bouncing phase). For these modes,
  if we ignore the $k^2$ term in Eq.~(\ref{eom_scalar}), the solution can be approximated by
  \bq
  \mu_k(\eta) \simeq a_k a(\eta)+b_k/a(\eta), \; (k^2 \ll |a''/a|),
  \eq
  which has been considered  in detail in  \cite{bolliet_comparison_2015}. However, these modes are beyond our interest because they are still outside of our currently observable universe.

\section{Observational constraints on the  effects of the pre-inflationary dynamics}
\renewcommand{\theequation}{6.\arabic{equation}} \setcounter{equation}{0}
\lb{MCMC}

\begin{table*}
\caption{Best fit values of the six cosmological parameters and the  constraints on $k_\text{B}/a_0$ and $r$ at 95\% C.L for different cosmological models from different data combinations.}
\lb{bestfit}
\begin{ruledtabular}
\begin{tabular}{ccccc}
 Parameter  & Planck TT+lowP & Planck TT,TE,EE+lowP & Planck TT+lowP+$r$ & Planck TT,TE,EE+lowP+$r$ \\
\hline
$\Omega_{\rm b}h^2$ &  $0.022355$ & $ 0.022193$ & $0.022322 $ & $0.022064$\\
$\Omega_c h^2 $  & $0.11893$ & $ 0.12000 $ & $0.11908$ & $0.12071 $\\
$100\theta_{\rm MC}$& $1.04115$ & $ 1.04065 $ & $ 1.04080$ & $1.04057$ \\
$\tau$ &  $0.077835$ & $0.089272 $ & $0.081955$ & $0.085259$   \\
$\ln(10^{10}A_s)$ &  $3.088$ & $3.112 $ & $3.101$ & $3.104 $ \\
$n_s$ &  $0.9662$ & $ 0.9647 $ & $ 0.9658$ & $ 0.9607 $ \\
\hline
$k_{\rm B}/a_0 $ & $<3.12\times 10^{-4}$& $ <3.05\times 10^{-4}$& $<3.14\times 10^{-4}$& $< 3.14 \times 10^{-4}$ \\
$ r$ &  $ ---- $& $---$ & $ < 0.113 $& $ < 0.107 $\\
\end{tabular}
\end{ruledtabular}
\end{table*}

The quantum corrections (\ref{pw})  are $k$-dependent and expected to be constrained by observations. In the following, we perform the CMB likelihood analysis by using the Planck 2015 data
\cite{planck_collaboration_planck_2015-4}, with the MCMC code developed in \cite{lewis_cosmological_2002}.
In order to carry out the CosmoMC code let us parameterize the primordial scalar and tensor spectra described in Eqs.~(\ref{scalar_pw}) and  (\ref{tensor_pw}) as,
\bqn
\mathcal{P}_\mathcal{R}(k) = (1+\delta_\text{Pl}) \mathcal{P}^{\text{GR}}_{\mathcal{R}}(k),\\
\mathcal{P}_h(k) = (1+\delta_\text{Pl}) \mathcal{P}^{\text{GR}}_{h}(k),
\eqn
where $\delta_\text{Pl}$ is given by \footnote{The oscillating terms in Eq.~(\ref{pw}) oscillate very fast and have negligible effects when integrating them out  with time. So, they can be safely ignored observationally.
In addition, in \cite{tao_zhu_universal_2016}  $\delta_\text{Pl}$ was denoted by  $\delta_{\cal{P}}$. }
\bqn
\delta_\text{Pl} = \left[1+\cos \left(\frac{\pi }{\sqrt{3}}\right)\right] \text{csch}^2\left(\frac{\pi  k}{\sqrt{6} k_{\rm B} }\right),
\eqn
and the standard power-law spectra $\mathcal{P}^{\text{GR}}_{\mathcal{R}}(k)$ and $\mathcal{P}^{\text{GR}}_{h}(k)$ are parameterized  in their standard  forms,
\bqn
\mathcal{P}^{\text{GR}}_{\mathcal{R}} &=& A_{s} \left(\frac{k}{k_*}\right)^{n_s-1+\cdots},\nb\\
\mathcal{P}^{\text{GR}}_{h} &=& A_{t} \left(\frac{k}{k_*}\right)^{n_t+\cdots}.
\eqn
 Here $A_s (A_t)$ is the scalar (tensor) amplitude, $n_s (n_t)$  the scalar (tensor) spectral index, and $k_*=0.05 {\rm Mpc}^{-1}$ denotes the pivot scale.

We assume the flat cold dark matter model with the effective number of neutrinos $N_{\text{eff}}=3.046$ and fix the total neutrino mass $\Sigma m_\nu=0.06 eV$.
Let us first  consider the scalar spectrum and vary the following seven parameters,
\bqn
(\Omega_{\rm b}h^2, \Omega_\text{c}h^2, \tau, \Theta_s, n_s, A_s, k_\text{B}/a_0),
\eqn
where $\Omega_{\rm b}h^2$ and $\Omega_\text{c}h^2$ are, respectively,  the baryon and cold dark matter densities, $\tau$ is the optical depth to reionization, $\Theta_s$ is the ratio (multiplied by $100$)
of the sound horizon at decoupling to the angular diameter distance to the last scattering surface. In addition, we have one more parameter $k_\text{B}/a_0$, which is related to the effects of the pre-inflationary dynamics.
For the six cosmological parameters ($\Omega_{\rm b}h^2, \Omega_\text{c}h^2, \tau, \Theta_s, n_s, A_s$), we use the same prior ranges as in \cite{planck_collaboration_planck_2013}, while for the parameter $k_{\rm B}/a_{0}$,
which is related to the bouncing effects, we set the prior range as $k_{\rm B}/a_0 \in [10^{-8}, 0.002] {\rm Mpc}^{-1}$.

\begin{figure*}
\includegraphics[width=8.1cm,height=5.5cm]{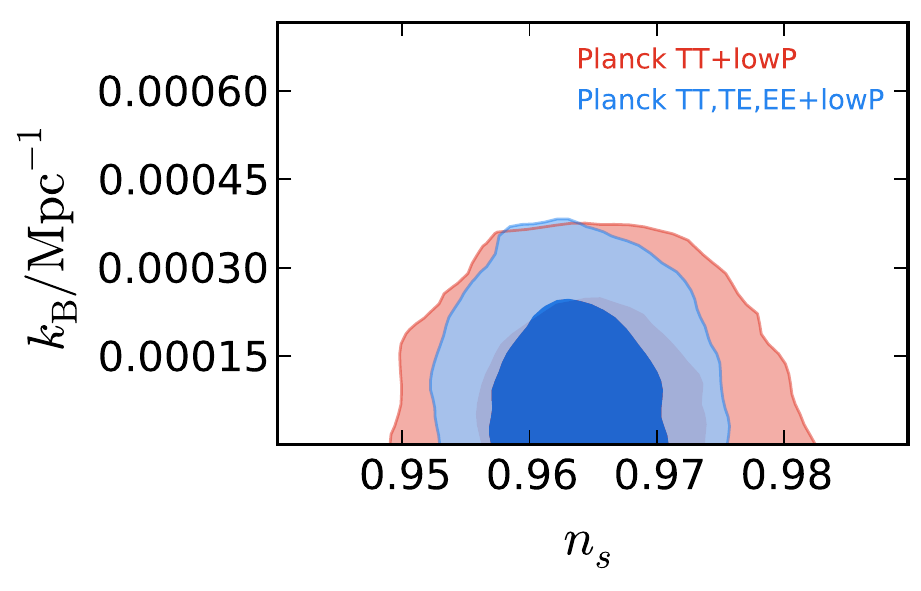}
\includegraphics[width=8.1cm,height=5.5cm]{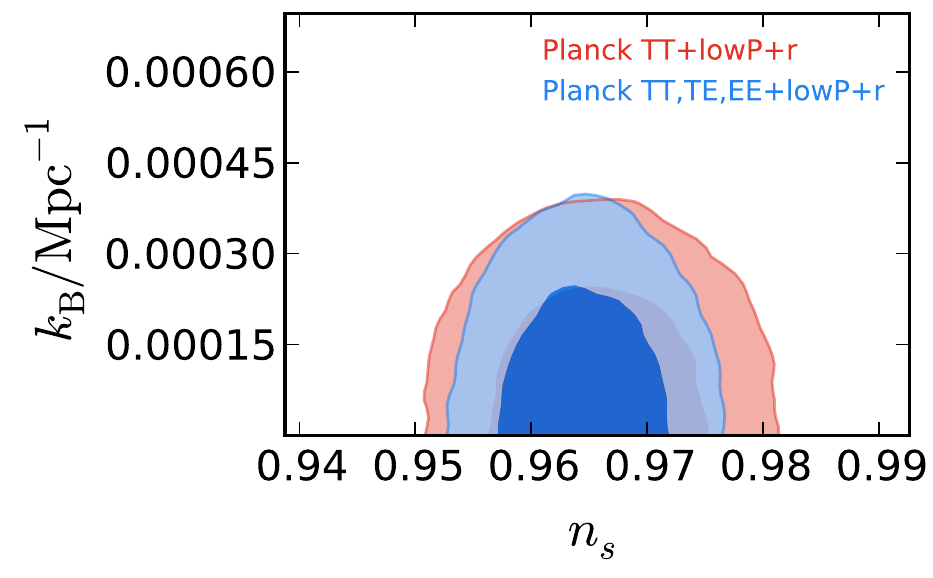}
\caption{The CMB likelihood analysis  in the ($k_B,\; n_s$)-plane with a robust fitting  $n_s \simeq 0.965$.
The observational constraints on $(n_s, k_\text{B}/\text{Mpc}^{-1})$ are obtained at 68\% and 95\% C.L. by using Planck 2015 TT+lowP and TT, TE, EE+lowP data. The upper panel only considers the scalar spectrum,
while the bottom one includes the non-zero tensor contributions.  Note that we set $a_0=1$. } \label{CMB_Likelihood}
\end{figure*}

In particular, we use the high-$l$ CMB temperature power spectrum (TT) and the polarization data (TT, TE, EE) respectively with low-$l$ polarization data (lowP) from Planck2015. In Table.~\ref{bestfit},
 we list the best fit values of the six cosmological parameters and constraints on $k_{\rm B}/a_0$ and $r$ at $95\%$ C.L. for different cosmological models from different data combinations.
 Marginalizing other parameters, we find that $k_\text{B}/a_0$ is constrained by the Planck TT+lowP (Planck TT,TE,EE+lowP) to  [cf. the left panel of Fig.~\ref{CMB_Likelihood}],
\bqn
\frac{k_\text{B}}{a_0} < 3.12 \times 10^{-4}\text{Mpc}^{-1} (3.05 \times 10^{-4}), \;\; {\rm at \; 95\% \; C.L}.\nb\\
\eqn
When we add one more parameter, the tensor-to-scalar ratio $r=A_{(t)}/A_{(s)}$, to include the tensor spectrum, the Planck TT+lowP (Planck TT,TE,EE+lowP) data yields [cf. the right panel of Fig.~\ref{CMB_Likelihood}],
\bqn
\frac{k_\text{B}}{a_0}  < 3.14\times 10^{-4}\text{Mpc}^{-1} (3.14 \times 10^{-4}),\;\; {\rm  at \;95\% \; C.L.}\nb\\
\eqn
These upper bounds shows that the observational constraints on the pre-inflationary dynamics effects are robust to different data sets (without/with polarization data included) and whether the tensor spectrum is included.

\begin{figure*}
\includegraphics[width=14.5cm]{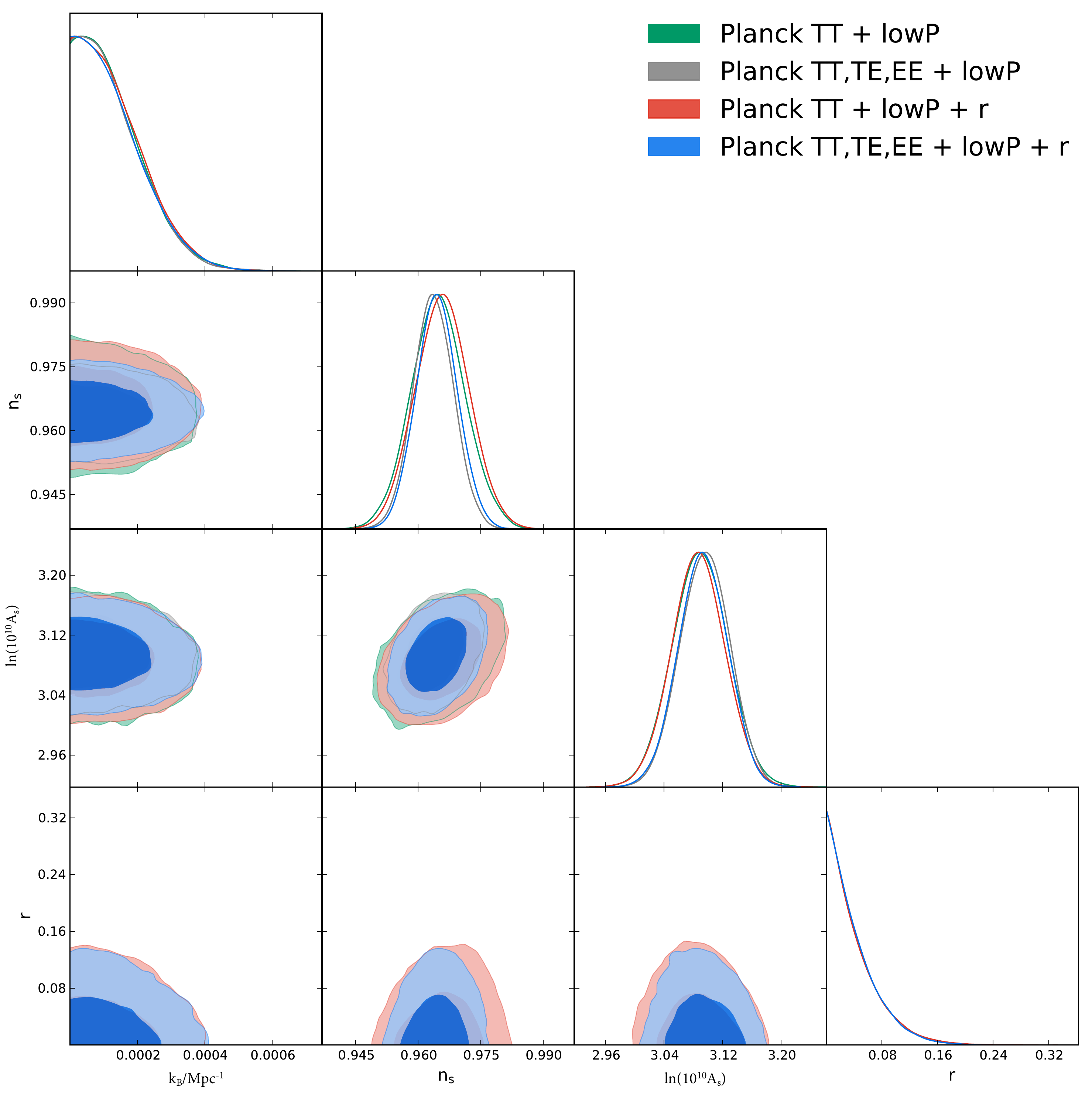}
\caption{Observational constraints on various pairs of parameters ($68\%$ and $95\%$ contour lines) and the probability distributions for $\ln(10^{10} A_s)$, $n_s$, $k_{\rm B}/a_0$, and $r$ by using Planck 2015 data.
Note that in the numerical simulations we set $a_0=1$.
} \label{triA}
\end{figure*}

In Fig.~\ref{triA} we show constraints on various pairs of the cosmological parameters and their respective probability distributions for the CosmoMC runs described above and for the results from Planck 2015 data.
We notice that the colored curves which represent the probability distributions of $k_{\rm B}/a_0$ are almost perfectly superposed, which strongly indicates again that the constraints on $k_{\rm B}$ derived in this paper are robust.

Using the relation
\bqn
\frac{k_\text{B}}{a_0} = \sqrt{\frac{\gamma_{\rm B}}{3}} \frac{a_{\rm B}}{a_0} m_{\rm Pl}=\sqrt{\frac{\gamma_{\rm B}}{3}} m_{\rm Pl} e^{-N_{\rm tot}},
\eqn
where $N_{\rm tot} \equiv \ln{(a_0/a_{\rm B})}$ denotes the total e-folds from the quantum bounce until today, the above upper bounds on $k_{\rm B}/a_0$ can be translated into the constraint on the  total $e$-folds
$N_{\text{tot}} $ as
\bqn
N_{\rm tot}>141  \;\; (95\% {\rm C.L.}),
\eqn
where we have taken $\rho_\text{c}=0.41 m_\text{Pl}^4$. This in turn leads to a lower bound $\delta N_* > N_{\text{tot}}-N_*-N_\text{after}$, where
$\delta N_* \equiv \ln{(a_*/a_B)}$, $N_* \equiv \ln{(a_{\text{end}}/a_*)}$, and $N_\text{after} \equiv \ln{(a_0/a_{\text{end}})}$, where  $a_*$ denotes
 the expansion factor at the moment that our current horizon exited the Hubble horizon  during the slow-roll inflation, and $a_{\text{end}}$ is that at the end of inflation. Taking $N_*\simeq 60\simeq N_\text{after}$,
we find
\bq
\delta N_* \gtrsim 21.
\eq
It should be noted that  our above results are based on the hypothesis: (1) The universe is filled with a scalar field with an inflationary potential $V(\phi)$ and the background evolution is dominated by the kinetic energy of the inflaton  at the
quantum bounce. (2) We impose the BD vacuum initial conditions  at the contracting phase right before the quantum bounce.

\section{Summary and Ourlook}
\renewcommand{\theequation}{7.\arabic{equation}} \setcounter{equation}{0}
\lb{conclusions}

In this paper, we have provided a detailed and systematical study of the  evolutions of the background and linear scalar and tensor perturbations of a flat FLRW
universe in the framework of  {\em the dressed metric approach} of LQC \cite{agullo_quantum_2012, agullo_extension_2013, agullo_pre-inflationary_2013}.
A remarkable feature  is the replacement of the big bang singularity by a quantum bounce \cite{agullo_quantum_2012,agullo_extension_2013,agullo_pre-inflationary_2013}.
In addition, slow-roll inflation is an attractor in the phase space of the initial conditions and most likely to happen with generic initial conditions \cite{agullo_detailed_2015, ashtekar_quantum_2017}.

To study the universal properties of the pre-inflationary dynamics in the framework of  {\em the dressed metric approach}, in this paper we have mainly focused on models, the dynamics of which is dominated
by the kinetic energy of the inflaton at the quantum bounce, i.e.,
\bq
\lb{BCs}
 \frac{1}{2}  \dot\phi^2(t_B) \gg  V(\phi(t_B)),
\eq
 as these models are the only ones found so far that inevitably lead to the slow-roll inflation in the late stage of the evolution of the universe
\cite{bonga_inflation_2016,bonga_phenomenological_2016}.

\subsection{Evolution of the Background}

For all the models that satisfy the initial condition   (\ref{BCs}), we have found the following for the evolution of the background of the flat FLRW universe:

\begin{itemize}

\item The evolution of the universe prior to the preheating can be divided universally  into three different phases (see Fig.~\ref{EoS}):
 \bq
 \lb{phases}
 {\mbox{\em
 bouncing,  transition,  and   slow-roll inflation}}.\nb
 \eq

During the bouncing phase, the evolution of the universe is  dominated by the kinetic energy of the inflaton, so the equation of state $w(\phi)$ defined by Eq.~(\ref{EOS}) remains
practically $w(\phi) \simeq 1$ during this whole phase. However, at $t/t_\text{Pl} \simeq 10^4 \sim 10^5$, the kinetic energy suddenly decreases, and $w(\phi)$ soon decreases from $w(\phi) \simeq 1$ to
$w(\phi) \simeq - 1$. This transition phase is very short in comparison with the other two phases. Afterwards,    the universe enters an accelerating phase, where $\ddot{a} > 0$. At the beginning of this phase, the
absolute value of the slow-roll parameter $\epsilon_H$ defined by Eq.~(\ref{epsilonH}) is still large, but soon settles down to zero, whereby the slow-roll inflation starts, as  shown explicitly in Section II.

\item During the bouncing phase, {\em the evolutions of the expansion factor $a(t)$ and the scalar field $\phi$ are independent of the inflationary potential, and can well be approximated by the analytical solutions
of Eq.~(\ref{scalar_analytical}) and (\ref{phi_sol}), respectively}.

The main reason is that  the potential  $V(\phi)$ remains very small and the kinetic energy is  dominant during this whole phase. For example, for the potential $V(\phi) = V_0\phi^2$,  we find that
$V(\phi)/m_\text{Pl}^4 \in(2\times 10^{-11}, 4.5\times 10^{-11})$; for $n=1/3$,  $V(\phi)/m_\text{Pl}^4 \in(9 \times 10^{-12},
1.2\times 10^{-11})$; and for the Starobinsky potential,  we have $V(\phi)/m_\text{Pl}^4 \in (7\times 10^{-13}, 7.3\times 10^{-13})$. Clearly, in this whole phase, we can safely ignore the effects of
the potential and set it to zero, such that the solution Eq.(\ref{scalar_analytical}) for $a(t)$ immediately follows.

\item During the transition phase, the expansion factor $a(t)$ and scalar field $\phi(t)$ can be well approximated  analytically by Eq.~(\ref{phieq_expansion}) with $a_c, \; \phi_c$ and $\dot\phi_c$
being given by Eqs.~(\ref{ac_2a}) -(\ref{phidotc_2a}), where $t_c$ is the moment
when the kinetic energy is equal to the potential energy, so that $w(\phi_c) = 0$ [cf. Fig. \ref{EoS}]. Then, the e-folds $N_c [\equiv \ln(a_c/a_B)]$ can be calculated analytically  by Eq.~(\ref{Nc}).

From the moment $t_c$ until the beginning of the slow-roll inflationary phase, denoted by the time $t_i$, the e-fold that the universe expands is  $\Delta N \equiv N_i -N_c = \ln(a_i/a_c) \simeq 0.1$,
as shown both numerically and analytically in Tables I-IV, where $N_i$ is analytically given by Eq.~(\ref{N_i}).

 Once the universe enters the slow-roll inflationary phase, the e-folds $N_\text{inf} [\equiv \ln(a_\text{end}/a_i)]$ from $t_i$ to the end of the inflation can be calculated by the standard formula given by Eq.~(\ref{Ninf_general}).

 \item To be complete, in Section II we have also studied the evolution of the universe in which the total energy of the inflaton is dominated by the potential energy $V(\phi)$  at the quantum bounce,
 that is, $\frac{1}{2}  \dot\phi^2(t_B) \ll V(\phi(t_B))$, and in particular,  for the Starobinsky potential
 we have shown  that slow-roll inflation never happens, which is consistent with the results obtained in \cite{bonga_inflation_2016,bonga_phenomenological_2016}.

 \end{itemize}

As noticed previously,   the modified Friedmann equation (\ref{friedmann}) and the Klein-Gordon equation (\ref{klein-gordon}) are also derived in {\em the deformed algebra approach} 
\cite{BHKS09,MCBG12,CMBG12,CBGV12,CLB14,BBCGK15}. So, all the results obtained for the evolution of the background  in this paper are equally applicable to this approach. In fact, 
they are applicable to any theory of gravity in which the background is governed by  Eqs.(\ref{friedmann}) and   (\ref{klein-gordon}).

 \subsection{Perturbations and Observational Constraints}

With the above understanding of the background evolution of the universe, we then turned to study the linear scalar and tensor perturbations during the above mentioned three different phases, and mainly found the following:

\begin{itemize}

\item During the bouncing and transition phases, the potential term $U(\phi)$ given by Eq.~(\ref{up}) for the scalar perturbations of Eq.~(\ref{eom_scalar}) is always negligible in comparison with the term $a''/a$, as shown
 in Fig. \ref{UvsA}. As a result, during these two phases, the scalar and tensor perturbations satisfy the same equation of motion, given by Eq.~(\ref{eomt}).

\item During the  bouncing phase, the effective potential   $\mathscr{V}(\eta)[\equiv a''/a]$ given by Eq.~(\ref{potential}) (for both scalar and tensor perturbations) can be well approximated by the PT potential (\ref{PT}),
for which {\em an analytical expression for  the mode functions $\mu_k^{(s, t)}$ exists, and is given by Eq.~(\ref{sol_PT})}.

 During the transition and slow-roll inflationary  phases,  the mode functions $\mu_k^{(s, t)}$ are also known analytically, and are given, respectively, by Eqs.~(\ref{transition_solution}) and (\ref{hankel}).

 \item After matching the three solutions for the mode functions $\mu_k^{(s, t)}$ together, the coefficients $\alpha_k$ and $\beta_k$ of the mode functions during the slow-roll inflationary phase are given by
 Eqs.~(\ref{alphak}) and
 (\ref{betak}) in terms of $a_k$ and $b_k$ of the mode functions appearing in Eq.~(\ref{sol_PT}), which  describe the pre-inflationary dynamics of the mode functions during the bouncing phase, and will be determined by the
 initial conditions. In general,  it is expected that $\beta_k \not= 0$ at the onset of the slow-roll inflation, that is, {\em particles are generically created due to the pre-inflationary dynamics}. Note that in GR we normally impose
 the BD vacuum, $\left(\alpha_k^{\rm GR}, \; \beta_k^{\rm GR}\right) = (1, 0)$,  at the onset of inflation \cite{baumann_tasi_2009}.

\item It is exactly because the particle creations during the pre-inflationary phases that {\em the Bogoliubov coefficients $\alpha_k$ and $\beta_k$ in general depend on the wavenumber $k$ via Eqs.~(\ref{alphak}),  (\ref{betak}) and (\ref{acs})}.
It further implies that the power spectra, both scalar and tensor, will  in general depend on $k$, that is, the power spectra are no longer scale-invariant, which provides a great opportunity to test the theory observationally.
Such a dependence can be seen from Fig. \ref{length}, from which we can see that at the quantum bounce the modes with $k > k_B$ are all within the Hubble horizon, while the ones with $k < k_B$ are all outside  the Hubble horizon.
Depending on the ratio $k/k_B$, the modes have different dynamics during the pre-inflationary phases, although after the moment $t_s$ all the modes will be inside  the horizon.
Certainly, this dependence cannot be very strong. Otherwise, it will be in conflict with current observations that show that the spectra are almost scale-invariant \cite{planck_collaboration_planck_2014-1,planck_collaboration_planck_2015-4}.

\item To determine $a_k$ and $b_k$, we have considered two commonly used sets of initial conditions in Section V. One is the BD vacuum state imposed in the contracting phase \cite{LB13}, right prior to the bounce, as shown in Fig.  \ref{length}.
For $t < - t_s$, all the modes are within the Hubble horizon, so   the BD vacuum  state is a natural choice in this case. The other set is imposed at the bounce \cite{ashtekar_loop_2010,CK11}. As shown above, at this moment,
some modes are inside the Hubble horizon, and some are outside of it. So, in this case the BD vacuum  state is no longer a choice. Instead, we imposed  {\em the fourth-order adiabatic vacuum state}    \cite{agullo_quantum_2012, agullo_extension_2013, agullo_pre-inflationary_2013}. Within the validity of the latter, however, we found that they essentially lead to the same results of the parameters $a_k$ and $b_k$, all of which are given by Eq.~(\ref{akbk}).

\item It is remarkable to note that the parameters $a_k$ and $b_k$ for both scalar and tensor perturbations are all  given by Eq.~(\ref{akbk}). 
Then, the power spectra of the scalar and tensor perturbations are proportional to
the same factor $|\alpha_k+\beta_k|^2$, as shown by Eqs.~(\ref{scalar_pw}) and (\ref{tensor_pw}). As a result, {\em the ratio $r$ between the tensor and scalar perturbations is the same as that given in GR}.

\item  In addition, after the effects of the pre-inflationary dynamics are taken into account, as shown above, the power spectra now are generically scale-dependent.
 Therefore, by measuring the $k$-dependence of the power spectra, one can directly test LQC. Fitting the power spectra to the Planck 2015 temperature (TT+lowP) and polarization (TT,TE,EE+lowP) data,
 we found the total e-folds from the quantum bounce to the current time must be,
 \bq
 N_\text{tot}=\ln\left(\frac{a_0}{a_\text{B}}\right) > 141,\;  ({\rm at 95\% C.L.}),
 \eq
 that is, to be consistent with observations, the universe must have expanded at least $132$ e-folds from the bounce until now, so the scale-dependent features
  are well diluted. Otherwise, it will be in conflict with current  observations.

 \end{itemize}

 \subsection{Outlook}

With the above main results, we would like to note the following.
%
%
%
The first issue is about the initial conditions of the perturbations. As we explained in Sec.~\ref{initial_conditions}, this is a subtle issue during the bouncing phase because in
general there is not a preferred initial state for a quantum field in arbitrarily curved spacetime \cite{winitzki_cosmological_2005, agullo_unitarity_2015, agullo_preferred_2015}
(see also \cite{ashtekar_initial_2017,MBS17} for a recent discussion). The general solution of the Bogoliubov coefficients $\alpha_k$ and $\beta_k$  given by Eqs.~(\ref{alphak}) and (\ref{betak})
are  not limited to any specific set of the initial conditions. Instead, they are given in terms of the two parameters $a_k$ and $b_k$, which are uniquely determined by initial conditions.
So, in principle one can use these expressions  to study the effects of the initial conditions.   Recently, Ashterkar and Gupt \cite{ashtekar_quantum_2017,ashtekar_initial_2017} have proposed other
 new initial states, some of which can explain the observed scalar spectrum suppression at large scales and thus fits the observations better than the standard power-law
spectrum. It is also interesting to consider these initial conditions and its observational implications analytically, by using the formulas presented in this paper.


Since  inflationary models with a single scalar field have been extensively studied in the framework of GR \cite{baumann_tasi_2009,MRV}, it would be very interesting to see if the effects of the pre-inflationary
dynamics in other inflationary models could lead to any  observational signatures,  in addition to the ones found so far.  In the standard  inflationary models, one in
general imposes the initial conditions for the background evolution simply by hand at the onset of the slow-roll inflation, while in LQC, as we discussed in Sec.~\ref{numerical},
the initial values of $\phi_\text{B}$ and $\dot \phi_\text{B}$
have to be chosen so that $H(t_\text{B})=0$ at the bounce, which imposes one additional constraint, so that    only one degree of freedom for the scalar field is left.
Moreover, as  shown in Sec.~\ref{MCMC},
current observations already impose  constraints on the total e-folds $N_\text{tot}$, so the $k$-dependent features are well diluted, in order to be consistent with current observations.
 Because the analytical formulas developed in this paper
for  both background and perturbation evolutions  are  general, it would be very interesting  to apply them to other   inflationary models in order to see  if further constraints can be obtained.


  In addition,  when we considered the Starobinsky potential, our analysis  was limited to the Einstein frame. In the framework of GR, it was shown that they are equivalent.
 However,  whether this is also true or not    in LQC is still an open question. According to \cite{artymowski_comparison_2013}, in general the Einstein  and Jordan frames are no longer
 equivalent  at the quantum level. Thus, it is interesting to explore the Starobinsky model and its corresponding cosmological perturbations   directly in the
Jordan frame,  based on the quantization  proposed in \cite{zhang_extension_2011, zhang_loop_2013}.


 Yet, once particles are created, it is usually expected that  non-Gaussianity will also raise \cite{baumann_tasi_2009,MRV}. Therefore, it would be very interesting to study
 non-Gaussianity of primordial scalar and tensor perturbations by using the analytical solutions presented  in this paper. Numerical studies were already carried out in \cite{Agullo15}.

 We hope to return to the above issues soon in other occasions.


\section*{Acknowledgements}

We would like to thank A. Barrau, S. Brahma, J. Lewandowski, A. Marciano and P. Singh for valuable comments and suggestions. 
Part of the work was reported in {\em the Informal Spring Meeting on Quantum Gravity,
Shanghai 2017}, Fudan University, Shanghai, May 11-12, 2017, and  {\em LOOPS' 17}, Warsaw, Poland, July 3 - 7, 2017. This work is supported in part by
the Chinese NSF Grants, Nos. 11375153 (A.W.), 11675145(A.W.), 11675143 (T.Z.), and 11105120 (T.Z.). K.K. was supported by 
the Baylor University Summer Sabbatical Programme.

\appendix

\section{Derivation of $t_c$ for the Starobinsky potential}
\lb{teqST}
\renewcommand{\theequation}{A.\arabic{equation}} \setcounter{equation}{0}

For the Starobinsky potential, the time $t_c$ is determined by the equation
\bqn
\frac{1}{2}\dot \phi_c^2 = \frac{3}{32\pi }M^2 M_\text{Pl}^2 \left(1-e^{-\sqrt{2/3}\phi_c/M_\text{Pl}}\right)^2,
\eqn
from which we find,
\bqn
\pm \dot \phi_c = \sqrt{\frac{3}{16 \pi}}M m_\text{Pl}\left(1-e^{-\sqrt{16 \pi /3} \phi_c/m_\text{Pl}}\right).
\eqn
In the following we divide our discussions into two parts:  $\dot \phi_c >0$ and  $\dot \phi_c <0$.

\subsection{ $\dot \phi_c>0$}
In this case, we  find
\bqn
\dot \phi_c = \sqrt{\frac{3}{16 \pi}}M m_\text{Pl}\left(1-e^{-\sqrt{16 \pi /3} \phi_c/m_\text{Pl}}\right),
\eqn
where we only consider the case  $\phi_c>0$, so   the right-hand side of the above equation is always positive
 \footnote{Even when we consider a negative value of $\phi_\text{B}$ as initial input, the scalar field will finally reach a positive value at $t=t_c$.}, and $\phi_c$ and $\dot \phi_c$ are given, respectively,  by
\bqn
\phi_c &\simeq&  \phi_\text{B} + \frac{m_\text{Pl}}{2\sqrt{3\pi}} \ln \left(2 \sqrt{\gamma_\text{B}} \frac{t_c}{t_\text{Pl}}\right),\\
\dot \phi_c &\simeq& \frac{m_\text{Pl}^2}{\sqrt{12\pi}} \frac{t_\text{Pl}}{t_c}.
\eqn
Since
\bqn
&&\exp\left(- \sqrt{\frac{16\pi}{3}} \frac{\phi_c}{m_\text{Pl}}\right)\nb\\
&&~~~=2^{-2/3} \left(\sqrt{\gamma_\text{B}} \frac{t_c}{t_\text{Pl}}\right)^{-2/3}e^{-\sqrt{16\pi/3} \phi_\text{B}/m_\text{Pl}},\nb\\
\eqn
we have
\begin{widetext}
\bqn
 \frac{m_\text{Pl}^2}{\sqrt{12\pi}} \frac{t_\text{Pl}}{t_c} =  \sqrt{\frac{3}{16 \pi}}M m_\text{Pl}\left[1-2^{-2/3} \left(\sqrt{\gamma_\text{B}} \frac{t_c}{t_\text{Pl}}\right)^{-2/3}e^{-\sqrt{16\pi/3} \phi_\text{B}/m_\text{Pl}}\right].
\eqn
Solving this equation we obtain
\bqn\lb{tc_staro_a}
t_c = \frac{2}{3M}+\frac{e^{-\frac{2 \sqrt{3 \pi } \phi _B}{m_{\text{Pl}}}} \left[\left(\sqrt{36 \gamma_B m_{\text{Pl}}^2 e^{\frac{4 \sqrt{3 \pi } \phi _B}{m_{\text{Pl}}}}-3 M^2}+6 \sqrt{\gamma _B} m_{\text{Pl}} e^{\frac{2 \sqrt{3 \pi } \phi _B}{m_{\text{Pl}}}}\right)^{2/3}+(3M^2)^{1/3}\right]}{2 (9M)^{1/3} \sqrt{\gamma _B} m_{\text{Pl}} \left({\sqrt{36 \gamma _B m_{\text{Pl}}^2 e^{\frac{4 \sqrt{3 \pi } \phi _B}{m_{\text{Pl}}}}-3 M^2}+6 \sqrt{\gamma _B} m_{\text{Pl}} e^{\frac{2 \sqrt{3 \pi } \phi _B}{m_{\text{Pl}}}}}\right)^{1/3}}.
\eqn
\end{widetext}
We can  immediately   see that,  when $\phi_\text{B}$ becomes large, the above expression approaches
\bqn
t_c \rightarrow \frac{2}{3M},
\eqn
which implies that when the value of $\phi_\text{B}$ increases, $t_c$ will approach a value that only depends on the parameter $M$.

\subsection{$\dot \phi_c<0$}

In this case, we find
\bqn
\dot \phi_c = - \sqrt{\frac{3}{16 \pi}}M m_\text{Pl}\left(1-e^{-\sqrt{16 \pi /3} \phi_c/m_\text{Pl}}\right),
\eqn
where we also only consider the case $\phi_c>0$, thus the right-hand side of the above equation is always negative and $\phi_c$ and $\dot \phi_c$ are given, respectively,  by
\bqn
\phi_c &\simeq&  \phi_\text{B} - \frac{m_\text{Pl}}{2\sqrt{3\pi}} \ln \left(2 \sqrt{\gamma_\text{B}} \frac{t_c}{t_\text{Pl}}\right),\\
\dot \phi_c &\simeq& - \frac{m_\text{Pl}^2}{\sqrt{12\pi}} \frac{t_\text{Pl}}{t_c}.
\eqn
Since
\bqn
&&\exp\left(- \sqrt{\frac{16\pi}{3}} \frac{\phi_c}{m_\text{Pl}}\right)\nb\\
&&~~~=2^{2/3} \left(\sqrt{\gamma_\text{B}} \frac{t_c}{t_\text{Pl}}\right)^{2/3}e^{-\sqrt{16\pi/3} \phi_\text{B}/m_\text{Pl}},\nb\\
\eqn
we have
\begin{widetext}
\bqn\lb{equation}
 \frac{m_\text{Pl}^2}{\sqrt{12\pi}} \frac{t_\text{Pl}}{t_c} =  \sqrt{\frac{3}{16 \pi}}M m_\text{Pl}\left[1-2^{2/3} \left(\sqrt{\gamma_\text{B}} \frac{t_c}{t_\text{Pl}}\right)^{2/3}e^{-\sqrt{16\pi/3} \phi_\text{B}/m_\text{Pl}}\right].
\eqn
\end{widetext}
As we have shown  numerically, in order to produce sufficient e-folds ($\ge 60$) during the slow-roll inflation, we have to require $\phi_\text{B} \in (3.61, +\infty)$. This leads to a simplification in the above equation, as
 the second terms in the square bracket can be neglected in comparison to the first term. Then,  we  find
\bqn
t_c \simeq \frac{2}{3M},
\eqn
which implies that when the value of $\phi_\text{B}$ increases, $t_c$ will approach a value that only depends on the parameter $M$. In order to determine the weak dependence of $t_c$ on $\phi_\text{B}$,
we can consider a small perturbation $\delta t_c$. Then,  considering only the first-order expansion in terms of $\delta t_c$, we find
\bqn
\delta t_c = \frac{2}{3M} \left(\frac{4\sqrt{\gamma_\text{B}}m_\text{Pl}}{3M}\right)^{2/3} e^{-\sqrt{16\pi/3} \phi_\text{B}/m_\text{Pl}}.
\eqn
Thus,  finally we obtain
\bqn\lb{tc_staro_b}
t_c = \frac{2}{3M} \left[1+\left(\frac{4\sqrt{\gamma_\text{B}}m_\text{Pl}}{3M}\right)^{2/3} e^{-\sqrt{16\pi/3} \phi_\text{B}/m_\text{Pl}} \right].\nb\\
\eqn

\section{Asymptotic expansion of $|\Gamma(x+i y)|$ for a large $|y|$}
\renewcommand{\theequation}{B.\arabic{equation}} \setcounter{equation}{0}
\lb{gamma_function}

In this appendix we are going to derive the asymptotic behavior of the Gamma function $|\Gamma(x+iy)|$ when $|y|$ is large. Note that the following asymptotic formula of the $\Gamma(z)$ function
at infinity (i.e., $|z| \to +\infty$) is known,
\bqn\lb{Gammaz}
\Gamma(z)=\sqrt{2\pi}e^{-z} z^{z-1/2} \left(\sum_{k=0}^{\infty}\frac{g_k}{z^k}\right),
\eqn
where
\bqn
&&g_0=1,\;\;\;g_1=\frac{1}{12},\;\;\;g_2=\frac{1}{288}, \nb\\
&& g_3 = - \frac{139}{51840},\;\;\; g_4 = - \frac{571}{2488320}, \;\;\;\cdots.
\eqn
Now we are going to use the above formula to calculate $\Gamma(x+iy)$. For this purpose, let us first consider the term $e^{-z}$ with $z=x+iy$, which yields
\bqn
e^{-z}=e^{-x} e^{-iy}.
\eqn
Turning to the term $z^{z-1/2}$, we have
\bqn
(x+iy)^{x-1/2+iy},
\eqn
and so we find,
\bqn
&&(x-1/2+iy)\ln\left(x+iy\right)\nb\\
&&~~~~~~=(x-1/2+iy) \ln\left[\sqrt{x^2+y^2}\left(\cos{\theta}+i \sin{\theta}\right)\right]\nb\\
&&~~~~~~=(x-1/2+iy) \left(\ln{\sqrt{x^2+y^2}}+i\theta \right)\nb\\
&&~~~~~~=(x-1/2)\ln\sqrt{x^2+y^2}- \theta y\nb\\
&&~~~~~~~~~~+i\left[ \theta (x-1/2)+y \ln\sqrt{x^2+y^2}\right],
\eqn
where $\theta\equiv \arccos{\frac{x}{\sqrt{x^2+y^2}}}$. Thus,  we finally get
\bqn
z^{z-1/2}&=&(x^2+y^2)^{\frac{1}{2}\left(x-\frac{1}{2}\right)} e^{-\theta y} \nb\\
&&\times e^{i \left[ \theta (x-1/2)+y \ln\sqrt{x^2+y^2}\right]}.
\eqn
We need to expand the above expression about $y \to \infty $ up to the fourth-order. First, for $(x^2+y^2)^{\frac{x}{2}-\frac{1}{4}}$, we find
\bqn
&&(x^2+y^2)^{\frac{x}{2}-\frac{1}{4}}\nb\\
&&~~~~ =y^{x-\frac{1}{2}} \left(1+\frac{x^2}{y^2}\right)^{\frac{x}{2}-\frac{1}{4}}\nb\\
&&~~~~=y^{x-\frac{1}{2}} \Bigg[1+\left(\frac{x^3}{2}-\frac{x^2}{4}\right)\frac{1}{y^2}\nb\\
&&~~~~~~~~ + \left(\frac{5x^4}{32}-\frac{3x^5}{8}+\frac{x^6}{8}\right)\frac{1}{y^4}+\mathcal{O}\left(\frac{1}{y^4}\right)\Bigg].\nb\\
\eqn
Considering $e^{-\theta y}$, first we have
\bqn
\theta&=&\arccos\left(\frac{x}{\sqrt{x^2+y^2}}\right)\nb\\
&=& \arccos\left(\frac{x}{y}\frac{1}{\sqrt{1+\frac{x^2}{y^2}}}\right)\nb\\
&=& \frac{\pi}{2}-\frac{x}{y}+\frac{x^3}{3} \frac{1}{y^3}-\frac{x^5}{5}\frac{1}{y^5}+\mathcal{O}\left(\frac{1}{y^7}\right),
\eqn
thus,  we get
\bqn
-\theta y =-\frac{\pi y}{2} +x - \frac{x^3}{3} \frac{1}{y^2}+\frac{x^5}{5} \frac{1}{y^4}+\mathcal{O}\left(\frac{1}{y^6}\right),
\eqn
and
\bqn
e^{-\theta y}&=&e^{-\frac{\pi y}{2}+x} e^{-\frac{x^3}{3}\frac{1}{y^2}}\nb\\
&\simeq&e^{-\frac{\pi y}{2}+x} \left[1-\frac{x^3}{3}\frac{1}{y^2}+\left(\frac{x^5}{5}+\frac{x^6}{18}\right)\frac{1}{y^4}\right].\nb\\
\eqn

For the terms in the bracket of Eq.~(\ref{Gammaz}), if we only consider the first four terms in the expansion
\footnote{The fifth or higher terms only contribute to the order $\mathcal{O}(1/y^6)$ in the expansion.},
we have
\begin{widetext}
\bqn
\left(\sum_{k=0}^{\infty}\frac{g_k}{z^k}\right)&=&g_0+\frac{g_1}{z}+\frac{g_2}{z^2}+ \frac{g_3}{z^3} + \frac{g_4}{z^4}+\mathcal{O}\left(\frac{1}{z^5}\right)\nb\\
&=& \Bigg[g_0^2 + \frac{g_1^2-2 g_0 g_2+ 2 g_0 g_1 x}{x^2+y^2}+ \frac{g_2^2-2g_1g_3+2g_0g_4+ 2(g_1g_2-3 g_0 g_3)x+4 g_0 g_2 x^2}{(x^2+y^2)^2} \nb\\
&&\;\;\; + \frac{g_3^2-2 g_2 g_4 + 2(g_2 g_3- 3 g_1 g_4) x+ 4 (g_1 g_3- 4 g_0 g_4) x^2 + 8 g_0 g_3 x^3}{(x^2+y^2)^3}\nb\\
&&\;\;\; +\frac{g_4^2+ 2 g_3 g_4 x+ 4 g_2 g_4 x^2 +8 g_1 g_4 x^3 + 16 g_0 g_4 x^4}{(x^2+y^2)^4}\Bigg]^{1/2} e^{i \vartheta}\\
&\simeq & \Bigg\{g_0+ \left(\frac{g_1^2}{2 g_0}-g_2 + g_1 x\right)\frac{1}{y^2} \nb\\
&&+\left[g_1 x^3+\left(\frac{g_1^2}{g_0}-3 g_2\right) x^2+\left(\frac{g_1^3}{2 g_0^2}-\frac{2 g_1 g_2}{g_0}+3 g_3\right) x+\frac{g_1^4}{8 g_0^3}-\frac{g_1^2 g_2}{2 g_0^2}
+\frac{g_1 g_3}{g_0}-g_4\right]\frac{1}{y^4}+\mathcal{O}\left(\frac{1}{y^6}\right) \Bigg\}e^{i\vartheta}.\nb\\
\eqn

Finally,  combining all the above expansions together, we find
\bqn\lb{gamma}
|\Gamma(x+iy)|=\sqrt{2\pi} |y|^{x-\frac{1}{2}} e^{-\frac{\pi y}{2}}  \left[1+\left(\frac{x^3}{6}-\frac{x^2}{4}+\frac{x}{12}\right)\frac{1}{y^2}
+ \left(\frac{x^6}{72}-\frac{11 x^5}{120}+\frac{49 x^4}{288}-\frac{5 x^3}{48}+\frac{x^2}{288}+\frac{x}{120}\right) \frac{1}{y^4}+\mathcal{O}\left(\f{1}{y^6}\right)\right].\nb\\
\eqn
\end{widetext}

\end{document}